\documentclass[aps,epsf]{revtex4}

\usepackage{amsmath}
\usepackage{amssymb}
\usepackage{graphicx}

\newcommand{\beq}{\begin{equation}}
\newcommand{\eeq}{\end{equation}}
\newcommand{\beqa}{\begin{eqnarray}}
\newcommand{\eeqa}{\end{eqnarray}}
\newcommand{\bmat}{\begin{displaymath}}
\newcommand{\emat}{\end{displaymath}}

\newcommand{\eq}[1]{Eq.~(\ref{#1})}

\newcommand{\Dy}{\Delta}
\newcommand{\Dyeff}{(\Delta_{\rm eff})}
\newcommand{\Dyeffs}{\Delta_{\rm eff}}

\newcommand{\lan}{\langle}
\newcommand{\ran}{\rangle}

\newcommand{\tav}[1]{\left\lan #1 \right\ran}
\newcommand{\sav}[1]{\left [\hspace*{-.13cm}\left| #1 \right|\hspace*{-.13cm}\right ]}
\newcommand{\lsav}{\Biggl [\hspace*{-.13cm}\Biggl | }
\newcommand{\rsav}{\Biggr |\hspace*{-.13cm}\Biggr ]_{\rm av} }

\newcommand{\Tc}{T_{\rm c}}

\newcommand{\tf}{\tilde{f}}
\newcommand{\ty}{\tilde{y}}
\newcommand{\tY}{\tilde{Y}}
\newcommand{\tpsi}{\tilde{\psi}}

\newcommand{\hw}{h_{\rm w}}
\newcommand{\hc}{h_{\rm c}}
\newcommand{\hs}{h_{\rm s}}

\newcommand{\tO}[1]{O_{#1}\frac{\partial}{\partial O_{#1}}}

\newcommand{\spsum}{\sideset{}{'}{\sum}_{\rm SP}}
\newcommand{\ssum}{\sideset{}{'}{\sum}}
\newcommand{\nt}{n_{\rm tot}}
\newcommand{\spav}[1]{\left \langle #1 \right \rangle_{\rm SP}}

\begin{document}

\title{Step-wise responses in mesoscopic glassy systems: a mean field approach}

\author{Hajime Yoshino$^{1,2}$ and Tommaso Rizzo$^{3,4}$}

\affiliation{
$^1$Laboratoire de Physique Th{\'e}orique  et Hautes {\'E}nergies, Jussieu, 
5\`eme {\'e}tage,  Tour 25, 4 Place Jussieu, 75252 Paris Cedex 05, France\\
$^2$Department of Earth and Space Science, Faculty of Science,
Osaka University, Toyonaka 560-0043, Japan\\
\small $^{3}$Laboratoire de Physique Th\'{e}orique de l'Ecole
Normale Sup\'{e}rieure, 24 rue Lhomond, 75231 Paris, France\\
$^4$"E. Fermi" Center, Compendio del Viminale, Via Panisperna 89 A, 00184, Rome Italy
\\
{\tt yoshino@ess.sci.osaka-u.ac.jp,  tommaso.rizzo@inwind.it}
}

\begin{abstract}
We study statistical properties of peculiar responses in glassy systems
at mesoscopic scales based on a class of mean-field  spin-glass models
which exhibit 1 step replica symmetry breaking. Under variation of a
generic external field, a finite-sized sample of such a system exhibits
a series of step  wise responses which can be regarded as a finger print
of the sample. We study in detail the statistical properties
of the step structures based on a low temperature expansion approach
and a replica approach.
The spacings between the steps vanish
in the thermodynamic limit so that arbitrary small but finite 
variation of the field induce infinitely many level crossings
in the thermodynamic limit leading to a static chaos effect which
yeilds a self-averaging, smooth macroscopic response. 
We also note that there is a strong analogy 
between the problem of step-wise responses in glassy systems 
at mesoscopic scales and intermittency in turbulent flows due to shocks.
\end{abstract}

\maketitle

\section{Introduction}
Glassy systems sometimes exhibit anomalously large macroscopic response 
to external perturbations. In the case of spin-glasses, the most well 
known effect is the so called {\it anomaly}: field cooled magnetization 
is significantly larger than zero field cooled magnetization \cite{Nagata79}. 
Because of the anomaly, aging effects clearlly appear in the magnetization
in spin-glasses \cite{Vincent,Nordblad}.
Another striking effect is the so called {\it rejuvenation}. 
For example spin-glasses aged (relaxed) for a long time can rejuvenate, 
i.~e. restart aging, through small changes of the temperature 
\cite{rejuvenation-temperature} or of the applied magnetic field 
\cite{rejuvenation-field}. Similar phenomena have been observed 
in polymer glasses \cite{sturik,Waldron,bellon,Fukao}, 
relaxor ferro-electrics \cite{relaxor-ferroelectirc},
deutron glass \cite{deutron-glass} and other glassy systems.

In order to understand these anomalous macroscopic effects, it is 
desirable to study what is going on at {\it mesoscopic} length and 
time scales. Indeed there is a class of phenomenological theories which
{\it assumes} that mesoscopic excitations called droplets dictate physical 
xoproperties of glassy systems \cite{BM87,FH88,FH91,KTW-scaling}. 
It remains as a challenge to find these droplets directly by 
experimental and theoretical studies. To this end,
one must find sensible ways to work directly at mesoscopic
scales escaping from the inevitable {\it self-averaging} mechanism 
working at larger scales.

Experimental studies of magnetoresistence in mesoscopic spin-glass samples \cite{weissman} have proved that magnetoresistence provides a {\it fingerprint} of a given sample reflecting its frozen-in spin-pattern. Quite interestingly changes
of statistical weights of a few low lying states under magnetic field
has been observed \cite{weissman-field}, which suggests level crossings
among low lying excited states due to the variation of the magnetic field.
 Mesoscopic measurements 
also appear promising for other glassy systems such as vorticies in 
super-conductors\cite{Bolle} and polymer glasses \cite{russel}.

To make a step forward from a theoretical side, we study in the present 
paper statistical properties of responses in a class of mean field 
systems of {\it finite sizes} $N$. A great advantage of the mean field 
approach is that it allows us to obtain results analytically 
from a given microscopic hamiltonian. It will  gives us a 
generic {\it mean-field} picture for peculiar properties at mesoscopic 
scales in glassy systems. The latter may provide us hints 
to develop {\it droplet pictures} for broader classes of glassy systems.

Somewhat surprisingly statistical properties of finite sized systems 
have not been explored extensively in the studies of mean-field models. 
The present work is motivated by an interesting numerical observation 
by Kirkpatrik and Young \cite{KY} who showed that the magnetization 
of a finite sized sample of the Sherrington-Kirkpatric (SK) model,
that is given by a certain realization of random interaction bonds
between the spins, grows in a step-wise manner as a function of 
the applied magnetic field. Note that in the thermodynamic limit 
the magnetization curve (per spin), 
which is a thermodynamic quantity, 
must converge to a unique limit independent samples.
Later it was suggested by Young, Bray and Moore \cite{YBM}
that the fact that linear-susceptibilities 
are non self-averaging in mean-field models reflects 
the step-wise responses. Quite interestingly similar 
observation  has been done numerically in directed polymer 
in random media \cite{M90} by M\'{e}zard,
suggesting that the step-wise response is not a pathology only found
in mean-field models. Our work is also motivated by a seminal work 
by Krzakala and Martin (KM) \cite{KM} who considered an extended 
version of Derrida's random-energy-model (REM) \cite{REM} to analyze 
chaotic resuffling of low lysing states under variation of a generic 
external field. An important basis of our present work is the idea of 
generalized complexity introduced in our previous work \cite{RY}
which is motivated by the work by KM.

In the present paper we focus on a class of mean-field spin-glass models
which exhibit 1 step replica symmetry breaking (RSB), which is analytically
and physically more tractable than the systems with full RSB
such as the SK model mentioned above. By a combination 
of a low temperature expansion approach and a replica approach we analyze 
statistical properties of the step-wise responses of finite sized systems.
We will identify three characteristic scales for the steps, namely
its height (per spin) $\Dy/ \sqrt{N}$ , width $\hw \sim T/\Dy \sqrt{N}$ 
at temperature $T$ and spacing $\hs \sim T_{c}/\Dy \sqrt{N}$ with $T_{c}$
being the critical temperature. In the thermodynamic limit $N \to \infty$, 
the steps become invisible leading to apparently smooth macroscopic response.

The static step-wise response reflects level crossings among low lying 
metastable-states, which are the solutions of the Thouless-Anderson-Palmer
(TAP) equation \cite{tap} in the mean-field models,
under variation of the external field. In the systems 
with 1 step RSB, the metastable states have zero overlap with respect to each
other meaning that (spin) configurations are completely different on different
metastable states. Thus variation of the external field completely
change the energy-landscape, called the static chaos effect 
\cite{MBK,BM87,FH88,kondor}.
It is worth to note that the static chaos effect,
which appear at macroscopic scales \cite{P83,RY}, emerge as step-wise, 
intermittent responses at mesoscopic scales. 
Actually step-wise response itself is not at all 
new in frustrated systesm which oftenly exibit sequence of
phase transitions. Most well known example may be the Devil's 
stair cases in the magnetization curve of 
the ANNI (axial-next-nearest-neighbour Ising) models \cite{selke}.
In this respect, a distinct feature of the case of static chaos 
effect is that there is a {\it continuous} sequence of steps 
(phase transitions) in the thermodynamic limit. It is also amusing to note
that there is an intimate analogy among the present problem of step-wise 
responses in glassy systems at mesoscopic scales,
jerky effective energy landscape of pinned elastic manifolds
\cite{BBM}, 
and intermittency in turbulent flows due to shocks \cite{Kida,BMP,BM}, 
as we explain in the end of the present paper.

The organization of the paper is as follows.
In sec.~\ref{sec-step-wise-response} we introduce our models and
sketch basic features of their step-wise responses.
In sec.~\ref{sec-gc} we revisit the generalized 
complexity introduced in \cite{RY}. Based on the latter we first discuss
macroscopic responses from a general point of view. Then we analyze 
statistical properties of low lying metastable states which will serve
as a basis for the analysis of mesoscopic responses.
In sec.~\ref{sec-lowt-expansion} we analyze statistical properties
of the mesoscopic responses by a low temperature expansion scheme.
Then in sec.~\ref{sec-replica} we analyze the problem again by 
a replica approach. In sec.~\ref{sec-connection} we discuss connections
between our problem and 
related problems including the intermittency in tubulent flows.
In sec.~\ref{sec-conclusions} we summarize our result and discuss 
some perspectives. In the appendices we report some details.

\section{Step-wise responses in finite sized mean-field models}
\label{sec-step-wise-response}

In the present paper we consider a generic class of mean field spin-glass models which exhibit 
static glass transitions with 1 step RSB.
By now it is well known that the generic phenology of such 
a class of models capure quite well static and dynamical properties 
of real glassy systems \cite{p-spin,BCKM,CC}. 

\subsection{p-spin mean-field spin-glass model}
\label{subsec-p-spin}

\begin{figure}[t]
\begin{center}
\includegraphics[width=0.5\columnwidth]{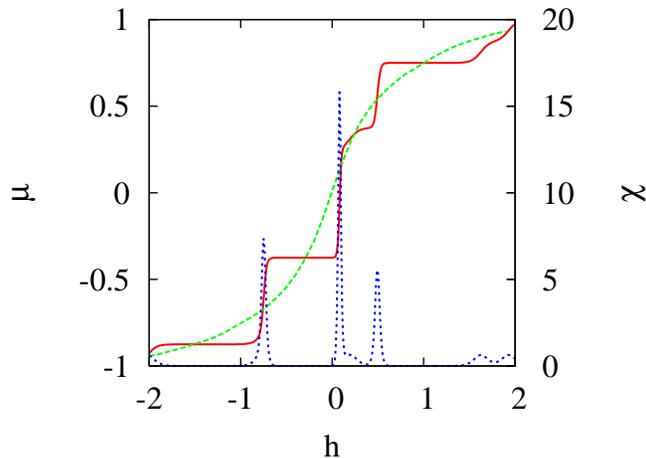}
\end{center}
\caption{Magnetization curve in a small sample of 
a $p=3$ Ising mean-field spin-glass model. The solid step-wise curve
is the equilbrium values of the 
magnetization $m=(1/N)\sum_{i}\langle S_{i} \rangle_{h}$
at $T/J=0.1$ and at various values of the external field $h$
obained by an exact enumerations of the partition function
of a $N=16$ system.  The $S$ shaped line (cross)
is obtained by taking averages over large number of samples, 
which actually has only very weak finite size effects so that it is very close
to the thermodynamic one $N \to \infty$.
The spiky curves is the linear susceptibility
$\chi=(\beta/N)\sum_{ij} (\langle S_{i}S_{j} \rangle_{h} - \langle S_{i} \rangle_{h} \langle S_{j} \rangle_{h})$ with $\beta=1/T$.
Similar features have been observed in the SK model ($p=2$) see \cite{KY}
and directed polymer in random media \cite{M90}.
}
\label{fig-pspin}
\end{figure}

As a representative model let us take the p-spin Ising mean-field spin-glass
model \cite{REM,GM} whose hamiltonian is given by,
\beq
H=-\sum_{i_1 < i_2 < \ldots i_p} J_{i_1,\ldots,i_{p}} \sigma_{i_1}\ldots \sigma_{i_p}
-h \sum_{i} \sigma_{i}.
\label{eq-hamiltonian-pspin}
\eeq
Here $\sigma_{i}$ $(i=1,\ldots,N)$ are Ising variables which are coupled with each other by 
$p$-body random interactions $J_{i_1,\ldots,i_{p}}$ and subjected to an external field of strength $h$.
The coupling $J_{i_1,\ldots,i_{p}}$ follows a Gaussian distribution with zero mean and variance
$J \sqrt{\frac{p!}{2N^{p-1}}}$,
\beq
\sav{J_{i_1,\ldots,i_{p}}}=0 \qquad 
\sav{J^{2}_{i_1,\ldots,i_{p}}}= J^{2} \frac{p!}{2N^{p-1}}
\eeq
Here and in the following we denote averages 
over different realizations of the quenched randomness as $\sav{\ldots}$.

For $p>2$ the $p$-spin Ising model exhibists a static phase transition
from the paramagnetic (liquid) phase to glassy phase characterized by 
1 step RSB. We note the transition temperature, which is usually interperted
as the so called Kauzmann temperature,  as $T_{c}$ in the
present paper.

We wish to examine changes of the magnetization
\beq
\mu=\frac{1}{N}\sum_{i} \tav{\sigma_{i}}
\eeq
under variation of the extenral field $h$ in {\it finite sized systems} of $N$ spins at temperatures below $T_{c}$.
Here and in the following we denote thermal avarages as $\tav{\ldots}$.
In section \ref{sec-replica} we derive the generating functional for the mesoscopic magnetic responses in this model.

For an illustration we show in Fig.~\ref{fig-pspin} an example of 
magnetization curve observed numerically in a certain small sample 
of $p=3$ model. By a {\it sample} we mean a specific realization 
of the $J_{ij}$ bonds in \eq{eq-hamiltonian-pspin}.
Note the step-wise increase of the magnetization $\mu$ 
under increase of the magnetic field $h$. The pattern of the steps
depend on each realization of samples so that it can be regarded as
a {\it fingerprint} of a given sample. Note also that 
the linear susceptibility $\chi$ is quite spiky and significant only 
in the visinity of the steps. 

On the other hand we know that in the thermodynamic limit $N \to \infty$ 
the magnetization curve {\it must} converge to a unique limit 
independently of samples. Thus the steps obserbed in mesoscopic 
(finite sized) samples must somehow disappear in $N \to \infty$ limit.
But how?  

A similar observation has been made also in the SK model by Kirkpatrick 
and Young \cite{KY}. Young, Bray and Moore \cite{YBM} have argued that 
2the step-wise response reflects crossings of free-energies
of low lying metastable states under variation of the magnetic field. 
M\'{e}zard has discussed similar phenomena in the directed polymer in
random media \cite{M90}. 

In our present paper we consider the 1 step RSB mean-field models instead of the full RSB models
like the SK model. The 1 step RSB models are techinically and conceptually much simpler 
than the full RSB models and allow us to perform much more detailed, direct analysis 
on the mesossopic responses.


\subsection{effective random energy model}
\label{subsec-rem}

In most parts of the present paper we consider more simplied 
models called random energy models (REM) which are suited for our purposes
for various reasons.

Formally, Derrida's original REM \cite{REM,GM} can be derived by considering 
the limiting case $p \to \infty$ of the $p$-spin model given above. In this limit the model
allows analytical calucurations to a great extent.

Physically the mesoscopic step-wise response arise from jumps between different
metastable states (inter-state response). In this respect, responses within each 
metastable states (intra-state response) are irrelvant from our point of view.
In the REM the 
intra-state responses automatically vanish so that we can avoide to work hard to disentabgle 
intra-state and inter-state responses. 

The partition function of the REM at temperature $T=\beta^{-1}$ under the field $h$ can be written  as,
\beq
Z_{\rm REM}=\sum_{i=1}^{e^{Nc}}e^{-\beta(F_{i}- h
Y_{i})}.
\label{eq-def-mod-REM}
\eeq
Here $i=1 \ldots e^{Nc}$ ($c>0$) is the label for the metastable states. $F_{\alpha}$
is the free-energy of the metastable state $\alpha$
at $h=0$ and $Y_{\alpha}$ is a variable which is
conjugate to the external field $h$. 

In the case of magnetic systems, as the original REM by Derrida \cite{REM}, the variable $Y$ corresponds 
to the magnetization which takes random values at different metastable states. However in the present paper we wish to 
leave it as a {\it generic} extensive random variable conjugated to a generic external field $h$, 
in the same sprit of the seminal work by Krzakala and Martin (KM) \cite{KM}. We may call such a model
as an {\it effective} REM. As far as the inter-state responses are concerned, 
the effective REM should correctly simulate a given original model which exhibits 1 step RSB.

Then the eseential ingredient we need to know is the density of metastable states at a give value of the free-energy $F$ and the variable $Y$. 
The metastable states are obtained as the solutions of the equation of states called Thouless-Anderson-Palmer
(TAP) equations \cite{tap}. In a previous work \cite{RY} we have given an analytic 
recipe to compute the logarithm of such a density of states, which we called {\it generalized complexity}
from the TAP equation of a given microscopic model.

\begin{figure}[h]
\begin{center}
\includegraphics[width=\textwidth]{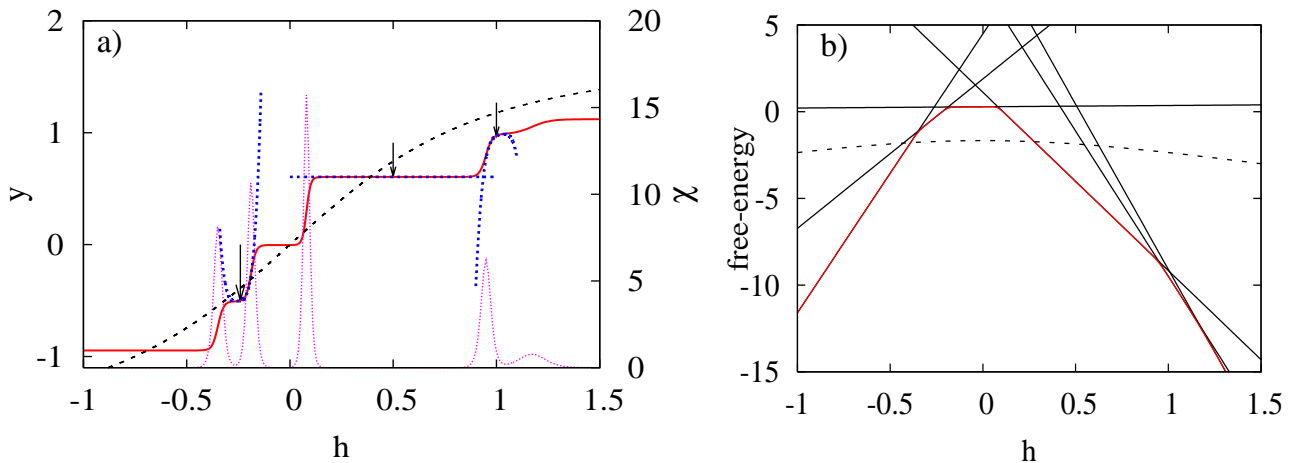}
\end{center}
\caption{
Step wise responses and level-crossings of the free-energies
in an effective REM. Here we used the toy model proposed by
Krzakala and Martin \cite{KM}. (See sec. \ref{sec-gc} and appendix
\ref{sec-KM} for the details.)
In a) the solid (red) line which exhibits a sequence of steps 
is the response curve $y(h)$ computed numerically
in a certain finite size ($N=17$) sample of a model specified
by the density of state \eq{eq-gc-KM} with $c=2\ln 2$, $\Dy=1$ at temperature $T/\Tc^{0}=0.1$. 
Here $\Tc^{0}=1/\sqrt{2c}$ is the critical temperature of the model under $h=0$ (See appendix \ref{sec-KM}).
The linear susceptibility $\chi(h)$,
and all non-linear susceptibilities $\chi_{n}(h)$ with $n \geq 2$ (not shown),
exhibit spikes at the edges of the staircases. The step-structures become
finer as $N$ is increased.
The doted (blue) lines represent local Taylor expansions of the response curve 
${\rm const}+ \chi_{1} \delta h + \chi_{2} (\delta h)^{2}+ \chi_{2} (\delta h)^{3}+ \chi_{2} (\delta h)^{4}$ wit $\delta h$ being distances from the points indicated by the arrows. 
The (black) dashed line is the response curve in the
thermodynamic limit obtained by taking first derivative of
the total free-energy density in the thermodynamic limit
$(f') ^{*} (h)$ given in \eq{eq-y-av-glass} with respect to $h$. 
The qualitative difference between
{\it thermodynamic susceptibilities} obtained by 
$\lim_{\delta h \to 0}\lim_{N \to \infty}$ and {\it susceptibilities
associated with fluctuations} (FDT) obtained by
$\lim_{N \to \infty}\lim_{\delta h \to 0}$ is clear.
In b) the evolution of
the free-energy densities $f'_{l}=f_{l}-h y_{l}$ under varying  $h$
for several low lying states are shown by solid (black) lines.
The free-energy density of the total system is shown by the solid (red) line.
The dashed line is the free-energy density in the thermodynamic limit
$(f') ^{*} (h)$ given in \eq{eq-y-av-glass}. 
}
\label{fig-steps-fdt}
\end{figure}

In Fig.~\ref{fig-steps-fdt} a) we show an example of the profile of the response observed in a REM.
In Fig.~\ref{fig-steps-fdt} b) it can be seen that free-energies of low lying
states exhibit level crossings precisely at the 'critical fields' $h_{c}$ corresponding to the
edges of the staircases of the response curve $y(h)=\tav{Y}/N$.

\subsection{Some basic features of the step-wise responses}

In the present paper we wish to characterize statistical properties 
of the step-wise responses in {\it mesoscopic samples} explicitely. By 
{\it mesoscopic} we mean that {\it 
$N$ is significantly larger than $1$ but not 
infinitely large}. Our main result is that there are three important characteristic
scales of the steps in the 1 step RSB mean-field models:  
1) typical height of a step  $\Delta Y$ , 
2) typical spacing $\hs$ between the steps and 
3) typical width $\hw$ of a step over which the step is rounded 
at finite temperatures $T$. Their orders of magnitudes are obtained as,
\beq
\Delta Y \sim \Dyeffs(h) \sqrt{N}
\qquad
\hs \sim \frac{T_{c}(h)}{\Dyeffs(h) \sqrt{N}}
\qquad 
\hw \sim \frac{T}{\Dyeffs(h) \sqrt{N}}
\label{eq-three-scales-steps}
\eeq
where $\Dyeffs(h)$ characterize strength of the fluctuation of the quantity 
$Y=N y$  conjugated to the external field $h$ over 
different metastable states and
$T_{c}(h)$ is the critical temperature under external field of strength $h$.
In the thermodynamic limit $N \to \infty$, the spacing $\hs$ between the steps vanish
and the response curve converges to a unique limit.

We arrive at the above conlcusion by analyzing anomalous scaling properties of
thermal fluctuations associated with the steps, distribution function of the response
in the zero temperature limit and decorrelation of two real replica systems subjected
to different fields. We study these properties by two approaches :
a low temperature expansion approach in section \ref{sec-lowt-expansion} and 
a replica approach in section \ref{sec-replica}.

\subsubsection{Anomalous thermal fluctuations}
\label{subsec-thermal-fluctuation}

A set of useful quantities which characterize the step-wise responses
are linear and non-linear susceptibilities
\beq
\chi_{n}(h)=\left. \frac{\partial^{n} \langle y(h+\delta h)\rangle}{d \delta h^{n}}  
\right |_{\delta h=0} \qquad (n=1,2,\ldots,)
\eeq
which exhibit series of spikes along the $h$ axis as shown in 
Fig.~\ref{fig-pspin} and Fig.~\ref{fig-steps-fdt}. 
Note that the susceptibilities become significant 
only in close vicinities of the edges of the staircases, 
\beq
\chi_{n} \sim O\left (\Delta Y \hw^{-n} \right) 
\qquad \mbox{(around the edges of the steps)}
\label{eq-sus-edge}
\eeq
and remain almost zero elsewhere. Thus a given sample under a given value of $h$
may or may not have large susceptibilities: it will have large 
susceptibilities only if it {\it happens} to have a 'critical field' $h_{c}$ (edge of a step)
close enough to $h$.
Thus we naturally expect that linear and all non-linear 
susceptibilities are {\it not} self-averaging, as first pointed out
by Young, Bray and Moore \cite{YBM} in the SK model.

The susceptibilities are related to connected correlation functions
of the $Y$ variable through the static fluctuation dissipation theorem (FDT).
FDT is derived by expanding the free-energy $N f_{J}(T,h+\delta h)$ 
of a given sample, say $J$, in power series of an {\it infinitesimal} 
shift in the field $\delta h$,
\beqa
 -N \beta f_{J}(T,h+\delta h)  - (- N \beta f_{J}(T,h))
= \ln \langle e^{\beta \delta h Y} \rangle 
=  \beta \delta h \kappa_{1}(Y) 
+ \frac{(\beta \delta h)^{2}}{2!} \kappa_{2}(Y) 
+ \frac{(\beta \delta h)^{3}}{3!} \kappa_{3}(Y) 
+ \ldots
\label{eq-f-expansion-h}
\eeqa
where $\langle \cdots \rangle$ stands for weighted averages within
the 'unperturbed system' ($\delta h=0$).
We denote the $n$-th thermal cumulant moments (or connected 
correlation functions) of the variable $Y$ as $k_{n}(Y)$.
The first few moments
read as $\kappa_{1}(Y)=\tav{Y}$,
$\kappa_{2}(Y)=\tav{Y^2}-\tav{Y}^{2}$,
$\kappa_{3}(Y)=\tav{Y^3}-3 \tav{Y^2}\tav{Y}+2 \tav{Y}^{3}$,  
$\kappa_{4}(Y)=\tav{Y^4}-4\tav{Y}\tav{Y^3}+6\tav{Y}^2\tav{Y^2}-3\tav{Y}^4$.
The above power series yields the usual {\it static} FDT;
a change of $Y$ induced by an infinitesimal increment 
of the field $\delta h$ can  be written as
\beqa
\delta Y_{J} = \chi_{1} \delta h + \chi_{2} (\delta h)^{2} + \chi_{3} (\delta h)^{3} + \ldots
\label{eq-Y-expansion-h}
\eeqa
with
\beq
\chi_{1}=\beta \kappa_{2}(Y) \qquad \chi_{2}=(\beta^{2}/2!)\kappa_{3}(Y)  \qquad   \chi_{3}=(\beta^{3}/3!)\kappa_{4}(Y)  \qquad  \ldots
\label{eq-FDT}
\eeq
where $\chi_{1}$ is the linear susceptibility and $\chi_{2}$, $\chi_{3}$
,$\ldots$ are non-linear susceptibilities of a given sample $J$. 
If the thermal fluctuation of $Y$ is Gaussian, as it happens 
at  $T> \Tc$, higher connected correlation functions vanishes
$\kappa_{n}(Y)=0$ for $n \geq 3$ and thus the non-linear susceptibilities
vanish.

At around the edge of the steps where the response is significant
(See \eq{eq-sus-edge}) the response may be described by a Taylor expansion,
\beq
\delta Y_{J} = \Delta Y \sum_{n=1}^{\infty} c_{n}
\left(\frac{\delta h}{\hw}\right)^{n} 
\label{eq-y-fdt-expansion}
\eeq
with $c_{n}$ being $O(1)$ numerical coefficients. 
Note that here the small parameter for the expansion is 
$h/\hw$ with $\hw$ being the width of the thermal rounding of the steps
given in \eq{eq-three-scales-steps}.
This is indeed evident
in the plot of Fig.~\ref{fig-steps-fdt} a) where
we show some examples of truncated Taylor expansions of the
response curve $y(h)$ at several points along the $h$ axis (marked by
arrows). 
Around the edges of the steps, the local Taylor expansion provides a good approximation 
only for sufficiently close neighbourhood of width $\hw$. 
Thus ``linear response theories'' can {\it not} be invoked
to estimate responses beyond the scale of a single step $\hs$
which actually vanishes in the thermodynamic limit $N \to \infty$. (See \eq{eq-three-scales-steps})

Then one can notice that
{\it thermodynamic susceptibilities} obtained in the limit
$\lim_{\delta h \to 0}\lim_{N \to \infty}$  i.e. 
taking derivatives of 'smooth thermodynamic free-energy density', 
and {\it susceptibilities associated with fluctuations} (FDT) 
obtained in the other way around $\lim_{N \to \infty}\lim_{\delta h \to 0}$,
which directly reflect 'rugged free-energy landscape', 
are completely different objects. 
In short the two limits are not commutative {\it everywhere} in the glass phase,
\beq
\lim_{\delta h \to 0}\lim_{N \to \infty} \neq \lim_{N \to \infty}\lim_{\delta h \to 0}.
\label{eq-non-commute}
\eeq
The non-commutativity of the two limits signals the presence of the steps.

On passing let us note that {\it disorder-average} of the {\it linear}
susceptibility defined via static fluctuation dissipation theorem (FDT) 
must agree with the 1st derivative of the thermodynamic response curve 
in a rather {\it accidental} manner as discussed 
in appendix \ref{sec-note-linear-sus}. Importantly the equivalence does 
{\it not} hold for the non-linear susceptibilities.
Although the value of linear susceptibility obtained via FDT coincides,
after averaging it over disorder, with the 'thermodynamic linear susceptibility',
it must be emphasized that the meaning of 'linear responses' is totally 
different in the two cases. As can be seen in Fig.~\ref{fig-steps-fdt},
linear response approximation is apparently not so bad for the thermodynamic 
response curve while we realize it is almost a nonsense in a given finite sized sample.

\subsubsection{Sharp response in $T \to 0$ limit }
\label{subsec-sharp-response}

The expansion of the response via FDT cannot go beyond the scale
of a single step. In particular, in the close vicinity of the edges of the steps
the Taylor expansion is valid  only within the thermal width of
the step $\hw$ (See \eq{eq-y-fdt-expansion}). Moreover the thermal width 
$\hw$ vanishes in the zero temperature limit $T \to 0$ at which
the step structure becomes most clear.
Thus it is tempting to look for another way to analyze 
the mesoscopic step-wise responses without relying on the FDT, which  
hopefully works at zero temperature.

By both the low temperature expansion method and the replica method, 
we will demonstrate that it is indeed possible to take $T \to 0$ limit {\it first} and 
study mesoscopic responses consisting of successive sharp steps within 
small increments of the field $\delta h$.
Interestingly enough, we will otbain another series expansion 
of the mesoscopic response in which the relvant small parameter
is $\delta h/\hs$ instead of $\delta h/\hw$ which appears in the FDT case 
(Note that $\hw < \hs$ since $T< T_{c}(h)$). 
Thus the two limts $T \to 0$ and $\delta h \to 0$ are not commutative,
\beq
\lim_{\delta h \to 0}\lim_{T \to 0} \neq \lim_{T \to 0}\lim_{\delta h \to 0}.
\label{eq-non-commute-zeroT}
\eeq
Somewhat unexpectedly, the approach $\lim_{T \to 0}\lim_{\delta h \to 0}$
allows us to go beyond the limit of single step.
We will find that the $O(\delta h/\hs)$ term in the series 
corresponds to the case that
only single step is present in a given interval $\delta h$, 
$O(\delta h/\hs)^{2}$ term to the case of two steps, and so on.

\subsubsection{Cusp in the overlap: a mesoscopic aspect of the chaos effect}

In addition to the response of the variable, say $Y$ cojugated to a given external field $h$,
we examine the change of the corelation function between two replicas subjected to slightly
different fields,
\beq
C(\delta h) \equiv \frac{\sav{\tav{Y(h-\delta h)}\tav{Y(h+\delta h)}}}{\sav{\tav{Y(h)}^{2}}}
\label{eq-def-overlap}
\eeq
under variation of $\delta h$. This kind of correlation function or {\it overlap} function
is oftenly used to analyze the static chaos effects 
known in spin-glasses and related systems 
\cite{MBK,BM87,FH88,kondor,FH91,SY,RC,SHYT}.

We are sure that the correlation function must vanish $C \to 0$ for any small but finte variation 
of the field in the thermodynamic limit $N \to \infty$, i.e. chaos. This is because
in such a circumstance there will certainly be level crossings between different metastable states
which have zero overlap with respect to each other in 1 step RSB models \cite{RY}.

Our interest here is to look at {\it how} it decays as $\delta h$ is increased in mesoscopic 
(finite sized) samples. To this end we analyze \eq{eq-def-overlap} at finite temperatures using
the FDT approach and also directly at $T \to 0$ limit without relying on the FDT. We will find that the overlap $C(\delta h)$ 
exhibists a {\it cusp} singularity at $\delta h=0$ in $T \to 0$ limit, which is rounded at finite temperatures
over the scale $\hw$, and that the width of the cusp is given by $\hs$ (See \eq{eq-three-scales-steps}). 
Somewhat unexpectedly, as we discuss in section \ref{sec-connection}, the overlap function
has counterparts which describe the jerky effective energy landscape of pinned elastic manifolds 
\cite{BBM} and intermittencies due to shocks in turbulence \cite{BMP}.

\section{Generalized complexity}
\label{sec-gc}

\begin{figure}[b]
\begin{center}
\includegraphics[width=0.5\columnwidth]{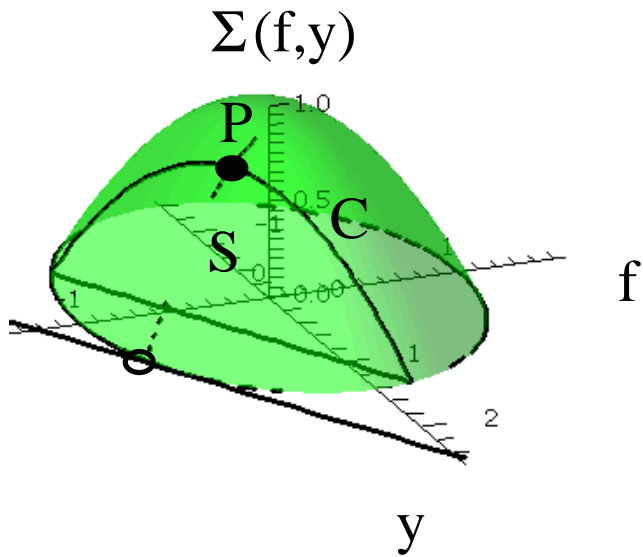} 
\end{center}
\caption{The generalized complexity $\Sigma=\Sigma(f,y)$ have
a shape like a surface of a bell cut by $\Sigma=0$ plane. 
The dashed line on the $\Sigma=0$ plane is the zero complexity 
curve $\Sigma(f,y)=0$. The surface $S(f')$ is normal to the
axis of the 'total free-energy' $f-\delta hy$ and cut the latter at
$f'=f-\delta hy$. Cross section between the plane $S(f')$ and the 
complexity surface $\Sigma=\Sigma(f,y)$ defines a curve $C$ 
on which the complexity is maximized at a peak $P=P(f')$.
}
\label{fig-gc}
\end{figure}

Now let us discuss some generic features of the density of the metastablestates (TAP states).
As we noted in in sec \ref{subsec-rem}, the effective REM can be constracted based only on this information
(See \eq{eq-def-mod-REM}).

We consider an ensemble of metastable states $i=1,2,\ldots$ 
each of which is a solution of the TAP equations of states at a given temperature $T$ and an external field $h$.
We denote the free-energy of a TAP state, say $\alpha$,
as $F_{\alpha}=Nf_{\alpha}$ and the extensive variable 
conjugated to $h$ as 
$Y_{\alpha}=N y_{\alpha}=-N\partial f_{\alpha}/\partial h$.

We rewrite the partition function of a given sample 
of the effective REM \eq{eq-def-mod-REM} at temperature $T$ 
and under slightly shifeted external field $h+\delta h$ with small
enough $\delta h$ as \cite{DeDY,P83},
\beq
\hspace*{-1cm}
Z_{\rm REM}(T,h+\delta h) =\sum_{\alpha} e^{-N\beta (f_{\alpha}- \delta h y_{\alpha})}
=\int df dy  e^{N [\Sigma(f,y)-\beta (f- \delta h y)]}
\label{eq-Z-TAP}
\eeq
where $\beta=1/T$. Here $\Sigma(f,y)$ is the generalized complexity \cite{RY}
defined as the logarithm of the density of states with respect to the
two variables $f$ and $y$,
\beq
\Sigma(f,y) = \frac{1}{N}\log \sum_{\alpha} \delta(f-f_\alpha)\delta(y-y_\alpha).
\label{eq-gc-def}
\eeq
In general \cite{RY} we expect that the generalized complexity $\Sigma(f,y)$ 
can be presented as a surface $\Sigma=\Sigma(f,y)$ 
which is convex upward as shown schematically in Fig.~\ref{fig-gc}. 
See \cite{RY} for an analytical recipe to compute the generalized complexity of a given microscopic model. Note that in general $\Sigma(f,y)$ can depend both
on the temperature $T$ and external field $h$.

As a toy model we can use the model propose by Krzakala and Martin (KM) \cite{KM}
in which the generaized complexity takes a particulary simple form,
\beq
\Sigma(f,y)= c - \frac{f^{2}}{2}-\frac{y^{2}}{2 \Dy^{2}},
\label{eq-gc-KM}
\eeq
which depends neither on the temperautre $T$ nor the field $h$.
In appendix \ref{sec-KM} we summarize some basic properties of the model.
We will find that it gives generic results concerning
inter-state responses common for all 1 step RSB models in the glass phase.

\subsection{Thermodynamic response}
\label{subsec-thermodynamic-response}

In the thermodynamic limit $N \to \infty$, 
at a given temperature $T$ and under an external field $h+\delta h$, 
the partition function \eq{eq-Z-TAP} is dominated by contribution from 
a certain saddle point $(f^*,y^*)$ which parametrize a particular 
set of TAP states. Then thermodynamic value of $Y$ 
variable is given by $Y=N y^{*}$ and the thermodynamic free-energy 
is given by $F'=N (f^{*}- \delta h y^{*})$.

The saddle point $(f^*,y^*)$ can be located as the following.
Let us consider the three-dimensional space $(f,y,\Sigma)$
as shown in Fig.~\ref{fig-gc}.
Consider a plane $S=S(f')$ on which the total free-energy density 
takes a constant 
value $f-\delta h y= f'$. 
Cross section between $S(f')$ and the complexity surface 
$\Sigma=\Sigma(f,y)$ 
defines a line $C=C(S)$ along which the complexity takes 
a maximum value at a certain point $P=P(C)$. 
Then for each value of $f'$ we have a unique point $P=P(f')$ 
associated with it. Now we look for a special 
point $P^{*}=(f^{*},y^{*},\Sigma^{*})$ 
such that it satisfies,
\beq
\left. \frac{\partial \Sigma(f,y)}{\partial f'} \right |_{P=P^{*}}=\frac{1}{T}.
\label{eq-saddle}
\eeq
It is easy to verify that $(f^*,y^*)$ 
is the saddle point which dominates the partition function \eq{eq-Z-TAP}
and $\Sigma^{*}$ is the value of the complexity at the saddle point.

As the temperature $T$ is lowered under a fixed value of the
external field $h$, the point $P^*$ moves downward
on the complexity surface. At a certain critical temperature
$T=\Tc(h)$, the value $\Sigma^{*}$ of the complexity  at the point $P^*$
becomes $0$. As the temperatures is lowered further the point 
$P^*$ remains pinned there. Thus the system is understood to be 
in the glass phase where $\Sigma^{*}=0$ at lower temperatures $T < \Tc(h)$.

Within the glass phase $T < T_{c}$, the saddle point moves smoothly along
the zero complexity line $\Sigma(f,y)=0$ 
under variation of the extenal field $h$.
The equilibrium values of the parameters $(f^*(h+\delta h),y^*(h+\delta h))$ 
under a given external field
$h+\delta h$ is simply obtained by looking for a line $f-\delta h y=f'$ 
with slope $\delta h$ which is tangent to the zero complexity curve 
$\Sigma(f,y)=0$. The touch point is nothing but the saddle point 
$(f^*(h+\delta h),y^*(h+\delta h))$ \cite{RY}. 

By varing $h$, the equilbrium value of $y^*(h)$ will change smoothly.
Consequently all linear and non-linear suscepibilities
\beq
\chi_{n}(h)=
\left. \frac{d^{n} y^{*}(h+\delta h)}{d \delta h^{n}} \right |_{\delta h=0}
\eeq
will be smooth functions of $h$.
Note that these are {\it thermodynamic susceptibilities}
which are obtained by taking thermodynamic 
limit $N \to \infty$ {\it before} taking derivatives with respect to $h$.

The fact that $(f^*,y^{*})$ varies 
with $h$ means that strong chaos is induced by extensive level crossings
in the 1 step RSB systems \cite{RY}: 
a set of TAP states which dominates the equilibrium states at a given
value of $h$ becomes completely out-of-equilibrium at $h+\Delta h$.
However, thermodynamic response itself is completely smooth and featureless, 
which ironically provides us no hint of 
the underlying chaos effect. Thus we are naturally lead 
to go down to smaller scales to look for some traces of the chaos effect.

In appendix \ref{sec-KM} we rephrase the above discussion based
on the toy model \eq{eq-gc-KM}.

\subsection{Low lying states close to $\Sigma=0$ line}
\label{subsec-low-lying-states}

We are interested in the glassy phase $T < \Tc(h)$ so that the shape of the 
complexity around the saddle point $(f^*(h),y^*(h))$ 
on the zero complexity line
$\Sigma=0$ will be particularly important. 
A generic shape around the saddle point close to $\Sigma=0$ 
may be cast into a form like,
\beq
\Sigma(f,y) \simeq \frac{f-f^{*}(h)}{\Tc(h)}
-\frac{(y-y_{P}(f))^{2}}{2\Dyeffs(h)^{2}} 
\label{eq-gc-around-saddle}
\eeq
where $y_{P}(f)$ describes the position of the peak $P$ 
(see Fig.~\ref{fig-gc}) at a given value of $f$. 
The latter may be described by a Taylor expansion,
\beq
y_{P}(f)=y^{*}(h)-a(h)(f-f^{*}(h))+ O((f-f^{*}(h))^{2}).
\label{eq-yp}
\eeq
with $a(h)$ being a certain constant.

The functional form \eq{eq-gc-around-saddle} suggests us to introduce 
a new variable $\tY$,
\beq
\tY = N \ty \qquad  \ty \equiv y-y_p(f).
\label{eq-def-ty}
\eeq
 Then we can rewrite the generalized complexity 
\eq{eq-gc-around-saddle} as,
\beq
N \Sigma \simeq  \frac{F-F^{*}(h)}{\Tc(h)}
-\frac{\tilde{Y}^{2}}{2N \Dyeffs(h)^2}
\label{eq-gc-around-saddle-2}
\eeq
Note that this means the density of states $e^{N \Sigma(\Xi)}$ is
decoupled into two parts: exponential distribution function for the free-energy
and Gaussian distribution function for the transverse variable $\tY$.
In the present paper we call $\tY$  as {\it transverse variable}.

In appendix \ref{sec-KM} we rephrase the above discussion based
on the toy model \eq{eq-gc-KM}. In particular we find that the
complexity of the model behaves as \eq{eq-gc-around-saddle-2}
close to the $\Sigma=0$ plane so that it should give generic
results common in all 1 step RSB models concerning inter-state responses
below $T_{c}(h)$.


\subsubsection{Level spacings among the low-lying states}
\label{subsubsec-level-spacing}

From \eq{eq-gc-around-saddle-2} we find that the total free-energy
$F$ of low lying states $(F-F^{*})/\Tc (h) \approx O(1)$ follow an
exponential distribution \cite{MPV}, 
\beq
\rho(F) \simeq \exp((F-F^*)/\Tc(h)).
\label{eq-p-f}
\eeq

\begin{figure}[t]
\begin{center}
\includegraphics[width=0.5\textwidth]{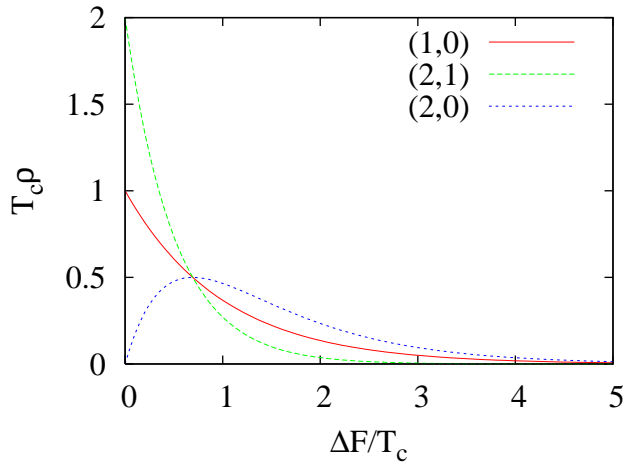}
\end{center}
\caption{Spacing between low lying states. 
Adjacent free-energy levels $(1,0)$,$(2,1)$,...are {\it not} repulsive
to each other. More generically, 
free-energies of $l$-th and $m$-th states
with $|l-m | > 1$ are ``repulsive'' to each other.
}
\label{fig-level-spacing}
\end{figure}

Now let us order the states as $l=0,1,2,\ldots$ 
according to the value of their total free-energies 
$F_{l}$ such that $F_{0} \leq F_{1} \leq F_{2} \ldots$.
We will need in particular the information of the statistics of
level-spcacings: differences between free-energies,
\beq
\Delta F_{(l,m)} \equiv F_{l}-F_{m} \qquad l > m.
\label{eq-def-delta-f}
\eeq
Recently Bertin \cite{bertin} 
have shown that spacing between random numbers drawn 
from an exponential distribution function
exactly obey a class of exponential distribution laws.
Following his proof we find from \eq{eq-p-f}
that free-energy gaps of adjacent levels $\Delta F_{(l,l-1)}$ 
follow an $l$-dependent exponential distribution\footnote{To avoid confusions we should note that the free-energies $F_{l}$
themselves follow the Gumbell distributions \cite{BM,bertin}.},
\beq
\rho_{(l,l-1)}(\Delta F_{(l,l-1)})
=\frac{l}{\Tc} e^{-l\frac{\Delta F_{(l,l-1)}}{\Tc}}.
\label{eq-p-df-norepulsion}
\eeq

The exponential form \eq{eq-p-df-norepulsion}
implies absence of repulsions between adjacent levels
in the sense that  $\lim_{x \to 0}\rho_{(l,l-1)}(x) > 0$ (See Fig.~\ref{fig-level-spacing}).
On the other hand we find distribution of the free-energy
difference between $l+1$-th and $l-1$-th states using \eq{eq-p-df-norepulsion} as,
\beq
\rho_{(l+1,l-1)}(\Delta F_{(l+1,l-1)})
=\frac{l(l+1)}{\Tc}
e^{-(l+1) \frac{\Delta F_{(l+1,l-1)}}{\Tc}}
\left(
1-e^{-l \frac{\Delta F_{(l+1,l-1)}}{\Tc}}
\right).
\label{eq-p-df-repulsion}
\eeq
This implies a certain {\it effective} level repulsion between
$l-1$-th and $l+1$-th states in the sense that 
$\lim_{x \to 0}\rho_{(l+1,l-1)}(x)=0$. (See Fig.~\ref{fig-level-spacing}).
Thus the free-energy levels are not {\it too much} degenerate: only a few
low lying states dominate the static properties in the glassy phase
in contrast the liquid phase $T> T_{c}(h)$ where exponentially many states
contribute.

\subsubsection{Distribution of the transverse variable $\tilde{Y}$}

A remarkable feature of the transverse variable $\tY$ 
is that {\it it is statistically independent from the free-energy}. 
From \eq{eq-gc-around-saddle-2}, 
we find $\tY_{l}$ follows a simple Gaussian distribution,
\beq
P(\tilde{Y}_{l})=\frac{1}{\sqrt{2 \pi N (\Dyeffs)^{2}}}
\exp \left(-\frac{\tilde{Y}_{l}^{2}}{2 N (\Dyeffs (h))^{2}}\right).
\label{eq-p-tY}
\eeq
which is independent of $l$, i. e. independent of the free-energy.

The original $Y$ variable depends both 
on the transverse variable $\tilde{Y}$ and the free-energy $F$,
\beq
Y -  Y^{*} = \tilde{Y} +a(h) (F - F^{*})
\label{eq-Y-old-new}
\eeq
which follows from \eq{eq-def-ty}. Here it is evident that 
thermal fluctuations and responses to the field of $Y$ is dominated
by that of $\tY$. 

Thus for the rest of this paper we will focus on thermal fluctuations 
of the variable $\tY$ instead of $Y$. Correspondintly we will
consider to apply a fictitious field conjugated to the tranverse
variable $\tY$ and examine its response. We will frequently denote
$\tY$ simply as $Y$ for simplicity.

\subsubsection{Discussions}

The most important features of the density of states specified by the
generalized complexity \eq{eq-gc-around-saddle-2} are the following two points:
1) In a finite fraction of  samples low lying states are almost degenerate
and 2) Existence of the transverse variable.
The combination of the two points yield the step-wise responses in the glass phase
at mesoscopic scales: a fraction samples can make a large response 
of order $\Dyeffs \sqrt{N}$ even by an infinitesimally small variation of the field.
Note that the situation must be distiguished from that in the liquid phase 
$T> T_{c}(h)$ where {\it all} samples respond. In contrast the 
response is {\it intermittent} in the glass phase 
because only some {\it finite fraction} of the samples are ready to respond to infenitesimal fields.

While the thermal fluctuations of the free-energy $F$ 
is necessarily of $O(T)$, thermal fluctuations of the transverse 
variable $\tY$ can be of order $O(\Dyeffs(h) \sqrt{N})$. Thus $\tY$ is
reminiscent of the Goldstone modes in the sense that it can 
make large thermal fluctuations 
and also large respons to small variations of the external field 
even at very low temperatures. (See \cite{M90,HF94} for  
related discussions in the problem of directed polymer in random media.)

In the case of the magnetic perturbation, the variance $\Dyeffs$ 
of the magnetization is related to the Edwards-Anderson (EA) 
order parameter $q_{\rm EA}$ as $\Dyeffs=\sqrt{q_{\rm EA}}$ 
under zero external magnetic field $h=0$ \cite{RY}. 
The latter has been also noticed by Franz and Virasoro
\cite{FV} in the context of dynamics.

The situation will change qualitatively if there are level-repulsions 
between adjacent free-energy levels as in the familiar Wigner's distribution. 
It will also depend on the  the parameter $\Dyeffs$. Chaos will be absent for
a perturbation with $\Dyeffs=0$.

Interestingly enough quite similar properties as 1) 2) are
{\it assumed} and plays prominent roles 
in the phenomenological Imry-Ma type droplet scaling arguments 
for glassy systems \cite{BM87,FH88,M90,FH91,SY} which assume 
thermally active {\it rare} low-energy excitations called {\it droplet excitations}.
The difference between the present mean field theory is that
the characteristic energy scale $\Tc$ is replaced by typical energy scale
of droplet excitations  which grows with the system size $L$,.
\bmat
T_{c} \hspace*{.5cm}\mbox{``mean field''} 
\qquad \to \qquad \Upsilon (L/L_{0})^{\theta} \hspace*{.5cm}\mbox{``droplet''}
\emat 
Here the exponent $\theta >0$ is usually called stiffness exponent
and $L_{0}$ is a certain length scale beyond which the power law
behaviour holds. Such an intriguing analogy has been pointed out 
in some literatures \cite{BD,BBM}.


\section{Low temperature expansion approach on mesoscopic responses}
\label{sec-lowt-expansion}

We now analyze responses at mesoscopic scales by a low temperature expansion
approach on the effective REM.
To this end, it is convenient to rewrite the partition function  
\eq{eq-def-mod-REM} of a given sample as,
\beq
Z_{J}=e^{-\beta F_{0}} \sum_{l=0}^{M-1}
\exp \left(-\beta \sum_{k=1}^{l} \Delta F_{k}+ \beta \delta h Y_{l} \right).
\label{eq-z-lowT}
\eeq
Here the index $l=0,1,\ldots,M-1$ label the states in an ascending order of the
values of their free-energies, i.e $F_{0} \leq F_{1} \leq \ldots \leq F_{M-1}$.
The free-energy gaps $\Delta F_{l}=F_{l+1}-F_{l}$ follow the
$l$-dependent exponential distribution \eq{eq-p-df-norepulsion}.
The variables $Y_{l}$ are understood as the {\it transverse variable} 
which is statistically independent from the free-energy at $\delta h=0$.
For simplicity we denote it as $Y$.

\subsection{A first look by a simple two-level model}
\label{subsec-two-level}

First let us look at a simple two-level model to get 
some basic insights. The two-level model is defined 
such that it consists of only the two lowest levels in a given sample
disregarding higher levels: the sum over the states $l=1,2,\ldots,M$ in  the partition function
\eq{eq-z-lowT} is truncated at $l=2$.
In the next section we will actually find that this simple two-level model
gives quantitatively correct results up to $O(T/\Tc(h))$ 
which is a good approximation at low enough temperatures. Contributions from higher levels
give only moderate correction terms of higher order in $T/T_{c}(h)$.

Let us consider a two-level system at temperature $T$ and under external field $h_{0}$.
It is easy to see that the response $\tav{y}_{h}-\tav{y}_{h_{0}}$ by varitaion of the field
can be expanded in power series of $\delta h=h-h_{0}$ as
\beq
\frac{\tav{y}_{h}-\tav{y}_{h_{0}}}{\Delta Y} = 
\sum_{n=0}^{\infty} \left (\frac{\delta h}{\hw} \right)^{n} 
\tilde{\chi}_{n} \left(\frac{h-\hc}{\hw}\right)
\label{eq-resp-two-level}
\eeq
with $\Delta F \equiv F_{1}-F_{0}$, $\Delta Y \equiv Y_{1}-Y_{0}$,
$\tilde{\chi}_{n}(y) \equiv \frac{\partial^{n} }{\partial y^{n}} \frac{1}{1+e^{-y}}$,
and
\beq
\hc \equiv h_{0}+\frac{\Delta F}{\Delta Y}  
\qquad \hw \equiv \frac{T}{\Delta Y} .
\label{eq-hc-hw}
\eeq

The above observation readily suggests the three important
characteristic scales of steps announced in \eq{eq-three-scales-steps}.
First, the scale of a single step
is given by $\Delta Y$ whose typical magnitude is,
\beq
\Delta Y  \sim  \sqrt{N}\Dyeffs.
\label{eq-def-Delta-Y}
\eeq
as \eq{eq-p-tY} implies.
Second, the distance from a given $h_{0}$ 
to the nearby 'critical field' $h_{c}$ where a level crossing take place
is given by $h_{c}-h_{0}=\Delta F/\Delta Y$.  Typical order of magnitude of the latter is,
\beq
\qquad \hs\sim \frac{\Tc(h)}{\sqrt{N}\Dyeffs}.
\label{eq-scale-hs}
\eeq
This follows from \eq{eq-p-df-norepulsion} which implies 
typically $\Delta F \sim \Tc(h)$.
Lastly $\hw$ introduced in \eq{eq-hc-hw}
defines the scale of {\it thermal width} over which which 
the edge of a step is {\it rounded} at temperature $T$. We readily find,
\beq
\hw \sim \frac{T}{\sqrt{N}\Dyeffs}.
\label{eq-scale-hw}
\eeq
Note that $\hw$ is the  scale up to which 
a local Taylor expansions of the response around
the edges of the steps converge as noted in \eq{eq-y-fdt-expansion}. 

\begin{figure}[t]
\begin{center}
\includegraphics[width=0.35\textwidth]{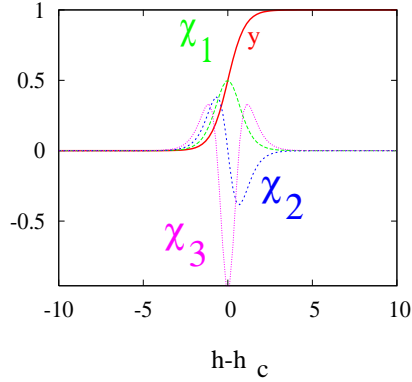}
\end{center}
\caption{Profiles of linear and non-linear susceptibilities 
associated with a step.
The solid (red) line represents a model function of a step 
$\tav{y}/\Delta Y=[1+e^{- (h-h_{c})/h_{\rm w}}]^{-1}$
with $h_{\rm w}=0.5$. The others are susceptibilities 
$\chi_{n}/\Delta Y=\partial^{n} (\tav{y}/\Delta Y)/\partial h^{n}$
which have significant amplitudes only in a narrow region 
of the {\it thermal width} $h_{\rm w}$ around the critical field' $h_{c}$.
Note also that sings of the non-linear susceptibilities oscillate.
}
\label{fig-one-step}
\end{figure}

From \eq{eq-resp-two-level} the susceptibilities are obtained as
\beq
\chi_{n}(h) = \Delta Y \frac{1}{\hw^{n}} \tilde{\chi}_{n} 
\left(\frac{h-\hc}{\hw}\right)
\eeq
which are displayed in Fig.~\ref{fig-one-step}.
Note that they are significant only within the thermal width 
$\hw$ around the critical field $\hc$. Since $\Delta F$
and $\Delta Y$ fluctuate from sample-to-sample, a given sample will have
significant susceptibilities under a given external field $h_{0}$
{\it only if the distance to the critical field $\hs=h_{c}-h_{0}$ happens to be
smaller than the thermal width $\hw$}. A simple minded estimate of 
the probably to actually have such a sample is,
\beq
\frac{\hw}{\hs}= O \left (\frac{T}{\Tc(h)} \right)
\label{eq-prob-rare}
\eeq
which becomes small at low temperatures  $T \ll \Tc(h)$.
With such a small probability the susceptibility of a given sample 
take significantly large values,
\beq
\chi_{n} \sim \Delta Y \hw^{-n} \sim (\Dyeffs)^{1+n} N^{(1+n)/2}
\label{eq-scale-chi}
\eeq
while they remain negligible otherwise.  

From \eq{eq-prob-rare} and \eq{eq-scale-chi} we find
sample-average of $p$-th moment of $\chi_{n}$ is dominated
by the {\it rare sample} and behave as,
\beq
\sav{\chi^{p}_{n}} \sim (\Delta Y \hw^{-n})^{p} \frac{\hw}{\hs} \sim 
\left[(\Dyeffs)^{1+n} N^{(1+n)/2}\right]^{p} \frac{T}{\Tc(h)}.
\label{eq-scaling-chi-n-p}
\eeq
Thus the sample-to-sample fluctuations of the susceptibilities
are very far from Gaussian and non-self averaging. For instance we find,
\beq
\frac{\sav{\chi^{4}_{n}}}{(\sav{\chi^{2}_{n}})^{2}} \sim
\left( \frac{T}{\Tc(h)}\right)^{-1}.
\eeq

The above argument based on the two-level model is surprisingly
similar to the typical Imry-Ma type scaling arguments in the droplet
phenomenology \cite{FH88,FH91}. Indeed let us note that a very similar 
Imry-Ma type scaling argument has been constracted for directed polymer
in random media subjected to tilt field \cite{M90}.

\subsection{Sample-to-sample fluctuations of linear and non-linear susceptibilities}
\label{subsec-thermal-fluctuation-lowT}

Now we study thermal fluctuations associated with the steps more 
systematically. Our task is to compute  $k$-th thermal cumulants of $Y$
(more precisely $\tY$) and examine its moments,
\beq
\sav{\kappa^{p_{1}}_{k_{1}}(Y)\kappa^{p_{2}}_{k_{2}}(Y)\cdots}= 
\sav{
\left.
\left (
\frac{\partial^{k_{1}}}{\partial (\beta \delta h)^{k_{1}}} \ln Z(h+\delta h)
\right)^{p_{1}}
\left (
\frac{\partial^{k_{2}}}{\partial (\beta \delta h)^{k_{2}}} \ln Z(h+\delta h)
\right)^{p_{2}}\cdots
\right |_{\delta h=0}
}.
\label{eq-def-kappa-k-p}
\eeq
where $Z((h+\delta h)$ is the partition function and $\sav{\cdots}$
denotes the average over different realizations of samples.
Here we compute them  by a low temperature expansion approach.
Later in section \ref{subsec-thermal-fluctuation-replica} we confirm the results 
by a replica approach. 

Omitting the global factor $e^{-\beta F_{0}}$ in \eq{eq-z-lowT}
which is irrelevant for our purposes, the partition function
can be formally rewritten as,
\beq
Z(\{O_{l},X_{l}\})= O_{0}+\sum_{l=1}^{M-1} O_{l} \prod_{k=1}^{l} X_{k},
\eeq
with
\beq
X_{l} \equiv e^{-\beta \Delta F_{l}} \qquad
O_{l} \equiv e^{\beta \delta h \tY_{l}}.
\eeq
The exponential distribution of $\Delta F_{k}$
given in \eq{eq-p-df-norepulsion} implies power law distribution of $X_{k}$,
\beq
p_{k}(X_{k}) =X^{k\frac{T}{\Tc(h)}-1}_{k}k\frac{T}{\Tc(h)}=k \frac{T}{T_{c}(h)}
\left \{1+\frac{T}{T_{c}(h)}\log X_{k} + O\left(\frac{T}{T_{c}(h)}\right)^{2} \right\}
\label{eq-p-x}
\eeq
defined in the range $0 \leq  X_{k} \leq 1$.
Now the sample average of the $p$-th moment of $k$-th thermal 
cumulant given by \eq{eq-def-kappa-k-p} can be expressed as,
\beqa
&& \sav{\kappa^{p_{1}}_{k_{1}}(\tY) \kappa^{p_{2}}_{k_{2}}(\tY)\cdots} \nonumber \\
&& =\sav{
\left. 
\prod_{r=1}^{p_{1}} \left \{
\left(
\sum_{l=0}^{M-1} Y_{l} O_{lr} \frac{\partial}{\partial O_{lr}}\right) ^{k_{1}}
 \ln Z(\{O_{lr},X_{l}\}) \right \} 
\prod_{r=p_{1}+1}^{p_{1}+p_{2}} \left \{
\left(
\sum_{l=0}^{M-1} Y_{l} O_{lr} \frac{\partial}{\partial O_{lr}}\right) ^{k_{2}}
 \ln Z(\{O_{lr},X_{l}\}) \right \}  \cdots \right |_{O=1}} 
\label{eq-kappa-k-p}
\eeqa

\begin{figure}[t]
\begin{center}
\includegraphics[width=0.5\textwidth]{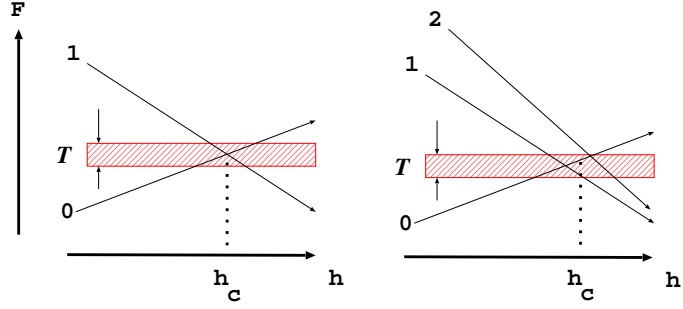}
\end{center}
\caption{Schematic pictures of free-energy levels close to a 'critical field' $\hc$.
(Left) A case when lowest two levels $l=0,1$ come close to each other with a separation 
of only $O(T)$. Rare samples which realize such a situation under a given $h$ make
$O(T/\Tc(h))$ contributions to the thermal cumulants under the given $h$.
$2$-level models take into account such samples correctly.
(Right) A case when lowest three levels $l=0,1,2$ come close to each other
simultaneously. Such rare samples make $O(T/\Tc(h))^{2}$ contributions.
$3$-level models are needed to take into account such samples correctly.
}
\label{fig-low-T-expansion-diagram}
\end{figure}

Let us emphasize that a $M$-th level model, 
which only takes into account the lowest $M$ levels, gives
correct results of the moments 
up to order $O(T/T_{c})^{M-1}$. We report the proof in 
appendix \ref{subsec-accuracy-m-level}.
This justifies the use
the simple two-level model discussed in the previous subsection \ref{subsec-two-level}
at low enough temperatures and	guarantees that correction terms at high orders
in $T/T_{c}(h)$, if needed, can be obtained systemacially by considering 
$3$,$4$,..-level models.

Now let us examine explicitely a few moments of $\kappa_{2}(Y)$
and $\kappa_{3}(Y)$ which are related to 
the linear $\chi_{1}=\beta \kappa_{2}(Y)$
and the 1st non-linear susceptibilities $\chi_{2}=\beta^{2}/2! \kappa_{3}(Y)$.
The details of the computation are reported in appendix \ref{sec-lowT-expansion-sus-appendix}
and we quote the results below.

For the linear susceptibility, we obtain by using a $3$-level model, 
\beq
\sav{\kappa_{2}(Y)}= (\sqrt{N}\Dyeffs)^{2} \left\{\frac{T}{\Tc(h)} +  O \left(\frac{T}{\Tc(h)}\right)^{3} \right\} 
\label{eq-kappa-2-1}
\eeq
\beqa
\sav{\kappa^{2}_{2}(Y)}= 
(\sqrt{N}\Dyeffs)^{4} 
\left \{\frac{T}{\Tc(h)} + O \left(\frac{T}{\Tc(h)}\right)^{3} \right\}.
\label{eq-kappa-2-2}
\eeqa

These  results demonstrates 
that the linear susceptibility is not self-averaging, i.e. 
$\sav{\chi^{2}_{1}} \neq (\sav{\chi_{1}})^{2}$ as expected.  
In sec. \ref{subsec-two-level} we have argued that 
probability to have a rare sample which happens to have a critical field $\hc$ 
close enough to a given $h$ is $O(T/\Tc(h)$ \eq{eq-prob-rare}.
Indeed in the above results 
we see that both the 1st and 2nd moment of the 2nd thermal cumulant
are proportional to $T/\Tc(h)$ and agree with 
the expected scaling form \eq{eq-scaling-chi-n-p}.

The sample average of the 1st moment of the 3rd thermal cumulant
is zero $\sav{\kappa_{3}(Y)}=0$ simply due to the symmetry reason.
(We again remind the readers that we are actually studying
the fluctuations of the transverse variable $\tY$.)
However its 2nd moment $\sav{\kappa^{2}_{3}(Y)}$ is non-zero.
Following the same steps as before we find after a lengthy algebra,
\beqa
\sav{\kappa^{2}_{3}(Y)}= 
(\sqrt{N} \Dyeffs)^{6} \;\;
2 \left\{\frac{T}{\Tc(h)} - \left(\frac{T}{\Tc(h)} \right)^{2}
+ O \left(\frac{T}{\Tc(h)}\right)^{3} \right\}.
\label{eq-kappa-3-2}
\eeqa
The result means that the non-linear susceptibility $\chi_{2}$
is strongly non-self averaging. The result agrees again with the 
expected scaling \eq{eq-scaling-chi-n-p} at low enough 
temperatures $T \ll \Tc(h)$. 
A remarkable feature is that it is {\it super} extensive,
\beq
\frac{\sqrt{\sav{\chi_{2}^{2}}}}{N} \sim \sqrt{N} \Dyeff^{3}
\eeq
which diverges in the thermodynamic limit $N \to \infty$.
Thus the non-linear susceptibility can be divergingly large 
either positively or negatively in a given sample.
This is an ambiguous signature of anomaly the associated with step-wise
responses which is invisible in thermodynamic susceptibilities.


\subsection{Distribution of mesoscopic responses in the zero temperature limit}
\label{sec-pdf-response-zero}

Here we take a different route to study the mesoscopic response 
by taking the zero temperature limit $T \to 0$ first.
As noted in \ref{subsec-sharp-response}, in this limit 
$\hw \to 0$, meaning that the steps become maximally sharp, 
so that the FDT approach cannot be invoked anymore.
While the FDT approach is limitted to desribe response only within the scale of single steps,
here we will find that we can actually go beyond this limit.

For each sample, we consider response at $T=0$
induced by the field increased from $h$ to $h+\delta h$ with $\delta h > 0$.
At the beginning the system stays at the ground state
under $h$. By increasing the field by $\delta h$ 
there may be level crossings by which
the variable $Y$  jump abruptly.
Thus at $T=0$ the induced response can be written simply as
\beq
\psi \equiv Y(h+\delta h)-Y(h)
\label{eq-resp-T0}
\eeq
where $Y(h)$ is the value of the $Y$ variable of the
ground state at $h$ and $Y(h+\delta h)$ is that under $h+\delta h$.
We are especially interested in the distribution function
of the induced response \eq{eq-resp-T0} over different realizations
of the samples,
\beq
p(\delta h,\psi) \equiv \sav{\delta(\psi-(Y(h+\delta h)-Y(h)))}.
\label{eq-p-psi}
\eeq
Fortunately we have complete information of the low lying states:
$Y_{l}$ (more precisely $\tY_{l}$ ) follows the Gaussian distribution 
given by \eq{eq-p-tY}
and the free-energy difference 
$\Delta F_{l,m}$ between $l$-th and $m$-th states
follow the distributions discussed in sec. \ref{subsubsec-level-spacing}.

\begin{figure}[h]
\begin{center}
\includegraphics[width=0.95\textwidth]{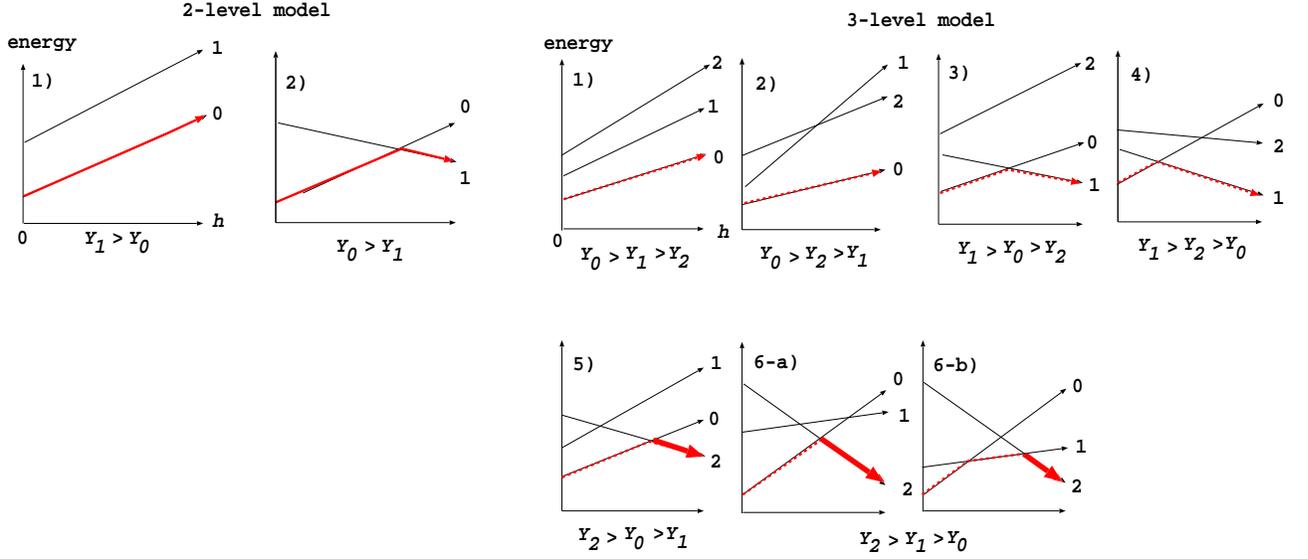}
\end{center}
\caption{(Left) Level crossing in the two-level system. In the
case 1) $Y_{1} > Y_{0}$ the two levels do not make a level 
crossing for $h >0$ so that there are no induced response
$\psi=0$.
On the other hand for the case 2) $Y_{1} < Y_{0}$ the two
levels  make a level crossing by which a step of the
response $\psi=Y_{1}-Y_{0}$ is induced.
(Right): Level crossing in the three-level systems. 
The cases 1)-4) are already included in the two-level model
(See Fig.~\ref{fig-resp-cross-2-3-level}).
In the cases 5),6-a) and 6-b), the third level comes into
play in the last stages represented by thick arrows.
}
\label{fig-resp-cross-2-3-level}
\end{figure}

\subsubsection{Two-level model}

Let us start from the two-level model. First we note
that the case 1) $Y_{1} > Y_{0}$ and 2) $Y_{1} < Y_{0}$ 
must be distinguished. As shown in Fig.~\ref{fig-resp-cross-2-3-level}
there are no level crossing and thus no response in the case 1)
for $\delta h > 0$.
In the case 2) the response remains zero $\psi=0$ until
the applied field reaches a critical field,
\bmat
h_{c}=\frac{\Delta F_{(1,0)}}{Y_{1}-Y_{0}}
\emat 
at which the free-energy level of the two states become 
the same and then jumps to $\psi=Y_{1}-Y_{0}$.
The distribution of the response $\psi$ may be decomposed
\beq
p^{\rm 2-level}(\psi,\delta h)=p^{\rm 2-level}_{0}(\psi,\delta h)+p^{\rm 2-level}_{1}(\psi,\delta h)
\label{eq-p-psi-decompose-2-level}
\eeq
where $p_{0}(\psi)$ is the contribution from the cases that
the ground state remains to be '0' and $p_{1}(\psi)$ is the
case that it is shifted to '1'. 
One can easily find (see appendix \ref{sec-direct-resp-appendix}),
\beq
p^{\rm 2-level}_{0}(\psi,\delta h)=\delta(\psi) \left \{ 
1-\int_{0}^{\infty} d\psi' 
\frac{e^{-\frac{(\psi')^{2}}{4 N\Dyeffs}}} {\sqrt{4 \pi N \Dyeffs^{2}}}
(1-e^{-\frac{\delta h}{\Tc(h)} \psi'})
\right\}
\qquad 
p^{\rm 2-level}_{1}(\psi,\delta h)=\frac{e^{-\frac{\psi^{2}}{4 N\Dyeffs}}}{\sqrt{4 \pi N \Dyeffs^{2}}}
(1-e^{-\frac{\delta h}{\Tc(h)} \psi}).
\label{eq-p-psi-2-level}
\eeq

If we take $N \to \infty$ limit with fixed $\delta h$ in 
\eq{eq-p-psi-decompose-2-level} and \eq{eq-p-psi-2-level},
the distribution function reduces
to a single delta function $\delta(\psi)$. 
On the other hand,the result also suggests us to 
consider the distribution of a scaled variable,
\beq
\tpsi\equiv \frac{\psi}{\Delta Y} \qquad \Delta Y =\sqrt{N}\Dyeffs
\label{eq-def-tpsi}
\eeq
where $\Delta Y$ is the scale of typical height of a step given
in \eq{eq-def-Delta-Y}. Its distribution function reads as,
\beq
\tilde{p}^{\rm 2-level} (\tpsi,\delta h/\hs)=
\tilde{p}_{0}^{\rm 2-level} (\tpsi,\delta h/\hs)
+ \tilde{p}_{1}^{\rm 2-level} (\tpsi,\delta h/\hs)
\label{eq-def-tpsi-2-level-scaling}
\eeq
with
\beqa
&& \tilde{p}_{0}^{\rm 2-level} (\tpsi,\delta h/\hs)
=\delta (\tpsi) \left[1-\int_{0}^{\infty} d\tpsi
\frac{e^{-\frac{\tpsi^{2}}{4}}}{\sqrt{4\pi}} 
\left(1-e^{-\frac{\delta h}{h_{s}}\tpsi}
\right)
\right]=\delta(\tpsi)\left (
1-\frac{\delta h}{h_{s}}
\right )+O\left(\frac{\delta h}{\hs} \right)^{2} \nonumber \\
&& \tilde{p}_{1}^{\rm 2-level} (\tpsi,\delta h/\hs)
=\frac{e^{-\frac{\tpsi^{2}}{4}}}{\sqrt{4\pi}}
\left(
1-e^{-\frac{\delta h}{\hs}\tpsi}
\right)=\frac{\delta h}{h_{s}}
\frac{e^{-\frac{\tpsi^{2}}{4}}}{\sqrt{4 \pi}}\tpsi  +O\left(\frac{\delta h}{\hs} \right)^{2}
\label{eq-p-psi-2-level-scaling}
\eeqa
where we find a natural scale of the field,
\beq
\hs=\frac{\Tc}{\sqrt{N}\Dy}.
\label{eq-scale-hs-2}
\eeq
Note that this is nothing but the scale of the 
typical spacing between steps given in \eq{eq-scale-hs}.

\begin{figure}[h]
\begin{center}
\includegraphics[width=0.5\textwidth]{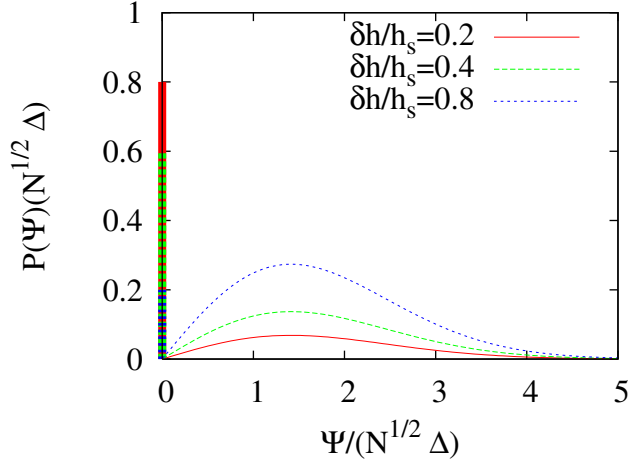}
\end{center}
\caption{Plot of the distribution function $\tilde{P}(\tpsi)$
of the scaled response $\tpsi=\psi/\sqrt{N} \Dy$. Here
contributions from the terms up to $O(\delta h/h_{s})$ in
\eq{eq-p-psi-2-level-scaling} are displayed.
For small fields $\delta h/\hs \ll 1$, most of the sample
remain intact by the presence of the field while
{\it rare samples} of a fraction of order $O(\delta h/\hs)$ make
large jumps of order $\sqrt{N}\Dy$.
}
\label{fig-plot-pof-resp-2-level}
\end{figure}

A notable feature of \eq{eq-p-psi-2-level-scaling}
is that it is an {\it analytic} function of the scaled 
field $\delta h/\hs$.
For sufficiently small field $\delta h/\hs \ll 1$, the terms
up to $O(\delta h/\hs)$ will be sufficient.
In Fig.~\ref{fig-plot-pof-resp-2-level} we display the 
distribution function using the terms up to $O(\delta h/\hs)$ 
in \eq{eq-def-tpsi-2-level-scaling}-\eq{eq-p-psi-2-level-scaling}.
We can understand the physical meaning of the profile as the following.
The term $\tilde{p}_{1}^{\rm 2-level} (\delta h/\hs,\tpsi)$, which is $O(\delta h/\hs)$, 
represent {\it rare samples} which jump by very small field
while the other term $\tilde{p}_{0}^{\rm 2-level} (\delta h/\hs,\tpsi)$ 
represent majority of the samples which remain
intact by the presence of the field. Thus the distribution function describes the
intermittency of the mesosocpic response.

\subsubsection{Three-level model}

Now we extend the analysis including the 3rd lowest state and 
consider a three-level model, which consists of
the ground state $0$, 1st excited state $1$ and the 2nd 
excited state $2$ under $h$. The pattern of level crossings
can be classified into $3!=6$ cases depending on the
ordering of $Y_{0}$, $Y_{1}$ and $Y_{2}$
as shown in Fig.~\ref{fig-resp-cross-2-3-level}.
In this model there can be at most $2$ level crossings within a given 
interval $\delta h$ while only $1$ level crossing was allowed
in the two-level model.

Similarly to \eq{eq-def-tpsi-2-level-scaling},
it is natural to decompose the scaled probability distribution function as,
\beq
\tilde{p}^{\rm 3-level}(\tpsi,\delta h/\hs)=\tilde{p}^{\rm 3-level}_{0}(\tpsi,\delta h/\hs)
+\tilde{p}^{\rm 3-level}_{1}(\tpsi,\delta h/\hs)+\tilde{p}^{\rm 3-level}_{2}(\tpsi,\delta h/\hs)
\label{eq-def-tpsi-3-level-scaling}
\eeq
where the 1st, 2nd and 3rd terms in r.h.s. corresponds to
the cases that the ground state at $h+\delta h$ is at '0', '1' and
'2' respectively. They are computed in the appendix \ref{sec-direct-resp-appendix}.
The results read as,
\beqa
&& \tilde{p}^{\rm 3-level}_{0}(\tpsi,\delta h/\hs)= 
\delta(\tpsi)\left [ 1- \int_{0}^{\infty} d\tilde{\psi} \left (
\tilde{p}^{\rm 3-level}_{1}(\tpsi)+\tilde{p}^{\rm 3-level}_{2}(\tpsi) \right)
\right ]  \nonumber \\
&& \tilde{p}^{\rm 3-level}_{1}(\tpsi,\delta h/\hs)=\frac{e^{-\frac{\tpsi^{2}}{4}}}{\sqrt{4\pi}}
\left( 1 - e^{-\frac{\delta h}{\hs}\tpsi} \right)
\left[ 
1-\int_{-\infty}^{\infty} \frac{dz_1}{\sqrt{4\pi}}e^{-\frac{z_{1}^{2}}{4}}
\int_{\frac{z_{1}+\tpsi}{2}}^{\infty}\frac{dz_{2}}{\sqrt{2\pi}}
e^{-\frac{z_{2}^{2}}{2}}
\left(
1-e^{-\frac{1}{2}\frac{\delta h}{\hs}(z_{2}-\frac{z_{1}+\tpsi}{2})}
\right)
\right] \nonumber \\
&& \tilde{p}^{\rm 3-level}_{2}(\tpsi,\delta h/\hs)=\left( 
1-2e^{-\frac{\delta h}{\hs}\tpsi}+e^{-2\frac{\delta h}{\hs}\tpsi}
\right)
\int_{-\infty}^{\infty} \frac{dz_1}{\sqrt{4\pi}} e^{-\frac{z_{1}^{2}}{4}}
\int_{\frac{z_{1}-\tpsi}{2}}^{\infty} \frac{dz_2}{\sqrt{2\pi}} e^{-\frac{z_{2}^{2}}{2}} \nonumber \\
&& + 
\left( 
1-2e^{-\frac{\delta h}{\hs}\tpsi}+e^{-2\frac{\delta h}{\hs}\tpsi}
\right)
\int_{-\infty}^{\infty} \frac{dz_1}{\sqrt{2\pi}} e^{-\frac{z_{1}^{2}}{2}}
\int_{z_{1}}^{\infty} \frac{dz_2}{\sqrt{2\pi}} e^{-\frac{z_{2}^{2}}{2}}
 \left( 1-e^{-2X} \right)_{X=\frac{\delta h}{\hs}(\tpsi-(z_{2}-z_{1}))}
\nonumber \\
&& +\int_{-\infty}^{\infty} \frac{dz_1}{\sqrt{2\pi}} e^{-\frac{z_{1}^{2}}{2}}
\int_{z_{1}}^{\infty} \frac{dz_2}{\sqrt{2\pi}} e^{-\frac{z_{2}^{2}}{2}}
\left[
-(1-e^{-2X})+\frac{4}{3}(1-e^{-3X})-\frac{1}{2}(1-e^{-4X})
\right]_{X=\frac{\delta h}{\hs}(\tpsi-(z_{2}-z_{1}))}.
\label{eq-p-tpsi-3-level-scaling}
\eeqa

Comparing \eq{eq-p-tpsi-3-level-scaling} and 
\eq{eq-p-psi-2-level-scaling} we find that following.
First $p_{2}(\psi,\delta h/\hs)$, which is absent in the 2-level model,
is $O((\delta h/\hs)^{2})$.  Second, 
$p_{0}(\psi,\delta h/\hs)$ and $p_{1}(\psi,\delta h/\hs)$
of the 3-level model agree with those of 2-level model
up to $O(\delta h/\hs)$. The situation is reminiscent of the
low temperature expansion where we found that $M$-level model
give correct results up to $O((T/T_{c})^{M-1})$ as we prooved in
section \ref{subsec-accuracy-m-level}. In the present context
we do not have an equivalent proof but we believe that $M$-level model
gives correct results up to  $O((\delta h/\hs)^{M-1})$. By this way, we can 
go beyond the limit of single steps and access successive steps.

\subsubsection{Discussions}

The above results suggest that distribution of ths scaled variable 
$\tpsi$ follows a generic scaling form,
\beqa
&&\tilde{P}(\tpsi,\delta h/\hs)= 
\delta (\tpsi) 
\left(1-\int_{0}^{\infty} d\tpsi \tilde{P}_{\rm jump}(\tpsi,\delta h/\hs) \right)
+ \tilde{P}_{\rm jump}(\tpsi,\delta h/\hs) \nonumber \\
&& \tilde{P}_{\rm jump}(\tpsi,\delta h/\hs) \equiv \sum_{n=1}^{\infty}
\left (\frac{\delta h}{\hs}\right)^{n} f_{n}(\tpsi) 
\qquad \mbox{for} \qquad \delta h/\hs  \ll 1
\label{eq-scaling-p-tpsi}
\eeqa
The term $\tilde{P}_{\rm jump}(\tpsi,\delta h/\hs)$ represent contributions of the
samples in which the ground state changes responding to the field.
We expect correct value of the coefficient $f_{n}$ will be obtained
by considering $n+1$-level model, which can take into account possible 
$n$ successive jumps in a given internal $\delta h$. 
We expect that the qualitative feature 
is explained well by Fig.~\ref{fig-plot-pof-resp-2-level} and
higher order terms of $\delta h/\hs$ give only mild modifications
without changing very much the shape of the distribution function.

The distribution function of the response is a {\it smooth} 
function of the scaled field $\delta h/\hs$ in spite of the fact that 
the underlying responses in each sample are highly non-linear processes.
It must be contrasted with the expansion of the {\it response itself} in 
a given sample in  power series of $\delta h/\hw$ (See \eq{eq-resp-two-level}).
The static FDT relies on the latter kind of expansion and thus it can 
{\it not} be applied beyond the scale of a single step $\hs$.
On the other hand, it must be noted that the power series of $\delta h/\hs$ 
will converge only for sufficiently small $\delta h/\hs$. Since $\hs$ given 
in \eq{eq-scale-hs-2} (or \eq{eq-scale-hs}) vanishes 
in the thermodynamic limit $N \to \infty$,
the new power series also describes not macroscopic but mesoscpic responses.

Let us call the regime $\delta h \ll \hs$, where the expansion converges, 
as {\it weakly perturbed regime}, and the other regime 
$\delta h \gg \hs$ as {\it strongly perturbed regime} following \cite{SY}.
\begin{itemize}
\item 
An important generic feature in the weakly perturbed regime 
is that even with such a small field 
$\delta h \ll \hs$, where most of the samples do not respond at all, 
there are a fraction of 'fragile' samples of order $\delta h/\hs$ 
which are able to respond strongly chaging its own ground state. 
Although they are {\it rare}, they dominate average behaviours. 
This means, static chaos effect does not appear all of a sudden 
at certain large sizes (or strong enough field) but appear strongly on 
rare samples even at small sizes (or weak field)
and thus appear gradually in the averaged quantities.
The weakly perturbed regime is examined numerically in detail 
in the static chaos problems of directed polymer in random
media in \cite{SY}. Recently it was also examined in detail
in a study of bond-chaos effect in the Edwards-Anderson spin-glass model
\cite{Krzakala-Bouchuad}. 
The weakly perturbed regime is also proposed to explain some features
observed in temperature-shift experiments on spin-glasses \cite{uppsala-osaka,ghost}.

\item In the strongly perturbed regime $\delta h \gg \hs$, where most of
the samples will respond, we still expect the distribution function
of the repsonse $\psi$ follows the generic scaling form 
as \eq{eq-scaling-p-tpsi} but with a different functional form,
\beq
\tilde{P}(\tpsi,\delta h/\hs) 
\sim \frac{\exp \left (-\frac{(\tpsi-\delta h/\hs)^2}{4} \right)}{\sqrt{4\pi}} 
\qquad \mbox{for} \qquad \delta h/\hs \gg 1
\eeq
which means
\beq
P(\psi,\delta h) \sim 
\frac{\exp \left(-\frac{(\psi-  N \delta h/T_{c}(h))^2}{4N (\Dyeffs(h)^{2})}\right)}{\sqrt{4\pi N (\Dyeffs(h))^{2}}} \qquad \mbox{for} \qquad \delta h/\hs \gg 1. 
\eeq
in terms of the original unscaled variable $\psi=Y(h+\delta h)-Y(h)$.

The reason for the above functional form is the following. Since there 
will be large number of steps $\delta h/\hs \sim O(\sqrt{N}) \gg 1$ 
witin a fixed interval $\delta h$, each step with an increment of 
order $\Delta Y \sim \Dyeffs(h) \sqrt{N}$,
the total response within $\delta h$ will be $O(N)$. 
This is consistent with the fact that the response must 
be {\it extensive} in the thermodynamic sense.
Since the total response is the sum of suc a large number of steps, we expect
its sample-to-sample fluctuation follows the Gaussiandistribution 
due to the central limit theorem. 
From the thermodynamic calcurations we know that
\beq
\sav{\psi}=\sav{Y}_{h+\delta h}-\sav{Y}_{h}=N  \frac{(\Dyeffs(h))^{2}\delta h}{T_{c}(h)}
\label{eq-psi-strong}
\eeq
as given in \eq{eq-y-av-glass}. Since there will be for sure
some steps in the strongly perturbed regime, $Y$ varianle 
of the ground state at $h$ and that at $h+\delta h$ 
must be uncorrelated. Thus the squared-variance of $\psi$ will be just the sum
of those at $h$ and $h+\delta h$, which amounts to $2N \Dyeffs(h)^{2}$.
In short, in the strongly perturbed regime we observe thermodynamic
response which is {\it self-averaging}, while in the weakly perturbed regime,
we observer mesoscopic response  which is {\it non-self averaging}. 

\end{itemize}

Lastly it is interesting to note that {\it average} response, obtained
in the weakly perturbed regime, matches precisely with that 
in the strongly perturbed regime \eq{eq-psi-strong}. 
From \eq{eq-def-tpsi-2-level-scaling}-\eq{eq-p-psi-2-level-scaling} we find
\beq
\sav{\psi}=\frac{\delta h}{\hs}
+ O \left(\frac{\delta h}{\hs} \right)^{2}
\eeq
which means,
\beq
\sav{\psi}= N \frac{(\Dyeffs(h))^{2}\delta h }{\Tc(h)}
+ O \left(\frac{\delta h}{\hs} \right)^{2}.
\label{eq-sav-psi-expansion-2-level}
\eeq
This agrees with the average in the strongly perturbed regime 
\eq{eq-psi-strong}. Both are linear in $\delta h$ so that 
they yeild the same linear susceptibility.
This coincidence is consistent with the expectation
that disorder-average of the linear susceptibility 
and 1st derivative of thermodynamic response should agree
(See appendix \ref{sec-note-linear-sus}). 

\subsection{Cusp in the overlap}
\label{subsec-overlap-lowT}

\begin{figure}[h]

\begin{center}
\includegraphics[width=0.6\textwidth]{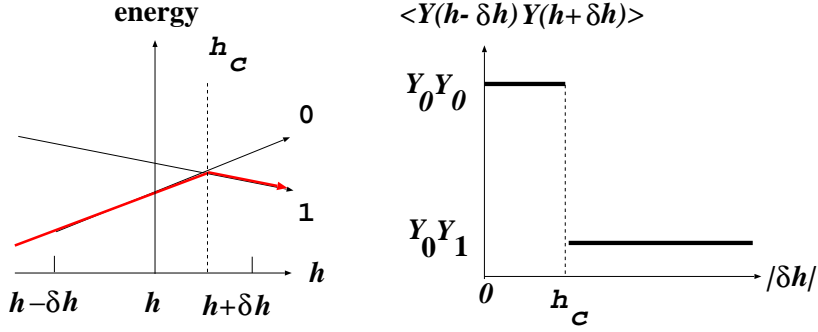}
\end{center}
\caption{Level crossing and decay of the overlap between
2-replicas subjected to $h-\delta h$ and $h+\delta h$. Here
an example of the case $Y_{0} > Y_{1} > 0$ is shown.
}
\label{fig-overlap-2-level}
\end{figure}

Finally we analyze the overlap function, given by \eq{eq-def-overlap},
between two replicas subjected to slightly
differnt fields which reads as,
\beq
C(\delta h)=\frac{
\sav{ \tav{Y(h+\delta h)} \tav{Y(h-\delta h)}}
}{
\sav{\tav{Y(h)}^{2}}.
}
\eeq

\subsubsection{FDT approach}

Let us consider a Taylor expansion of the correlation function 
$\tav{Y(h+\delta h)} \tav{Y(h-\delta h)}$ in power series of $\delta h$,
\beqa
\tav{Y(h+\delta h)} \tav{Y(h-\delta h)}
=(\kappa_{1}(Y))^{2}-(\beta \delta h)^{2}
((\kappa_{2}(Y))^{2}-\kappa_{3}(Y)\kappa_{1}(Y)) + O (\beta \delta h)^{4}
\label{eq-overlap-thermal-cumulants}
\eeqa
Note that terms of odd powers of $\delta h$ are absent smiply due to
the symmetry of the correlation function under 
$\delta h \leftrightarrow -\delta h$.

\begin{figure}[h]

\begin{center}
\includegraphics[width=0.4\textwidth]{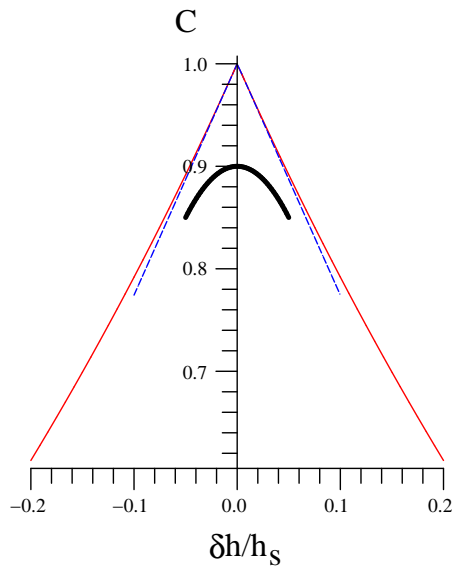}
\end{center}
\caption{The profile of the overlap function. It exhibits
a cusp at $T=0$, which is rounded at finite temperatures.
The doted line is \eq{eq-overlap-T=0-2-level} obtained at
zero temperature with the two-level model, which correctly 
captures the feature around the top of the cusp.
The solid line is the result obtained by the replica method at $T=0$
which yeilds the full profile. (See  sec. \ref{subsec-overlap-replica}.)
The dashed line is the curve at $T/T_{\rm c}=0.1$ \eq{eq-overlap-fdt}
obtained by the FDT approach (using the 2-level model), 
which describes thermally rounded region of wdith $\hw$ around
the center.
}
\label{fig-cusp}
\end{figure}

We have obtained $\kappa_{2}^{2}(Y)$ up to $O(T/T_{c})^{3}$ 
as \eq{eq-kappa-2-2}. To the same order we find,
\beq
\sav{\kappa^{2}_{1}(Y)}=(\Dy \sqrt{N})^{2}
\left \{ 1-  \frac{T}{T_{c}} + O\left(\frac{T}{T_{c}} \right)^{3}\right\}
\qquad 
\sav{\kappa_{3}(Y)\kappa_{1}(Y)}=-(\Dy \sqrt{N})^{4}
\left \{ \frac{T}{T_{c}} - \left(\frac{T}{T_{c}} \right)^{2}
+ O\left(\frac{T}{T_{c}} \right)^{3}\right\}
\eeq
as reported in  Appendix \ref{sec-lowT-expansion-sus-appendix}.
Combining these results we obtain,
\beq
C(\delta h)= 1 + O\left(\frac{T}{T_{c}} \right)^{3}
- \left (\frac{\delta h}{\hw} \right)^{2}
\left \{2 \frac{T}{T_{c}}
+\left(\frac{T}{T_{c}} \right)^{2}
+ O\left(\frac{T}{T_{c}} \right)^{3} \right\}
+O\left(\frac{\delta h}{\hw} \right)^{4} 
\label{eq-overlap-fdt}
\eeq
where $\hw=T/(\Dy \sqrt{N})$ as given by \eq{eq-three-scales-steps}.
As shown in Fig.~\ref{fig-cusp}, the series \eq{eq-overlap-fdt}
describes the thermally rounded region of the correlation function 
around $\delta h=0$.

We know that the thermal cumulants which appear 
\eq{eq-overlap-thermal-cumulants} are non-self averaging. Thus the
overlap must be non-self averaging as well.
  
\subsubsection{Zero temperature limit}

Now we turn to the case of zero temperature limit where the FDT approach
breaks down since $\hw \to 0$. 
For simplicity we only consider the 2-level model.
As sohwn in Fig. \ref{fig-overlap-2-level},
the overlap between the two replicas remains $Y^{2}_{0}$ in the range
$0 < |\delta h | < h_{c}$ and becomes $Y_{0}Y_{1}$ at $|\delta h| > \delta h_{c}$. The critical field is given by $\delta h_{c}=\Delta F/(Y_{0}-Y_{1})$.

Considering the cases $Y_{0} > Y_{1}$ and $Y_{1} > Y_{0}$,  we obtain
\beqa
&& \lim_{T \to 0} \sav{\tav{Y(h-\delta h)Y(h+\delta h)}}=
\int_{-\infty}^{\infty} dY_{0}\int_{Y_{0}}^{\infty} dY_{1}
P(Y_{0})P(Y_{1})
\left [
Y_{0}^{2}\int_{|\delta h| (Y_{1}-Y_{0})}^{\infty}d \Delta F \rho(\Delta F)
+Y_{0}Y_{1}\int_{-\infty}^{|\delta h| (Y_{1}-Y_{0})}d \Delta F \rho(\Delta F)
\right] \nonumber \\
&& + \int_{-\infty}^{\infty} dY_{0}\int_{-\infty}^{Y_{0}} dY_{1}
P(Y_{0})P(Y_{1})
\left [
Y_{0}^{2}\int_{|\delta h| (Y_{0}-Y_{1})}^{\infty}d \Delta F \rho(\Delta F)
+Y_{0}Y_{1}\int_{-\infty}^{|\delta h| (Y_{0}-Y_{1})}d \Delta F \rho(\Delta F)
\right]  \nonumber \\
&& = N \Dy^{2} 
\left[
1-\frac{\delta h}{\hs}
\left (
\int_{-\infty}^{\infty}\frac{dy_{0}}{\sqrt{2\pi}}
\int_{-\infty}^{y_{0}}\frac{dy_{1}}{\sqrt{2\pi}}
y_{0} (y_{0}-y_{1})^{2}
-\int_{-\infty}^{\infty}  \frac{dy_{0}}{\sqrt{2\pi}}
\int_{y_{0}}^{ \infty}\frac{dy_{1}}{\sqrt{2\pi}}
y_{0} (y_{0}-y_{1})^{2}
\right)
+ O \left(\frac{\delta h}{\hs} \right)^{2}
\right] 
\eeqa
from which we find,
\beq
\lim_{T \to 0} C(\delta h)=
1-\frac{4}{\sqrt{\pi}}\frac{|\delta h|}{\hs} 
+  O \left (\frac{|\delta h|}{\hs} \right)^{2}.
\label{eq-overlap-T=0-2-level}
\eeq

A remarkable feature of the correlation function at $T=0$ is that it
exhibits a cus singurality at $\delta h=0$. 
In Fig.~\ref{fig-overlap-2-level} it is evident that the cusp is
strongly non-self averaging.
The linear $|\delta h|$
profile is due to the presence of {\it rare samples} which have critical 
fields $\delta h_{c}$ close to $\delta h=0$.

\section{Replica approach}
\label{sec-replica}

In this section we analyze the mesoscopic responses by a replica approach. This complements the low temperature exapnsion approach discussed in the previous section. We construct a generating functional from which various correlation functions associated with the mesoscopic responses can be computed exactly.

\subsection{Generating functional}
\label{subsec-generating-functional}

To analyze statistical properties of mesoscopic static responses 
it is useful to consider the following object:
\beq
\frac{\partial^{k_{1}}}{\partial (\beta \delta h)^{k_{1}}}
\frac{\partial^{k_{2}}}{\partial (\beta \delta h)^{k_{2}}}
\cdots
\frac{\partial^{k_{p}}}{\partial (\beta \delta h)^{k_{p}}}
\sav{
(-\beta F (\delta h_{1}))
(-\beta F(\delta h_{2}))\ldots
(-\beta F(\delta h_{p}))}
\eeq
with $k_{1} \geq 1$, $k_{2} \geq 1$,...,$k_{p} \geq 1$.
Here $F(\delta h)$ is the free-energy of a system sujected to
a small probing field $\delta h$. 
The derivatives with respect to the probing fields
yeild disorder averages of various kinds of correlation 
functions of our interests.
Using the replica formalism the product of the free-energies
$\sav{F(\delta h_{1})F(\delta h_{2})\ldots F(\delta h_{p})}$
in the above expression can be replaced by
\beq
\lim_{n_{1},n_{2},\ldots,n_{p}\to 0}
\frac{1}{n_{1}n_{2}\cdots n_{p}}\overline{Z^{n_{1}}(\delta h_{1})
Z^{n_{2}}(\delta h_{1})\ldots Z^{n_{p}}(\delta h_{p})}
\label{eq-z-real-replicas}
\eeq
where $Z(\delta h)$ is the partition function of the system.
Thus we need to consider {\it real} replicas $r=1,2,\ldots,p$ which
are subjected to different probing fields 
$\delta h_{1},\delta h_{2},\ldots, \delta h_{p}$ 
and replicated further into  $n_{1},n_{2},\ldots,n_{r}$ replicas.

\subsubsection{p-spin Ising model and REM}

To be specific, let us consider the $p$-spin Ising mean-field spin-glass 
model \cite{REM,GM} given \eq{eq-hamiltonian-pspin}.
Following the standard steps \cite{GM} we obtain the replicated
partition function of $n$-replicas,
\beq
\sav{Z^{n}}={\rm Tr}_{\sigma}
\sav{e^{-\beta \sum_{a=1}^{n}H(\{\sigma^{a}_{i}\})}}
=e^{nN(\beta J)^{2}/4}
\int \prod_{a < b} dQ_{ab} \int \prod_{a < b} \frac{d \lambda_{ab}}{2\pi}
e^{-N G(\{Q_{ab}\},\{\lambda_{ab}\})}
\label{eq-replica-z-p-spin}
\eeq
where
\beq
G(\{Q_{ab}, \lambda_{ab} \})=-\frac{(\beta J)^{2}}{4}
\sum_{a \neq b} Q_{ab}^{p}+\frac{1}{2}\sum_{a \neq b}\lambda_{ab} Q_{ab}
- \ln {\rm Tr}_{\sigma} e^{{\cal L}}
\eeq
with
\beq
{\cal L} = \frac{1}{2} \sum_{a\neq b} \lambda_{ab} \sigma_{a} \sigma_{b} 
+ \sum_{a} \beta (h +\delta h_{a})\sigma_{a}
\eeq
Here $\lambda_{ab}$ are Lagrangian multipliers introduced to define the overlap,
\beq
Q_{ab}=\frac{1}{N}\sum_{i=1}^{N} \sigma^{a}_{i}\sigma^{b}_{i}.
\eeq
The integrals over $Q_{ab}$ and $\lambda_{ab}$ are evaluated using
the saddle point method.

Note that we have added small perturbing field $\delta h_{a}$ 
applied on each replica in addition to the uniform external field $h$. 
{\it We suppose that the perturbing fields are infenitesimally small 
such that they do not change the saddle point itself.}

\begin{figure}[t]
\begin{center}
\includegraphics[width=0.7\textwidth]{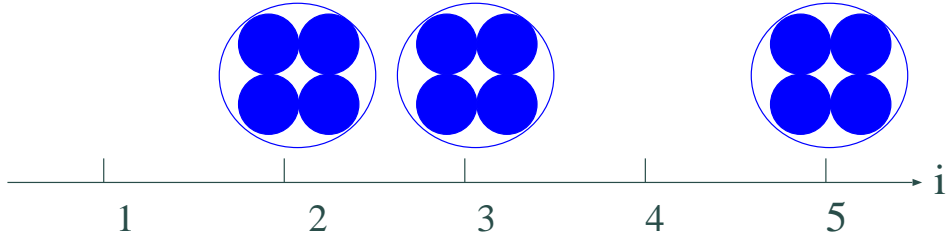}
\end{center}
\caption{A schematic representation of an 1 step RSB ansatz.
$n$ replicas ('atoms') are grouped in $n/m$ clusters ('molecules')
of size $m$. Different clusters occupy different metastable states $i$.
}
\label{fig-m-atom}
\end{figure}

\begin{figure}[t]
\begin{center}
\includegraphics[width=0.2\columnwidth]{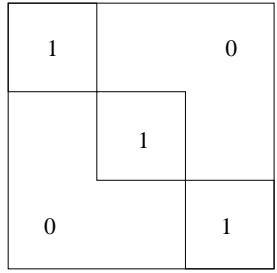}
\end{center}
\caption{Parisi's matrix $I$ for a 1 step RSB solution of size $n \times n$. 
The size of the blocks are $m\times m$. Different solutions,
 corresponding to different grouping of the atoms in
 Fig.~\ref{fig-m-atom}, are obtained by permutations of the lows and columns.
}
\label{fig-1rsb-matrix}
\end{figure}

Now we consider a particular 1 step RSB saddle point in which the $n$ 
replicas are groupd into clusters ${\cal C}=1,2,\ldots,n/m$ of size $m$.
(see Fig.~\ref{fig-m-atom}.)
The parameter $m$ is known to depend on the temperature
as (See \eq{eq-m-appendix} for the case of the REM),
\beq
m=\frac{T}{\Tc(h)}.
\label{eq-m}
\eeq
For such a saddle point the $n\times n$ matricies $Q_{ab}$ 
and $\lambda_{ab}$ can be put into block-diagonal forms with block size $m$
(see Fig. \ref{fig-1rsb-matrix}). We denote the matrix elements in the
block diagonal parts as $q_{1}$,$\lambda_{1}$ and the others 
as $q_{0}$ and $\lambda_{0}$. 
Then one obtains for sufficiently small $n$,
\beqa
\ln {\rm Tr}_{\sigma} e^{\cal L}=-n\frac{\lambda_{1}}{2}
+ \int Dz_{0} \sum_{{\cal C}=1}^{n/m}
\ln \int D_{z_{1{\cal C}}} \prod_{a \in {\cal C}} (2 \cosh \Xi(h+\delta h_{a}))
\eeqa
with $\int Dz \ldots \equiv \int_{-\infty}^{\infty} \frac{dz}{\sqrt{2\pi}}e^{-z^{2}/2}\ldots $ and
\beq
\Xi(h)=\beta h+\sqrt{\lambda_{0}}z_{0}+
\sqrt{\lambda_{1}-\lambda_{0}}z_{1{\cal C}}.
\eeq
Making an exansion about the perturbing field $\delta h_{a}$, we obtain
\beq
\ln {\rm Tr}_{\sigma} e^{\cal L}=  (\ln {\rm Tr}_{\sigma} e^{\cal L})_{\delta h=0}
+\mu \sum_{a=1}^{n} \beta \delta h_{a}
+ T\chi^{\rm EA} \sum_{a=1}^{n} \frac{(\beta \delta h_{a})^{2}}{2}
+\frac{\Dy^{2}}{2} \sum_{{\cal C}=1}^{n/m} \sum_{a \in {\cal C}, b \in {\cal C}}
\beta \delta h_{a} \beta \delta h_{b} + O (\delta h)^{3}
\eeq
with 
\beq
\mu=\int D_{z_{0}} \frac{\int D_{z_{1}} \cosh^{m}(h)\tanh \Xi(h)}{
\int D_{z_{1}} \cosh^{m}(h)
}
\qquad 
T \chi_{\rm EA}=1-\int D_{z_{0}} 
\left \{ \frac{\int D_{z_{1}} \cosh^{m}(h)\tanh \Xi(h)}{
\int D_{z_{1}} \cosh^{m}(h) 
}\right \}^{2}
\label{eq-chi-ea-h=0}
\eeq
and
\beq
\Dyeff^{2}=\int D_{z_{0}}
\left [
\frac{\int D_{z_{1}} \cosh^{m}(h)\tanh^{2} \Xi(h)}{
\int D_{z_{1}} \cosh^{m}(h)}
-
\left \{ \frac{\int D_{z_{1}} \cosh^{m}(h)\tanh \Xi(h)}{
\int D_{z_{1}} \cosh^{m}(h) 
}\right \}^{2}
\right].
\label{eq-delta-h=0}
\eeq

The replicated parition function \eq{eq-replica-z-p-spin} contains
contributions from other 1 step RSB solutions obtained by permutations
of replicas. One must also integrate over fluctuations around saddle points.
Then the replicated partition function is obtained as,
\beq
 \sav{Z^{n}} 
 =  c \sum_{\rm SP} \exp \left [ N\mu  \sum_{\alpha=1}^{n} \beta\delta h_{a}
+ N T\chi^{\rm EA} \sum_{\alpha=1}^{n} \frac{(\beta \delta h_{\alpha})^{2}}{2} 
 +\frac{( \sqrt{N} \Dyeffs)^{2}}{2} 
\sum_{{\cal C}=1}^{n/m}
\sum_{\alpha \neq \beta \atop \alpha,\beta \in {\cal C}} \beta \delta h_{\alpha}
\beta \delta h_{\beta}
 + O (\delta h)^{3}
\right]
\eeq
where  $\sum_{\rm SP}$ stands for summation over all possible permutations
of replicas, i.e. different ways of grouping the $n$ replicas into $n/m$
clusters of size $m$. The factor
$c$ stands for a common factor which does not depend on the probing field
$\delta h$'s.
Then for the case that we have $p$ replicas (See \eq{eq-z-real-replicas}) we
obtain the following expression,
\beqa
 \sav{Z_{1}^{n_{1}}Z_{2}^{n_{2}} \ldots Z_{p}^{n_{p}}} 
 = &&c \sum_{\rm SP} \exp \left [ 
\mu \sum_{a=1}^{n} \beta \delta h_{a}
+ T\chi^{\rm EA} \sum_{a=1}^{n} \frac{(\beta \delta h_{a})^{2}}{2} \right .
\nonumber \\
&& \left. +\frac{( \sqrt{N} \Dyeffs)^{2}}{2} \sum_{r,s=1}^{p}
\beta \delta h_{r} \beta \delta h_{s} 
\sum_{\beta=(r,1)}^{(r,n_{r})}
\sum_{\beta=(s,1)}^{(s,n_{s})}
 \delta_{i_{\alpha},i_{\beta}}
 + O (\delta h)^{3}
\right].
\label{eq-generating-functional-p-spin}
\eeqa
Here we have $p$ groups of replicas. We labeled the groups
as $r=1,2,\ldots,p$. $r$-th group consists of $n_{r}$ replicas
which we labeled as $(r,1),(r,2),\ldots,(r,n_{r})$. The total number of
replicas is
\beq
\nt=\sum_{r=1}^{p} n_{r}.
\eeq

Let us explain the physical meaning of the terms in the argument of the 
exponential function in \eq{eq-generating-functional-p-spin}.
\begin{itemize}
\item It is easy to veryfy that 
the parameter $\mu$ for the terms at order $O(\delta h)$
represents the expectation value of the magnetization 
$\mu = \sav{\tav{\sigma}}$.
\item We have devided the $O(\delta h)^{2}$ terms into two parts:
one which  involves single replicas and the other which envolves
pairs of replicas. Physically we understand the former as due
to {\it intra-state response} and the latter as due to 
{\it inter-state response}. As is well known, in the limit $p\to \infty$ 
the $p$-spin model reduces to the REM \cite{REM,GM}
in which the {\it intra-state} linear-suscepibility $\chi_{\rm EA}$
is zero by construction of the model. 
\end{itemize}

Here let us note that the evaluations of the 
$O(\delta h)^{2}$ terms must be done carefully. Actually the expressions 
for the parameters $\chi_{\rm EA}$ and $\Dyeffs$
\eq{eq-chi-ea-h=0} and \eq{eq-delta-h=0}
are valid only for zero external field $h=0$.
This is because for $h \neq 0$, the magnetization $\mu$ is non-zero
so that integrations over fluctuations around the saddle 
points yield additional $O(\delta h)^{2}$ terms. Such calcurations 
are very difficult to perform in general. An exception is the case of REM
($p \to \infty$ limit) in which we only need to consider
fluctuation of the {\it size}  of the clusters $m$ in the 1 step RSB solutions.
as we report in appendix \ref{appendix-m-fluctuation}.

For the rest of the present section we consider the case of
REM $p \to \infty$ which allows us to neglect intra-state responses.
As we noted in section \ref{subsec-rem},
REM can be regarded as an effective model for a generic 1 step RSB systems.

\subsection{Sample-to-sample fluctuations of linear and non-linear susceptibilities}
\label{subsec-thermal-fluctuation-replica}

We now analyze the statistics of the susceptibilities
using the replica approach to confirm the results obtained by the
low temperature expansion approach in sec. 
\ref{subsec-thermal-fluctuation-lowT}.
As we show below we obtain presumably exact
results for the sample average of the $p$-th moment of 
the thermal cumulants $\sav{\kappa_{k}^{p}(Y)}$ given by \eq{eq-def-kappa-k-p}.
Interestingly the results are expressed in power series of the parameter 
$m=T/\Tc(h)$ (sed \eq{eq-m}) which can actually be viewed
as  expansions in power series of $T$, i.e. low temperature expansion.

Sample-average of $p$-th moment of the thermal cumulant $\kappa_{k}(Y)$
given by \eq{eq-def-kappa-k-p} can be computed as,
\beqa
&& \sav{\kappa^{p}_{k}(Y)}= \left.
\lim_{ n_{1},n_{2},\ldots,n_{p} \to 0} 
\frac{\partial^{k}}{\partial (\beta \delta h_{1})^{k}}
\frac{\partial^{k}}{\partial (\beta \delta h_{2})^{k}}
\cdots \frac{\partial^{k}}{\partial (\beta \delta h_{p})^{k}}
\frac{\sav{Z_{1}^{n_{1}}Z_{2}^{n_{2}}\ldots
Z_{p}^{n_{p}}}}{n_{1}n_{2}\cdots n_{p}} \right |_{\delta h=0} \nonumber \\
&&= (\sqrt{N}\Dyeffs)^{kp}
\left.
\lim_{ n_{1},n_{2},\ldots,n_{p} \to 0} \frac{1}{n_{1} n_{2} \cdots n_{p}}
\frac{\partial^{k}}{\partial x_{1}^{k}}
\frac{\partial^{k}}{\partial x_{2}^{k}}
\cdots \frac{\partial^{k}}{\partial x_{p}^{k}}
\spav{
\exp \left[
\frac{1}{2}\sum_{r,s=1}^{p} x_{r}x_{s}A_{rs}
\right]
\right |_{x=0}}
\label{eq-kappa-n-p-replica}
\eeqa
where we introduced a scaled variable $x_{r}$ defined as
\beq
x_{r}=\frac{\beta \delta h_{r}}{\Dyeffs\sqrt{N}}=\frac{\delta h_{r}}{\hw}
\label{eq-def-h-over-hw}
\eeq
with $\hw$ given in \eq{eq-three-scales-steps} and
\beq
A_{rs}\equiv  \sum_{\alpha=(r,1)}^{(r,n_{r})} \sum_{\beta=(s,1)}^{(s,n_{s})}
\delta_{i_{\alpha},i_{\beta}}
\label{eq-def-A}
\eeq
and the bracket $\spav{\ldots}$ means to take an average over all possible
partitions of $\nt$ replicas into $\nt/m$ clusters of size $m$.

Let us begin from the linear susceptibility. 
As in the approach by the low temperature expansion
we study 1st and 2nd moment of the 2nd thermal cumulant
$\kappa_{2}(Y)$.
Using the prescription \eq{eq-kappa-n-p-replica} for
the sample-average of the 1st moment of the 2nd thermal
cumulant $\kappa_{2}(Y)$ we find,
\beq
 \sav{\kappa_{2}(Y)}= (N \Dyeff^{2})\lim_{n \to 0}\frac{1}{n} 
\spav{A}
\eeq
where 
\beq
A=\sum_{\alpha,\beta}^{n} \delta_{i_{\alpha},i_{\beta}}=
n+\sum_{\alpha \neq\beta}^{n} \delta_{i_{\alpha},i_{\beta}}
\eeq

We now need to evaluate $\spav{\delta_{i_{\alpha},i_{\beta}}}$ 
for $\alpha=\beta$ which is the probability that two distinct 
replicas belong to a common cluster. It is evaluated as,
\bmat
\spav{\delta_{i_{\alpha },i_{\beta}}}
=\frac{(n/m)\cdot m(m-1)\cdot(n-2)!}{n!}=\frac{m-1}{n-1}.
\emat
Here the denominator $n!$ is the total number of solutions
and the numerator is the number of solutions in which
two distinct replicas belong to a common cluster.
The first factor $n/m$ in the numerator is the number of ways 
to choose a cluster among $n/m$ clusters. The 2nd factor $m(m-1)$ 
is the number of ways to choose two replicas
out of $m$ replicas belonging to the cluster. 
The last factor $(n-2)!$ is the number
of ways to put the rest of replicas in the remaining $n-2$ seats.
Then we obtain
\bmat
\spav{\sum_{\alpha \neq \beta}^{n} \delta_{i_{\alpha },i_{\beta}}}
= n(n-1) \spav{ \delta_{i_{\alpha },i_{\beta}}}= n (m-1).
\emat

Using the above results we finally obtain a very simple
result,
\beq
\sav{\kappa_{2}(Y)}= (\sqrt{N}\Dyeff)^{2} m.
\label{eq-kappa-2-1-replica}
\eeq
Since $m=T/\Tc(h)$ as given in \eq{eq-m},
this result agrees with that obtained by the low temperature
expansion given by \eq{eq-kappa-2-1} obtained in 
sec \ref{subsec-thermal-fluctuation-lowT}.
There we have found that $O(T^{2})$ term is zero. The above result
means that actually all higher order terms are zero.

Next let us evaluate the sample-average of the 2nd moment of the 2nd thermal
cumulant $\kappa_{2}(Y)$.  Using \eq{eq-kappa-n-p-replica} we find,
\beq
\sav{\kappa^{2}_{2}(Y)}= 
(N \Dyeff^{2})^{2}\lim_{n_{1},n_{2} \to 0}\frac{1}{n_{1}n_{2}}
(2\spav{A^{2}_{12}} +\spav{A_{11}A_{22}}).
\eeq
We report the details of the computations
in appendix \ref{appendix-sample-fluctuation-replica}.
Here we only quote the result,
\beq
\sav{\kappa^{2}_{2}(Y)}= (\sqrt{N}\Dyeff)^{4} m,
\label{eq-kappa-2-2-replica}
\eeq
which again agree with the result obtained by the low temperature
expansion given by \eq{eq-kappa-2-2}. In sec 
\ref{subsec-thermal-fluctuation-lowT}
we have found that $O(T^{2})$ term is zero. The above result
means again that actually all higher order terms are zero.

Last let us study the non-linear susceptibility $\chi_{2}$. 
As in the approach by the low temperature expansion
we analyze the 2nd moment of the 3rd thermal cumulant $\kappa_{3}(Y)$.
We report the details of the computations in 
appendix \ref{appendix-sample-fluctuation-replica}. 
The result reads as,
\beq
\sav{\kappa^{2}_{3}(Y)}= 
(\sqrt{N} \Dyeff)^{6}\lim_{n_{1},n_{2} \to 0}\frac{1}{n_{1}n_{2}} 
(9\spav{A_{11}A_{12}A_{22}} + \spav{A^{3}_{12}})
=(\sqrt{N} \Dyeff)^{6} 2m(1-m),
\label{eq-kappa-3-2-replica}
\eeq
which agrees with \eq{eq-kappa-3-2}.

\subsection{Cusp in the overlap}
\label{subsec-overlap-replica}

Finally let us study the overlap \eq{eq-def-overlap} between two real replicas, 
'1' and '2', subjected to slightly different fields $h_{0}+\delta h_{1}$
and$h_{0}+\delta h_{2}$ by the replica approach.
As we discussed using the low temperature exapnsion 
approach in sec \ref{subsec-overlap-lowT},
the correlation between the two real replicas,
\beq
\tav{Y(h_{0}+\delta h_{1})Y(h_{0}+\delta h_{2})}
=\left. 
\frac{\partial^{2}}{\partial (\beta h_{1}) \partial (\beta h_{2})} 
\sav{\log Z(h_{1})\log Z(h_{2})}
\right |_{h_{1}=h_{0}+\delta h_{1}, h_{2}=h_{0}+\delta h_{2}} 
\eeq
should strongly fluctuate from sample-to-sample. Using the replica method its sample average
can be expressed formally as,
\beqa
&& \sav{\tav{Y(h_{0}+\delta h_{1})Y(h_{0}+\delta h_{2})}}
 = \left.  \frac{\partial^{2}}{\partial (\beta h_{1}) \partial (\beta h_{2})} \lim_{n_{1}, n_{2} \to 0}
\frac{1}{n_{1}n_{2}}
\sav{Z^{n_{1}}(h_{1}) Z^{n_{2}}(h_{2})} \right |_{h_{1}=h_{0}+\delta h_{1}, h_{2}=h_{0}+\delta h_{2}} \nonumber \\
&& =\frac{\partial^{2}}{\partial (\beta h_{1}) \partial (\beta h_{2})} \lim_{n \to 0}
\frac{4}{n^{2}}
\sum_{\rm SP}
\exp \left(\frac{(\sqrt{N}\Dyeffs)^{2}}{2}\sum_{r,s=1}^{2}\beta \delta h_{r} \beta \delta h_{s}
\sum_{\alpha=(1,1)}^{(1,n/2)}\sum_{\beta=(2,1)}^{(2,n/2)} \delta_{i_{\alpha},i_{\beta}}
\right).
\label{eq-overlap-replica}
\eeqa
In the last equation $\sum_{\rm SP}$ stands for the summation over all 1RSB solutions.
For simplicity we chose $n_{1}=n_{2}=n/2$ which will not change the result.

As in our previous analysis based on the low temperature expansion, here 
we can again analyze the correlation function in two ways 1) FDT approach and
2) $T \to 0$ limit approach.

For the 1) FDT approah it is useful to rewrite 
\eq{eq-overlap-replica} rescaling the fields by the width of the steps 
$\hw=T/\sqrt{N} \Dyeffs$ as we did in \eq{eq-def-h-over-hw}; 
\beq
\frac{\sav{\tav{Y(h_{0}+\delta h_{1})Y(h_{0}+\delta h_{2})}}}
{(\sqrt{N}\Dyeffs(h_{0}))^{2}}
=\frac{\partial^{2}}{\partial \left( \frac{\delta h_{1}}{\hw}\right) \partial \left( \frac{\delta  h_{2}}{\hw}\right)} \lim_{n \to 0}
\frac{4}{n^{2}}
\sum_{\rm SP}
\exp \left(\frac{1}{2}\sum_{r,s=1}^{2}  \frac{\delta h_{r}}{\hw} \frac{\delta h_{s}}{\hw}
\sum_{\alpha=(1,1)}^{(1,n/2)}\sum_{\beta=(2,1)}^{(2,n/2)} \delta_{i_{\alpha},i_{\beta}}
\right).
\label{eq-overlap-replica-fdt}
\eeq
On the other hand, for 2) $T \to 0$ approach we may rewrite it rescaling the fields by the
spacing between the steps $\hs=T_{c}(h_{0})/\sqrt{N} \Dyeffs$,
\beq
\lim_{T \to 0}\frac{\tav{Y(h_{0}+\delta h_{1})Y(h_{0}+\delta h_{2})}}
{(\sqrt{N}\Dyeffs(h_{0}))^{2}}
= \lim_{m \to 0}
\frac{\partial^{2}}{\partial \left( \frac{\delta h_{1}}{\hs}\right) \partial \left( \frac{\delta  h_{2}}{\hs}\right)} \lim_{n \to 0}
\frac{4}{n^{2}}
\sum_{\rm SP} m^{2}
\exp \left(\frac{1}{2m^{2}}\sum_{r,s=1}^{2}  \frac{\delta h_{r}}{\hs} \frac{\delta h_{s}}{\hs}
\sum_{\alpha=(1,1)}^{(1,n/2)}\sum_{\beta=(2,1)}^{(2,n/2)} \delta_{i_{\alpha},i_{\beta}}
\right).
\label{eq-overlap-replica-zero}
\eeq
Here we used $m=T/T_{c}(h_{0})$ given by \eq{eq-m}.

\subsubsection{FDT approach}

In the FDT approach we exapnd \eq{eq-overlap-replica-fdt} in 
power series of  $\delta h_{1}/\hw$ and $\delta h_{2}/\hw$
around $\delta h_{1}/\hw=\delta h_{2}/\hw=0$. We find for the case
$\delta h_{1}=-\delta h_{2}=-\delta h$,
\beqa
&& \frac{\sav{\tav{Y(h_{0}-\delta h)Y(h_{0}+\delta h)}}}
{(\sqrt{N}\Dyeffs(h_{0}))^{2}}
=\frac{\sav{\kappa^{2}_{1}(Y)}}{(\sqrt{N}\Dyeffs(h_{0}))^{2}}
-\left(\frac{\delta h}{\hw}\right)^{2}
\frac{\sav{\kappa^{2}_{2}(Y)}-\sav{\kappa_{3}(Y)\kappa_{1}(Y)}}{(\sqrt{N}\Dyeffs(h_{0}))^{4}}
+ O \left(\frac{\delta h}{\hw}\right)^{4} \nonumber \\
&& = (1-m) -m(2-m)\left(\frac{\delta h}{\hw}\right)^{2} 
+O \left(\frac{\delta h}{\hw}\right)^{4}
\eeqa
using $\sav{\kappa_{2}^{2}(T)}=(\sqrt{N} \Dyeffs(h))^{4}m$
given in \eq{eq-kappa-2-2-replica},
$\sav{\kappa^{2}(Y)}=(\sqrt{N} \Dyeffs)^{2}(1-m)$
and $\sav{\kappa_{3}(Y)\kappa_{1}(Y)}= -(\sqrt{N} \Dyeffs )^{4}  m(1-m)$
reported in appendix \ref{appendix-sample-fluctuation-replica}.
Thus we find the overlap function \eq{eq-def-overlap} as,
\beq
C(\delta h)=1-\frac{m(2-m)}{1-m}\left(\frac{\delta h}{\hw}\right)^{2} 
+O \left(\frac{\delta h}{\hw}\right)^{4}.
\eeq
Indeed the result agrees with that by the low temperature expansion
appraoch given in \eq{eq-overlap-fdt}.

\subsubsection{Zero temperature limit}

Techincal difficulty with the 2) $T \to 0$ 
approah is that we have to express
the result of the summation over 
all 1RSB solutions $\sum_{\rm SP}$ in a closed analytic form
so that we can take the $m \to 0$ limit analytically
\footnote{Note that in the 1) FDT approach, the situation 
was somewhat simpler since only some finte number of the 
solutions in the summation $\sum_{\rm SP}$ remained 
which could be collected diagramatically.}.
We report the calucrations in appendix \ref{appendix-replica-cusp}.
The results reads,
\beqa
 \lim_{T \to 0} \frac{C(\delta h)}{(\sqrt{N} \Dyeffs(h))^{2}} = &&
\sqrt{\frac{2}{\pi}}e^{-\frac{l^{2}}{8}}
\int_{0}^{\infty}dx 
\frac{e^{-\frac{x^{2}}{2}} (x^{2}-\frac{l^{2}}{4})}{
e^{-\frac{|l|}{2}}M_{0}( x-\frac{|l|}{2})+
e^{ \frac{|l|}{2}}M_{0}(-x-\frac{|l|}{2})
} \nonumber \\
&& -\frac{|l|}{\pi}e^{-\frac{l^{2}}{4}}\int_{0}^{\infty}dx 
\frac{e^{-x^{2}} e^{-x|l|}}{
\left\{
e^{-\frac{|l|}{2}}M_{0}( x-\frac{|l|}{2})+
e^{ \frac{|l|}{2}}M_{0}(-x-\frac{|l|}{2})
\right\}^{2}
} \nonumber \\
&& =1-\frac{4}{\sqrt{\pi}}\frac{|\delta h|}{\hs} 
+ O \left(\frac{|\delta h|}{\hs}\right)^{2}
\label{eq-cusp-replica}
\eeqa
with $l=2\delta h/\hs$.
Thus the overlap function exhibists a cusp as shown in Fig. \ref{fig-cusp} 
in agreement with the
result obtained by the low temperature exapnsion approach.

To derive the above result in appendix \ref{appendix-replica-cusp},
we used the approach of \cite{BM,BMP} which study
statistical properties of turbulant flows obeying the Burgers equation.
In the next section 
we discuss the intriguing connection between the intermittency 
in the turbulant flows and step-wise responses in mesoscopic glassy systems. 

\section{Mapping to a pinned elastic manifold problem 
and a  turbulence prpblem}
\label{sec-connection}

In this section we discuss the connecton between the present problem
of mesoscopic responses in glassy systems, the problem of
jerky effective energy landscape of pinned elastic manifolds
\cite{BBM} and intermittency in turbulent flows 
due to shocks \cite{Kida,BMP,BM}. 

Let us consider a simple pinned elastic manifold problem, namely a particle
on a one dimensional space connected to a Hookian spring and subjected
to a radom potential  \cite{SVBO}. 
The partitoin function and the Hamiltonian is defined as,
\beq
Z(Y_{0})=\int_{-\infty}^{\infty} \frac{dY}{\sqrt{2\pi/(\beta \kappa)}}e^{-\beta U(Y,Y_{0})} \qquad U(Y,Y_{0})=\frac{\kappa}{2}(Y-Y_{0})^{2}+F(Y).
\label{eq-def-toy}
\eeq
One end of the Hookian spring of spring constant $\kappa$ 
is connected to the particle at $Y$ and the other end is
fixed at $Y_{0}$. $F(Y)$ is a random potential whose values
are drawn from a Gaussian distribution with zero mean 
and a short-ranged spatial correlation,
\beq
\sav{F(Y)F(Y')}=N c(|Y-Y'|).
\eeq
Here $c(y)$ is a certain rapidly decaying function.
Comparing with the random energy model given by \eq{eq-Z-TAP}
with the simplest generalized complexity of the form \eq{eq-gc-KM},
we find the two models are related as,
\begin{itemize}
\item The extensive variable $Y$ becomes a {\it coordinate} of an 
one-dimensional space.
\item The Gaussian distribution of the $Y$ variables
over differnt metastable states with variance $\Dy \sqrt{N}$
amounts to an effective {\it entropic} spring force with spring constant,
\beq
\kappa=\frac{T}{\Dy \sqrt{N}}=\hw
\eeq
\item The random (free) energy $F$ at different metastable states 
amounts to a random potential $F(Y)$ with short-ranged correlation.
\item The external field $h$ is equivalent to,
\beq
Y_{0}=\frac{h}{\kappa}.
\eeq
The {\it effective force} that the system yeilds can be defined as,
\beq
f(Y_{0}) \equiv -T \frac{\partial \ln Z(Y_{0})}{\partial Y_{0}}
=\kappa (\tav{Y}-Y_{0})
\eeq
\end{itemize}

By moving the end point $Y_{0}$ of the spring, the position 
of the particle $\tav{Y}$ in equilibrium changes elastically 
up to a certain point beyond which it exibit a sudden jump 
to a different energy minimum. Such a process was studied in
detail by Bouchaud and M\'{e}zard \cite{BM}. 
The jump is of couse rounded at finite temperatures. 
This is what corresponds to 
the step-wise response we have discussed so far in the present paper. 
At such a jump,
the effective force $f(Y_{0})$ also changes discontnuously.
Thus the {\it effective energy landscape} $-T \ln Z(Y_{0})$ 
consists of parabolic wells matching with each other at
singular poionts where the effective force (slope) 
is discontinuous. Note that the overlap function
\eq{eq-def-overlap} we studied, which exhibit the same cusp
as shown in Fig.~\ref{fig-cusp}, corresponds to the correlation function
of the effective force $\sav{f(0)f(Y_{0})}$.

As pointed out by Bouchaud and M\'{e}zard \cite{BM}, 
the toy model \eq{eq-def-toy} can be mapped to a yet another 
problem of a one dimensional tubulent flow. By the so
called Cole-Hopf transformation,
\beq
v(Y_{0},t)=-\beta \frac{ \partial \ln Z(Y_{0})}{\partial Y_{0}} \qquad
\beta=\frac{1}{2 \nu} \qquad \kappa=\frac{1}{t} 
\label{eq-cole-hopf}
\eeq
we obtain the Burgers equatrion for the velocity field $v(Y_{0},t)$ at time $t$,
\beq
\frac{\partial v}{\partial t}+ v \frac{\partial v}{\partial Y_{0}}=
\nu \frac{\partial^{2} v}{\partial Y_{0}^{2}} 
\eeq
where $\eta$ is understood as the viscosity of the fluid.
The initial condition at time $t=0$ specified by,
\beq
v(Y_{0},0)=\frac{\partial E(Y_{0})}{\partial Y_{0}},
\eeq
so that the velocity field is completely randon at $t=0$.

The velocity field $v(Y,t)$ changes discontinuously
at certain points in the space, namely {\it shocks}. 
The intermittency of the flow due to the presence of the shock 
is manifested in the velocity-velocity correlation function, 
which corresponds to the overlap function \eq{eq-def-overlap}
in our probelm, as the cusp-like simgularity  shown in Fig, \ref{fig-cusp}.
To observe well defined shocks (steps) one need to consider fluid of high 
enough viscosity $\eta$ (low enough temperature $T$). 
The spacing between the shocks
(spacing between the steps $\hs$) decreases as the time $t$ (system size
$N$) increases.

The toy model \eq{eq-def-toy} is the simplest $0$ dimensional version
of the broad class of systems called {\it elastic manifolds in random media}.
The cusp like singularity of the force-force correlation function is
predicted for such a broad class of systems by a mean-field theoretical 
approach based on the replica method and also by a functional 
renormalization group approach \cite{BBM}. The thermally rounded region
of width $\hw$ around the top of a cusp in called 
{\it thermal boundary layer} \cite{BD}. Indeed the cusp is confirmed
by a recent extensive numerical study \cite{Middleton}. 
These problems can be mapped to turbulent flows under continous random
forcing \cite{BMP}. 

\section{Conclusions}
\label{sec-conclusions}

In the present paper we studied the statistical properties
of mesoscopic responses in glassy systems from a mean-field 
theoretical point of view. More specifically we have studied
a generic class on mean field models which exhibit 1 step replica
symmetry breaking. We conclude that the mesoscopic response consistes
of step-wise responses with typical height $\sqrt{N} \Dyeffs$, spacing  
$\hs \sim \Tc(h)/(\sqrt{N}\Dyeffs)$ and width $\hw \sim T/( \sqrt{N}\Dyeffs)$. 
This conclusion follows from observations of the mesosocpic response in two
different ways. One is based on the FDT: we studied statistic of
sample-to-sample fluctuations of the linear and non-linear
susceptibilities related to the spontaneous thermal fluctuations.
The latters are significant only in narrow regions of width $\hw$ 
around steps so that the averages of their moments are dictated 
by contributions from such rare regions. Our result proves
the intuitions developed by the pioneering works in the SK model a long
time ago \cite{KY,YBM}. We conclude that the FDT approach amounts 
to describe mesosocpic responses under variation of the external 
field $\delta h$ by power series of $\delta h/\hw$.
In the second  approach we did not rely on the FDT. Working directly
at $T=0$ we obtained series expansions of the mesoscopic response
in  power series of $\delta h/\hs$ going beyond the scale of single
steps. 
The presence of mesoscopic response means that the thermodynamic limit
$N \to \infty$ and $\delta h \to 0$ limit do not commute:
one observs thermodynamic susceptibilities by  $\lim_{\delta h \to
0}\lim_{N \to \infty}$ and mesoscopic responses by  $\lim_{N \to
\infty}\lim_{\delta h \to 0}$. Furthermore, the presence of the 
two expansions $\delta h/\hw$ and $\delta h/\hs$ means the two limits
$\lim T \to 0$ and $\lim \delta h \to 0$ do not commute either. 
The the non-commutativities of the limits reflect the step (or cusp)
singularity at mesoscopic scales.

We note that a techincal challenge remains to evaluate the parameter 
$\Dyeffs$ for a given perturbation from a given microscopic
Hamiltonian. It quantifies
the strength of a given perturbation.  Since it conceres with 
the {\it inter-state} responses, one must find ways 
to disentangle intra-state and inter-state responses in a given
response. For example, for the $p$-spin Ising model discussed at the
begining of  sec. \ref{sec-replica}, we could evaluate $\Dyeffs$ 
for general $p$ only under zero field.  For non-zero field, we could
evaluate it in the $p \to \infty$ limit by considering the
fluctuaion of the size of clusters $m$ of the 1 step RSB solutions as explained in 
Appendix. \ref{appendix-m-fluctuation} but remains as a challenge
for $p < \infty$.

Many features we obtained in the present paper can be rephrased 
by Imry-Ma type scaling argument, for instance the droplet theory
for spin-glasses \cite{FH88} and directed polymer in random media
\cite{FH91} proposed by Fisher and Huse,
as we noted from time to time in the present paper. 
In short, a low energy excitation of a mesoscopic
scale $L$ behaves much as a sample of a mean field model of size $N$.
In finite dimensions, not only thermodynamic responses but also 
susceptibilities related to spontaneous thermal fluctuations 
via static FDT should become self-averaging thanks to the
central limit theorem: independent mesoscopic 
excitations scattered over a given macroscopic sample
contribute simultaneously and independently from each other to the
total susceptibility. However it is of great interest to understand 
better properties at mesoscopic scales such as statistics of 
droplet-to-droplet fluctuations of the droplet excitations. 
We expect that our  mean-field approach may be useful to obtain 
{\it meanf field} approximations of such kind of statistical properties. 

In Imry-Ma type arguments one can incorporate
some refined knowledge of realistic low dimensional systems
obtained by other means. For example, our results can be refined
by the following adjustments. First the characteristic energy scale $\Tc$ 
may be replace by $\Upsilon (L/L_{0})^{\theta}$ where $L_{0}$
is a certain unit length scale, $\Upsilon$ is the stiffness 
constant and $\theta >0$ is the stiffness exponent. 
Then the variance of the random variable $Y$ 
may be generalized as $A (L/L_{0})^{\alpha}$ with $\alpha > 0$ 
replacing the simple $\Dy \sqrt{N}$ scaling.
Then one finds $\hw \sim (T/A)(L/L_{0})^{-\alpha}$, 
$\hs \sim (\Upsilon/A) (L/L_{0})^{-\zeta}$ where
$\zeta=\alpha-\theta$ is the so called chaos exponent \cite{BM87,FH88}.
If the chaos exponent is positive $\zeta >0$, the spacing between
the steps vanishes in the limit $L \to \infty$ leading to static chaos.
An interesting consequence is that the thermal width $\hw$ vanishes {\it
faster} than the spacing $\hs$ as $L \to \infty$ since $\theta > 0$.
It will be very interesting to develop futher the replica approach 
in order to verify these points theoretically.

\appendix

\section{On the mesoscopic and macroscopic linear-susceptibility}
\label{sec-note-linear-sus}

For a given sample, say $J$, of a finite size $N$ we can write,
\beq
y_{J}(h)=\int_{0}^{h} dh' \frac{\partial y_{J}(h)}{\partial h}
\label{eq-y-j}
\eeq
where we put the subscript $J$ in order to remind us that it is
associated with a single sample $J$. The {\it differential susceptibility}
which appears on the r.~h.~s. can be identified with
the correlation function through the static FDT,
\beq
\frac{\partial y_{J}(h)}{\partial h}=\beta (\tav{y^{2}_{J}(h)}-\tav{y_{J}(h)}^{2}).
\eeq
Taking average over different 
realizations of samples of both sides of \eq{eq-y-j} we obtain,
\beq
\sav{y_{J}(h)}=\int_{0}^{h} dh' \sav{\frac{\partial y_{J}(h')}{\partial h'}}
\qquad\mbox{or} \qquad 
\frac{\partial \sav{y_{J}(h)}}{\partial h} =
\sav{\frac{\partial y_{J}(h)}{\partial h}}.
\eeq
On the other hand we have 
\beq
\lim_{N \to \infty} y_{J}(h)= \lim_{N \to \infty} \sav{y_{J}(h)}
\eeq
since $Y_{J}=N y_{J}$ is a thermodynamic and thus self-averaging quantity.
Thus we obtain
\beq
\frac{\partial}{\partial h} \lim_{N \to \infty} y_{J}(h)
=\lim_{N \to \infty}\beta \sav{(\tav{y^{2}_{J}(h)}-\tav{y_{J}(h)}^{2})}.
\label{eq-equiv-sav-lin}
\eeq
at any $h$.
Thus 1st derivative of the thermodynamic response curve, say 'thermodynamic
linear susceptibility' 
and sample-average of the linear susceptibility obtained via static FDT
must coincide.
Note that the above argument cannot be repeated for non-linear susceptibilities
since linear-susceptibility may {\it not} be self-averaging, i.e. in general
$\lim_{N \to \infty}\partial y_{j}(h)/\partial h \neq \sav{\partial y_{J}(h)/\partial h}$.

\section{An effective REM}
\label{sec-KM}

We summarize below some basic properties of an effective REM 
with the simplest form of the generalized complexity given 
by \eq{eq-gc-KM}
\beq
\Sigma(f,y)= c - \frac{f^{2}}{2}-\frac{y^{2}}{2 \Dy^{2}} 
\label{eq-gc-erem}
\eeq
where $c ( \geq 0)$ is a constant which fixes the total number of states as,
\beq
M=e^{N c}.
\label{eq-M}
\eeq
The complexity  can be cast into the form like \eq{eq-gc-around-saddle}
as a function of the ``total'' free-energy $f'=f-hy$ 
and the transver variable $\ty$ given by \eq{eq-def-ty},
\beq
\Sigma(\tf,\ty)= c -\frac{\tf^{2}}{2} -\frac{\ty^{2}}{2(\Dyeffs)^{2}}
\label{eq-gc-erem2}
\eeq
where we introduced a 'renormalized variance',
\beq
\Dyeffs \equiv 
\frac{\Dy}{\sqrt{1+  (h \Dy})^{2}}
\label{eq-dy-eff}
\eeq 
and new variables defined as,
\beq
\tf \equiv  f' \sqrt{1-  (h \Dyeffs})^{2} \qquad 
\ty \equiv y+ h (\Dyeffs)^{2} f'.
\label{eq-tf-ty}
\eeq
In \eq{eq-gc-erem2} it is evident that  $\tf$ and $\ty$
are statistically independent from each other. In the present paper
the variable like $\ty$ which is statistically independent from the
free-energy is called as {\it transverse variable} (see \eq{eq-def-ty}). 

The shape of the generalized complexity close to the $\Sigma=0$ plane
can be obtained by rewriting $\Sigma$ as,
\beq
N \Sigma = \frac{F'-(F')^{*}(h)}{\Tc(h)}
-\frac{1}{2cN} \left(\frac{F'-(F')^{*}}{\Tc(h)}\right)^{2}
-\frac{\tilde{Y}^{2}}{2N \Dyeff^2}.
\eeq
Here $F'=N f'$, $\tilde{Y}=N\ty$ and $(F')^{*}=-2cN \Tc(h)$
with $T_{c}$ being the critical temperatrure obtained below as \eq{eq-Tc}.
At large enough system sizes $N$ the 2nd term on the r.h.s becomes
negligible compared with the 1st term and the functional form 
of the generalized complexity converges to the
expected generic form given by \eq{eq-gc-around-saddle} close to
the $\Sigma=0$ plane. Thus this effective REM should give generic
results common in all 1 step RSB models concerning inter-state responses.

It is straight forward to consider a slightly more generalized version
of the complexity including also an off-diagonal term 
like $- K  f \frac{y}{\Dy}$ in \eq{eq-gc-erem}. 
The latter represents a possible correlation between $f$ and $y$ 
of strength $\sim K$ for small enough $K$.
One can find again that the complexity can be cast into the quadratic form
\eq{eq-gc-erem2}.

\subsection{Thermodynamics}

Now let us determine thermodynamic state 
$(\tf^{*}, \ty^{*})$ at a given 
temperature $T$ and external field $h$.
In Fig.~\ref{fig-gc}, the variable $\tf$ is held constant on the line $C$ 
while the variable $\ty$ is varied. Along $C$ the complexity 
\eq{eq-gc-erem2} is maxized at the point $P$ where $\ty=0$.
Thus we fined $\ty^{*}=0$. On the other hand
$\tilde{f}^{*}$ is obtained through the condition \eq{eq-saddle} which reads 
as,
\beq
\frac{1}{T}= 
\left. \frac{\partial \Sigma}{\partial f'}  \right |_{\tf=\tf^{*}, \ty=0}
=- \tf^{*}\sqrt{1-(h \Dyeffs)^2}
\eeq
and the value of the complexity at the saddle point ($f^{*},y^{*}=0$) is
obtained as $\Sigma^{*}=c-(\tf^{*})^{2}/2$.

The saddle point value of the complexity $\Sigma^{*}$
becomes $0$  when $\tilde{f}^{*}=-\sqrt{2c}$.
Thus the critical temperature $\Tc$ is obtained as,
\beq
\Tc(h)=\frac{\Tc^{0}}{\sqrt{1-(h \Dyeffs)^2}}  \qquad \Tc^{0}=\frac{1}{\sqrt{2c}}
\label{eq-Tc}
\eeq
Here $h$ dependence comes from the $h$ dependence of $\Dyeff$
given in \eq{eq-dy-eff}. 

In the glass phase $T < \Tc$ the saddle point is fixed
at $(\tf^{*},\ty^{*})=(-\sqrt{2c},0)$. Then 
we find the total free-energy density $f'$ as
\beq
(f')^{*}(h)= - 2c \Tc(h)
\label{eq-f-glass}
\eeq
and the equilibrium value of the variable $y$ as
\beq
y^{*}(h)=\frac{\Dy^{2}}{T_{\rm c}(h)} h.
\label{eq-y-av-glass}
\eeq
using \eq{eq-tf-ty} , \eq{eq-Tc} and \eq{eq-dy-eff}.
Note that here $\Dy$ is the variance of the original $Y$ variable.
Taking a derivative of the last equation 
with respect to $h$ taking into account 
the $h$ dependence of $\Tc(h)$ given by \eq{eq-Tc} one finds,
\beq
\chi(h)=N\frac{\partial{y^{*}(h)}}{\partial h}=  N  \frac{\Dyeff^{2}}{\Tc(h)}\label{eq-chi-thermodynamic}
\eeq
where $\Dyeffs$ is now the variance of $Y$ given by \eq{eq-dy-eff}.

Taking a derivative of the free-energy $(f')^{*}$ 
given in \eq{eq-f-glass}
with respect to the temperature one finds the well known
result \cite{REM} that the heat-capacity is zero below $\Tc$,
\beq
C/N=0.
\label{eq-c-0}
\eeq

\subsection{Replica approach}

Here we scketch the construction of the replica approach for this model.
The partition function of a given sample is given by
\beq
Z(T,h,\delta h)=\sum_{i=1}^{M} e^{-\beta N (f'_{i}(h) -\delta h \ty_{i})}
\label{eq-Z-T-h-deltah}
\eeq
where $i$ is the label for each metastable state.
$f'(h)=f-h y$ is the total free-energy density under external
field $h$. The sum  runs over all states whose number 
is $M=e^{Nc}$ as given in \eq{eq-M}. We have also included 
{\it infinitesimal} probing field $\delta h$ coupled to 
the transverse variable $\ty$ defined in \eq{eq-tf-ty} which is statistically
uncorrelated with the free-energy density $f'$.

The distribution of the quenched random variables $f'$ and $\ty$ 
is specified by the generalized complexity given in \eq{eq-gc-erem2}
supplemented by the relation $f'=\sqrt{2c}\Tc(h)\tf$ 
which follows from \eq{eq-tf-ty} and \eq{eq-Tc}.
We obtain sample-averaged replicated free-energy as
\beq
\sav{Z^{n}}=\sum_{i_{1},i_{2},\ldots,i_{n}=1}^{M=e^{cN}}
\exp \left[N  \left\{ c (\beta \Tc(h) )^{2} 
+\frac{ (\beta \delta h)^{2}}{2} \Dyeff^{2} 
\right\} \sum_{\alpha,\beta =1}^{n} \delta_{i_{\alpha},i_{\beta}} 
\right].
\label{eq-replicated-z-KM}
\eeq

Let us summarize below the analysis of the thermodynamics
following \cite{REM,BM} which reproduces the results obtained above
without using the replica approach.
Here we switch off the probing field $\delta h=0$.
Within the replica symmetric ansatz (RS), one assumes that
$n$ replicas occupy different states. Then one obtains
\beq
\sav{Z^{n}} \sim \exp \left [c N n \{ 1+ (\beta \Tc(h))^{2} \} \right]
\eeq
from which one obtains the thermodynamic free-energy and entropy as
\beq
-\beta F_{\rm RS}/N= \lim_{n \to 0}\frac{\sav{Z}^{n}-1}{nN} = c \{ 1+ (\beta \Tc(h))^{2} \} \qquad
S_{\rm RS}/N = c \{ 1- (\beta \Tc(h))^{2} \} .
\eeq
The entropy vanishes as $T  \to \Tc(h)$ suggesting condensation
of the Boltzmann weight. Thus we find again that
$\Tc(h)$ given in \eq{eq-Tc} is the critical temperature.

Below $\Tc(h)$ we expect the 1 step replica symmetry breaking  (1 step RSB)
ansatz gives physically correct saddle points \cite{REM,BM}.
It amounts to assume that $n$ replicas are grouped into $n/m$ clusters
of size $m$ such that replicas belonging to a common cluster stay
at the same state while those belonging to different clusters stay
at different states. 

In Fig.~\ref{fig-m-atom} we show a schematic
picture of such a 1RSB solution. Note that there are many 1 step RSB solutions.
Each solution is determined by  1) choosing $n/m$ states to be occupied
out of $M$ possible states and 2) choosing a partition of $n$ 'atoms'
into $n/m$ 'molecules' out of $n!$ possibilities of such partitions.
The 1RSB ansatz yields,
\beq
\sav{Z^{n}} \sim \sum_{\rm SP} 
\exp \left [cN \frac{n}{m} \left\{ 1+ \left(\frac{\Tc(h)}{T/m}\right)^{2} \right\} \right]
\eeq
where $\sum_{\rm SP}$ stands for summation over all possible 1RSB solutions
explained above.  Here the summand is common for all 1RSB solutions
so that  $\sum_{\rm SP}$ just yield a multiplicative factor,
\beq
\sum_{\rm SP} 1 = \frac{M!}{(M-\frac{n}{m})!}\frac{n!}{(m!)^{n/m}},
\label{eq-spsum-org}
\eeq
which becomes $1$ in $n \to 0$ limit.
The 1RSB free-energy is related to the RS free-energy simply as
$F_{\rm 1RSB}(T,h)=F_{\rm RS}(T/m,H)$. Extremization with respect
to the parameter $m$, 
\beq
0 = \frac{\partial F_{\rm 1RSB}}{\partial m}
\label{eq-extremization-m}
\eeq
is formally identical to the condition $S_{\rm RS}(T/m,H)=0$
as first noted by Monasson \cite{Monasson}. Then one finds,
\beq
m=\frac{T}{\Tc(h)}.
\label{eq-m-appendix}
\eeq

At $T< \Tc(h)$, the thermodynamic free-energy is fixed to
\beq
F(T,h)/N=-2c \Tc(h).
\eeq
This agrees with \eq{eq-f-glass}.

\section{Low temperature expansions of the moments of the susceptibilities}
\label{sec-lowT-expansion-sus-appendix}

In this appendix we report some details of the 
computations by the low temperature expansion of the sample-averages of 
some moments of the thermal cumulants.

\subsection{Accuracy of $M$-level models}
\label{subsec-accuracy-m-level}

Here we proove that
a $M$-th level model, which only takes into account the lowest $M$ levels, give
correct results of the moments 
up to order $O(T/T_{c})^{M-1}$. 

We first note that \eq{eq-kappa-k-p} can be formally expanded as,
\beq
\sav{\kappa^{p_{1}}_{k_{1}}(Y) \kappa^{p_{2}}_{k_{2}}(Y)\cdots}= \sum_{l=1}^{M-1}
\sum_{0 < n_{1} < n_{2} \ldots < n_{l}} C_{n_{1},n_{2},\ldots,n_{l}} \sav{\prod_{k=1}^{l} X^{n_{k}}_{k}}. 
\label{eq-sav-pkappa-expansion}
\eeq
The coefficients $C_{p_{1},p_{2},\ldots,p_{l}}$ are to be obtained after averaging
over the $Y$ variables, which can be done interdependently of the $X$ variables
since they are independent from each other. We remind the readers that
we are considering the transverse $\tY$ variables. 

 From the latter we find,
\beq
\sav{X^{n}_{k}}=\int_{0}^{1}(X_{k})^{n} p_{k}(X_{k}) d X_{k}= 
\frac{k}{n}\frac{T}{\Tc(h)}\left[ 1+ \frac{k}{n}\frac{T}{\Tc(h)} \right]^{-1}.
\eeq
which yields,
\beq
\sav{\prod_{k=1}^{l} X^{n_{k}}_{k}}= \left(\frac{T}{\Tc(h)}\right)^{l}
\prod_{k=1}^{l} \frac{k}{n_{k}}
\left[ 1+ \frac{k}{n_{k}}\frac{T}{\Tc(h)} \right]^{-1}.
\label{eq-av-prod-x}
\eeq
Using \eq{eq-av-prod-x} in \eq{eq-sav-pkappa-expansion} we notice that
$M+1$-level model would give the same results as the $M$-level model 
except that it will have additional terms of order $O(T/\Tc(h))^{M}$
(and higher order terms). This in turn certifies that finite $M$-level model
should give correct results up to $O(T/\Tc(h))^{M-1}$.

Thus to collect $O(T/\Tc(h))$ terms a $2$-level model is sufficient.
A $3$-level model gives only 
$O(T/\Tc(h))^{2}$ correction terms which take care of rare samples
in which three levels happen to become degenerate as shown 
in Fig.~\ref{fig-low-T-expansion-diagram}.

Let us note that somewhat similar reasoning has been
used to justify the phenomenological Imry-Ma type 
scaling arguments for glassy systems 
by Fisher and Huse \cite{FH88,FH91} which assume some kinds of schematic 
low temperature expansions. 
The class of mean-field models we consider here enable explicit low temperature expansions.

The low temperature expansion can be implemented also
to compute the so called participation ratios,
\beq
\sum_{l=0}^{M-1} W^{p}_{l}= 
\sum_{l=0}^{M-1} 
\left \{
\frac{\prod_{k=1}^{l} X_{k}}{\sum_{l'=0}^{M-1} \prod_{k=1}^{l'} X_{k})}
\right \}^{p}.
\label{eq-participation-ratio}
\eeq
Evidently it can be formally expanded in the same way as
\eq{eq-sav-pkappa-expansion}. 
The participation ratios have been computed without the replica
method using the exponential distribution of random free-energies 
\eq{eq-p-f} reproducing the replica results \cite{MPV}
in the $M \to \infty$ limit. Our result implies finite $M$-model
gives correct result up to $O((T/\Tc(h))^{M-1})$.

\subsection{Linear susceptibility}
\label{appendix-lowT-linear}

Following the prescription \eq{eq-kappa-k-p} 
the sample-average of 1st moment of the 2nd thermal cumulant 
$\sav{\kappa_{2}(Y)}^{\rm 2-level}$ is computed within
the $2$-level model as the following.
\beqa
&& \sav{\kappa_{2}(Y)}^{\rm 2-level} = \sav{
\left.
\left(Y_{0}\tO{0}+Y_{1}\tO{1}\right)^{2}
 \ln Z(O_{0},O_{1},X_{1}) \right |_{O=1}}
 \nonumber \\
&&  = (\sqrt{N} \Dyeffs )^{2} \sav{\left. \left\{ 
\left(\tO{0} \right)^{2}+ \left(\tO{1}\right)^{2}
\right\} \ln (O_{0}+O_{1}X_{1}) \right |_{O=1}}
= (\sqrt{N} \Dyeffs )^{2} \int_{0}^{1} dX_{1} p_{1}(X_{1}) \frac{2X_{1}}{(1+X_{1})^{2}} \nonumber \\
&& =  (\sqrt{N} \Dyeffs )^{2} \left\{\frac{T}{\Tc(h)} 
-2 \ln 2\left(\frac{T}{\Tc(h)} \right)^{2}
+ O \left(\frac{T}{\Tc(h)}\right)^{3} \right\}
\eeqa
Here we used \eq{eq-p-x} and \eq{eq-p-tY} to evaluate the sample averages.
Then repeating the computation in $3$-level model we obtain,
\beqa
&& \sav{\kappa_{2}(Y)}^{\rm 3-level}= 
\left. \sav{
\left(Y_{0}\tO{0}+Y_{1}\tO{1}+Y_{2}\tO{2}\right)^{2}
\ln Z(O_{0},O_{1},O_{2},X_{1},X_{2})
 \right |_{O=1} } \nonumber \\
&& = 
(\sqrt{N}\Dyeffs)^{2}
\int_{0}^{1} dX_{1} p_{1}(X_{1})
 \frac{2X_{1}}{(1+X_{1})^{2}}  
+(\sqrt{N}\Dyeffs)^{2} 
\int_{0}^{1} dX_{1} p_{1}(X_{1})\int_{0}^{1} dX_{2} p_{2}(X_{2})
\left \{ \frac{2X_{1} (1+X_{2}+X_{1}X_{2})}{(1+X_{1}+X_{1}X_{2})^{2}}
-  \frac{2X_{1}}{(1+X_{1})^{2}}  \right \}
\nonumber \\
&& = 
\sav{\kappa_{2}(Y)}^{\rm 2-level} 
+ (\sqrt{N}\Dyeffs)^{2} \left\{2 \ln 2\left(\frac{T}{\Tc(h)} \right)^{2} 
+ O \left(\frac{T}{\Tc(h)}\right)^{3} \right\}  
\nonumber \\
&& = (\sqrt{N}\Dyeffs)^{2} \left \{\frac{T}{\Tc(h)} +  O \left(\frac{T}{\Tc(h)}\right)^{3} \right\} 
\eeqa
Note that in the 2nd equation we have slipt the integral into two parts,
one of which corresponds to that of the 2-level model.

By similar procedure
the sample-average of 2nd moment of the 2nd thermal cumulant 
$\sav{\kappa^{2}_{2}(Y)}^{\rm 3-level}$ is computed within
the $3$-level model as the following.
Following the prescription \eq{eq-kappa-k-p} 
the sample-average of 2nd moment of the 2nd thermal cumulant 
$\sav{\kappa^{2}_{2}(Y)}^{\rm 3-level}$ is computed within
the $3$-level model as the following.
\beqa
&& \sav{\kappa^{2}_{2}(Y)}^{\rm 3-level}= 
 \lsav
\left(Y_{0}\tO{01}+Y_{1}\tO{11}+Y_{2}\tO{21}\right)^{2}
\left(Y_{0}\tO{02}+Y_{1}\tO{12}+Y_{2}\tO{22}\right)^{2} \nonumber \\
&&  \left.
\ln Z(O_{01},O_{11},O_{21},X_{1},X_{2})
\ln Z(O_{02},O_{12},O_{22},X_{1},X_{2})
 \right |_{O=1} \rsav \nonumber \\
&=&  12(N\Dy^{2})^{2}\int_{0}^{1} dX_{1}p_{1}(X_{1})
\frac{X^{2}_{1}}{(1+X_{1})^{4}}\nonumber \\
&& + 12(N\Dy^{2})^{2} \int_{0}^{1} dX_{1}p_{1}(X_{1})\int_{0}^{1}dX_{2}p_{2}(X_{2})
\left \{
\frac{X^{2}_{1}(1+X_{2}+X^{2}_{2}+X_{1}X_{2}+X_{1}X_{2}^{2}+X_{1}^{2}X_{2}^{2})}{(1+X_{1}+X_{1}X_{2})^{4}} 
-\frac{X^{2}_{1}}{(1+X_{1})^{4}}
\right\}
\eeqa
In the last equation we have again slip the integral into two parts.
The 1st integral corresponds to that of 2-level model which is evaluated as
\bmat
\int_{0}^{1} dX_{1}p_{1}(X_{1})
\frac{X^{2}_{1}}{(1+X_{1})^{4}} = \frac{1}{12}\frac{T}{\Tc(h)}
+ \left(
\frac{1}{24}-\frac{1}{6} \ln 2 
\right)\left( \frac{T}{\Tc(h)}\right)^{2}Z
+ O\left( \frac{T}{\Tc(h)}\right)^{3}
\emat
while the 2nd integral corresponds to 
the contribution due to the 3rd level which is evaluated as,
\beqa
&& \int_{0}^{1} dX_{1}p_{1}(X_{1})\int_{0}^{1}dX_{2}p_{2}(X_{2})
\left \{
\frac{X^{2}_{1}(1+X_{2}+X^{2}_{2}+X_{1}X_{2}+X_{1}X_{2}^{2}+X_{1}^{2}X_{2}^{2})}{(1+X_{1}+X_{1}X_{2})^{4}} 
-\frac{X^{2}_{1}}{(1+X_{1})^{4}}
\right\}  \nonumber \\
&& = 
\left(
-\frac{1}{24}+\frac{1}{6} \ln 2 
\right)\left( \frac{T}{\Tc(h)}\right)^{2}
+ O\left( \frac{T}{\Tc(h)}\right)^{3} \nonumber
\eeqa
Combining the above results we finally obtain \eq{eq-kappa-2-2},
\beqa
\sav{\kappa^{2}_{2}(Y)}^{\rm 3-level}= 
 (\sqrt{N}\Dyeffs )^{4}\left\{\frac{T}{\Tc(h)} + O \left(\frac{T}{\Tc(h)}\right)^{3} \right\}.
\eeqa

\subsection{Non-linear susceptibility}
\label{appendix-lowT-nonlinear}

Using \eq{eq-kappa-k-p} 
the sample-average of 2nd moment of the 3rd thermal cumulant 
$\sav{\kappa^{2}_{3}(Y)}^{\rm 3-level}$ is computed within
the $3$-level model as the following.
\beqa
&& \sav{\kappa^{2}_{3}(Y)}^{\rm 3-level}
=  \lsav
\left(Y_{0}\tO{01}+Y_{1}\tO{11}+Y_{2}\tO{21}\right)^{3}
\left(Y_{0}\tO{02}+Y_{1}\tO{12}+Y_{2}\tO{22}\right)^{3} \nonumber \\
&&  \left.
\ln Z(O_{01},O_{11},O_{21},X_{1},X_{2})
\ln Z(O_{02},O_{12},O_{22},X_{1},X_{2})
 \right |_{O=1} \rsav \nonumber \\
&& = (N\Dyeff^{2} )^{3}
\int_{0}^{1} dX_{1} p_{1}(X_{1})\int_{0}^{1} dX_{2} p_{2}(X_{2})
F(X_{1},X_{2}).
\eeqa
where
\beqa
&& F(X_{1},X_{2}) \equiv 
12 X_{1}^{2}  \left[ 10+48 X_{1}^{2}X_{2}^{2}-7X_{1}^{2}X_{2}
-7X_{1}X_{2} \right. \nonumber \\
&& -7X_{1}^{3}X_{2}^{3}+10X_{1}^{4}X_{2}^{2}+10X_{1}^{2}
-20X_{1}X_{2}^{3}+10X_{1}^{2}X_{2}^{4}-20X_{1}^{4}X_{2}^{3}
+10X_{1}^{4}X_{2}^{4}-7X_{1}X_{2}^{2} \nonumber \\
&& \left. -7X_{1}^{2}X_{2}^{3}
+7X_{1}^{3}X_{2}^{4}-20X_{1}+7X_{2}-7X_{1}^{3}X_{2}^{2}
+7X_{1}^{3}X_{2}+10X_{2}^{2} \right]
(1+X_{1}+X_{1}X_{2})^{-6} 
\eeqa

Again the integral can be split into two parts.
\beqa
&& 
\int_{0}^{1} dX_{1} p_{1}(X_{1})\int_{0}^{1} dX_{2} p_{2}(X_{2})
F(X_{1},X_{2}) \nonumber \\
&& = 120 \int_{0}^{1} dX_{1} p_{1}(x_{1})
\frac{1-2X_{1}+X_{1}^{2}}{(1+X_{1})^{6}} 
+\int_{0}^{1} dX_{1} p_{1}(X_{1})\int_{0}^{1} dX_{2} p_{2}(X_{2})
\Biggl \{
F(X_{1},X_{2})
 - 120 \frac{1-2X_{1}+X_{1}^{2}}{(1+X_{1})^{6}}\ \Bigr \} 
\eeqa

The 1st integral on the r.h.s corresponds to that of 2-level model which is evaluated as
\bmat
\int_{0}^{1} dX_{1} p_{1}(x_{1})
\frac{1-2X_{1}+X_{1}^{2}}{(1+X_{1})^{6}} =\frac{1}{60} \frac{T}{\Tc(h)}
-\left (\frac{1}{240}+\frac{1}{30}\ln 2 \right)
\left( \frac{T}{\Tc(h)}\right)^{2}
+ O \left(\frac{T}{\Tc(h)}\right)^{3}
\emat 
while the 2nd integral on the r.h.s corresponds 
to the contribution due to the 3rd level which is evaluated as,
\bmat
\int_{0}^{1} dX_{1} p_{1}(X_{1})\int_{0}^{1} dX_{2} p_{2}(X_{2})
\Biggl \{
F(X_{1},X_{2})
 - 120 \frac{1-2X_{1}+X_{1}^{2}}{(1+X_{1})^{6}}\ \Bigr \} 
=-\left (\frac{3}{2}- 4\ln 2 \right)
\left( \frac{T}{\Tc(h)}\right)^{2}
+ O \left(\frac{T}{\Tc(h)}\right)^{3}
\emat
Combining the above results we finally obtain \eq{eq-kappa-3-2},
\beqa
\sav{\kappa^{2}_{3}(Y)}^{\rm 3-level}= 
(N \Dyeff^{2})^{3}\;\;
2 \left  [\frac{T}{\Tc(h)} - \left(\frac{T}{\Tc(h)} \right)^{2}
+ O \left(\frac{T}{\Tc(h)}\right)^{3} \right].
\eeqa

\subsection{Other moments}
\label{subsec-other-moments}

Here we present results of some other moments needed in the present paper
using the 3-level model.
The 2nd moment of the 1st thermal cumulant is evaluated as,
\beqa
&& \sav{\kappa^{2}_{1}(Y)}
=  \lsav
\left(Y_{0}\tO{01}+Y_{1}\tO{11}+Y_{2}\tO{21}\right)
\left(Y_{0}\tO{02}+Y_{1}\tO{12}+Y_{2}\tO{22}\right) \nonumber \\
&&  \left.
\ln Z(O_{01},O_{11},O_{21},X_{1},X_{2})
\ln Z(O_{02},O_{12},O_{22},X_{1},X_{2})
 \right |_{O=1} \rsav \nonumber \\
&&= (\sqrt{N} \Dyeffs)^{2}
)\int_{0}^{1} dX_{1} p_{1}(X_{1}) \int_{0}^{1} dX_{2} p_{2}(X_{2})
\frac{1+X_{1}^{2}+X_{1}^{2}X_{2}^{2}}{(1+X_{1}+X_{1}X_{2})^{2}} 
 = (\sqrt{N} \Dyeffs)^{2} \left\{ 1-
\frac{T}{T_{c}(h)} + \left ( \frac{T}{T_{c}(h)}\right)^{3}\right\}.
\eeqa

We also need $\sav{\kappa_{3}(Y)\kappa_{Y}}$ which is evaluated as,
\beqa
&& \sav{\kappa_{3}(Y)\kappa_{1}{Y}}
=  \lsav
\left(Y_{0}\tO{01}+Y_{1}\tO{11}+Y_{2}\tO{21}\right)^{3}
\left(Y_{0}\tO{02}+Y_{1}\tO{12}+Y_{2}\tO{22}\right) \nonumber \\
&&  \left.
\ln Z(O_{01},O_{11},O_{21},X_{1},X_{2})
\ln Z(O_{02},O_{12},O_{22},X_{1},X_{2})
 \right |_{O=1} \rsav \nonumber \\
&& = (\sqrt{N} \Dyeffs)^{4}
)\int_{0}^{1} dX_{1} p_{1}(X_{1}\int_{0}^{1} dX_{2} p_{2}(X_{2})
\frac{-6X_{1}(-2X_{1}+1+X_{1}^{2}+X_{2}-2X_{1}X_{2}^{2}-2X_{1}^{3}X_{2}^{2}+X_{1}^{3}X_{2}+X_{1}^{3}X_{2}^{3}+X_{1}^{2}X_{2}^{3})}{(1+X_{1}+X_{1}X_{2})^{4}}  \nonumber \\
&& = -(\sqrt{N} \Dyeffs)^{4}
\left \{ 
\frac{T}{T_{c}(h)}-\left(\frac{T}{T_{c}(h)}\right)^{2}
+O\left(\frac{T}{T_{c}(h)}\right)^{3}
\right\}
\eeqa

\section{Distribution of mesoscopic response at $T=0$}
\label{sec-direct-resp-appendix}

In this appendix we compute the distribution function
\eq{eq-p-psi} of the response $\psi \equiv Y(h+\delta h)-Y(h)$
at zero temperature.
For our convenience let us introduce some short hand notations,
\beq
\int DY_{l} \ldots \equiv \int \frac{d Y_{l}}{\sqrt{2\pi N \Dy^{2}}}
\exp \left ( 
-\frac{Y^{2}_{l}}{2N \Dy^{2}}
\right)  \ldots
\qquad 
\int DF_{(l,m)} \ldots \equiv \int d \Delta F_{l,m} 
\rho_{(l,m)} (\Delta F_{l,m}) \ldots.
\label{eq-def-shorthand}
\eeq
where $\rho_{(l,m)} (\Delta F_{l,m})$ is the distribution of
the free-energy difference $\Delta F_{l,m}$ between $l$-th and $m$-th states
under $h$ (See sec. \ref{subsubsec-level-spacing}).

For the two-level model it is easy to find,
\beq
p_{1}(\psi)=\int_{-\infty}^{\infty}D Y_{0} 
\int_{Y_{0}}^{\infty} DY_{1}
\int_{0}^{h(Y_{1}-Y_{0})}D \Delta F_{(1,0)} \delta(\psi-(Y_{1}-Y_{0}))
\qquad 
p_{0}(\psi)=\delta(\psi) \left(1- \int_{0}^{\infty} d\psi' p_{1}(\psi')\right)
\eeq
and thus
\beqa
p(h,\psi)= \delta(\psi) \left ( 
1-\int_{0}^{\infty} d\psi' 
\frac{e^{-\frac{(\psi')^{2}}{4 N\Dy}}} {\sqrt{4 \pi N \Dy^{2}}}
(1-e^{-\frac{h}{\Tc} \psi'})
\right)
+ \frac{e^{-\frac{\psi^{2}}{4 N\Dy}}}{\sqrt{4 \pi N \Dy^{2}}}
(1-e^{-\frac{h}{\Tc} \psi}).
\eeqa
using \eq{eq-def-shorthand},\eq{eq-p-tY} and \eq{eq-p-df-norepulsion}.
We again remind the readers that we are actually studying
the fluctuations of the transverse variable $\tY$.
Here the 1st and 2nd terms on the r.h.s correspond to 
$p_{0}(\psi)$ and $p_{1}(\psi)$ respectively.

Now  we extend the analysis including the 3rd state. 
Among the diagrams shown in Fig.~\ref{fig-resp-cross-2-3-level},
the cases that the state $1$ becomes the ground state 
appear in 3), 4) and 6-b). Then we can write it as,
\beqa
 p_{1}(\psi)=\int_{-\infty}^{\infty} DY_{0} \int_{Y_{0}}^{\infty} DY_{1}
\delta (\psi-(Y_{1}-Y_{0}))
\int_{0}^{h(Y_{1}-Y_{0})} D \Delta F_{10} 
 \left[
\int_{-\infty}^{Y_{1}} DY_{2}  \right. && \\ \nonumber
 \left.  + \int_{Y_{1}}^{\infty} DY_{2}
\int_{h(Y_{2}-Y_{1})}^{\infty} D \Delta F_{21} \right]. &&
\eeqa
Here the 1st term in $[ \ldots ]$ is the contribution from the case
3) and 4). The 2nd term $[ \ldots ]$ is due to the case 6-b).
Using \eq{eq-p-tY} and \eq{eq-p-df-norepulsion} we obtain,
\beqa
&& p_{1}(\psi)=\frac{e^{-\frac{\psi^2}{4N\Dy^{2}}}}{\sqrt{4\pi N \Dy^{2}}}
\left ( 1-e^{-\frac{h\psi}{\Tc}} \right)
\left [
1  \right. \nonumber \\ 
&& \left. -\int_{-\infty}^{\infty} \frac{dY_{+}}{\sqrt{4\pi N \Dy^{2}}} 
e^{-\frac{Y_{+}^{2}}{4N \Dy^{2}}}
\int_{\frac{Y_{+}+\psi}{2}}^{\infty} \frac{dY_{2}}{\sqrt{ 2 \pi N \Dy^{2}}}
e^{-\frac{Y_{2}^{2}}{2N \Dy^{2}}}
\left ( 1- e^{-\frac{h \left( Y_{2}-\frac{Y_{+}+ \psi}{2} \right)}{2\Tc}}
\right)
\right]
\eeqa
in the last equation we introduced new variables
$Y_{\pm}=Y_{1} \pm Y_{0}$ and explicitly integrated over $Y_{-}$.
The distribution of the scaled variable $\tpsi$ given in \eq{eq-def-tpsi}
is obtained as,
\beq
\tilde{p}_{1}(\tpsi)=\frac{e^{-\frac{\tpsi^{2}}{4}}}{\sqrt{4\pi}}
\left( 1 - e^{-\frac{h}{\hs}\tpsi} \right)
\left[ 
1-\int_{-\infty}^{\infty} \frac{dz_1}{\sqrt{4\pi}}e^{-\frac{z_{1}^{2}}{4}}
\int_{\frac{z_{1}+\tpsi}{2}}^{\infty}\frac{dz_{2}}{\sqrt{2\pi}}
e^{-\frac{z_{2}^{2}}{2}}
\left(
1-e^{-\frac{1}{2}\frac{h}{\hs}(z_{2}-\frac{z_{1}+\tpsi}{2})}
\right)
\right].
\label{eq-p-tpsi-3-level-scaling-1}
\eeq
where $\hs$ is given by \eq{eq-scale-hs-2}.

Next we examine 
the cases that the state $2$ becomes the ground state 
which can be found 5), 6-a) and 6-b) in
the diagrams in Fig.~\ref{fig-resp-cross-2-3-level} which yield,
\beqa
 p_{2}(\psi)=
\int_{-\infty}^{\infty} DY_{0} 
\int_{-\infty}^{Y_{0}} DY_{1} 
\int_{Y_{1}}^{\infty} DY_{2} \delta (\psi-(Y_{2}-Y_{0}))
\int_{0}^{h(Y_{2}-Y_{0})} D \Delta F_{20} 
&& \nonumber \\
+\int_{-\infty}^{\infty} DY_{0} 
\int_{Y_{0}}^{\infty} DY_{1} 
\int_{Y_{1}}^{\infty} DY_{2} \delta (\psi-(Y_{2}-Y_{0}))
\int_{0}^{h(Y_{2}-Y_{1})} D \Delta F_{21} 
\int_{\Delta F_{21}}^{h(Y_{2}-Y_{0})} D \Delta F_{20} &&
\eeqa
The  1st term on the r.h.s. of the above equation is due to the
case 5) while the 2nd term is due to the case 6-a) and 6-b).
The 1ast two integrals in the 2nd term ensure that both
two level crossings $(1,2)$ and $(0,2)$ have took place
in 6-a) and 6-b). (The order is reversed in the two cases.)

Using now \eq{eq-p-tY} and \eq{eq-p-df-norepulsion} for the $(0,2)$
level crossing, which is ``repulsive'', we obtain,
\beqa
&& p_{2}(\psi)=\left( 
1-2e^{-\frac{h \psi}{\Tc}}+e^{-2\frac{h \psi}{\Tc}}
\right)
\int_{-\infty}^{\infty} \frac{dY_{+}}{\sqrt{4\pi N \Dy^{2}}} e^{-\frac{Y_{+}^{2}}{4 N \Dy^{2}}}
\int_{\frac{Y_{+}-\psi}{2}}^{\infty} \frac{d Y_2}{\sqrt{2\pi N \Dy^{2}}} e^{-\frac{Y_{2}^{2}}{2 N \Dy^{2}}} \nonumber \\
&& + 
\left( 
1-2e^{-\frac{h \psi}{\Tc}}+e^{-2\frac{h \psi}{\Tc}}
\right)
\int_{-\infty}^{\infty} \frac{dY_{0}}{\sqrt{2\pi N \Dy^{2}}} 
e^{-\frac{Y_{0}^{2}}{2 N \Dy^{2}}}
\int_{Y_{0}}^{\infty} \frac{d Y_{1}}{\sqrt{2\pi N \Dy^{2}}} 
e^{-\frac{Y_{1}^{2}}{2 N \Dy^{2}}}
 \left( 1-e^{-2X} \right)_{X=\frac{h}{T}(\psi-(Y_{1}-Y_{0}))}
\nonumber \\
&& 
+\int_{-\infty}^{\infty} \frac{dY_{0}}{\sqrt{2\pi N \Dy^{2}}} 
e^{-\frac{Y_{0}^{2}}{2 N \Dy^{2}}}
\int_{Y_{0}}^{\infty} \frac{d Y_{1}}{\sqrt{2\pi N \Dy^{2}}}
e^{-\frac{Y_{1}^{2}}{2 N \Dy^{2}}}
f(X)_{X=\frac{h}{T}(\psi-(Y_{1}-Y_{0}))}
\label{eq-p-tpsi-3-level-2}
\eeqa
with
\beq
f(x) \equiv \left[ -(1-e^{-2X})+\frac{4(1-e^{-3X})}{3}-\frac{(1-e^{-4X})}{2} \right]
\eeq
In the 1st term on the r.h.s of \eq{eq-p-tpsi-3-level-2} we used
$Y_{\pm}=Y_{2} \pm Y_{0}$ and explicitly integrated over $Y_{-}$.

From the above result 
the distribution of the scaled variable $\tpsi$ given in \eq{eq-def-tpsi}
is obtained as,
\beqa
&& \tilde{p}_{2}(\tpsi)=\left( 
1-2e^{-\frac{h}{\hs}\tpsi}+e^{-2\frac{h}{\hs}\tpsi}
\right)
\int_{-\infty}^{\infty} \frac{dz_1}{\sqrt{4\pi}} e^{-\frac{z_{1}^{2}}{4}}
\int_{\frac{z_{1}-\tpsi}{2}}^{\infty} \frac{dz_2}{\sqrt{2\pi}} e^{-\frac{z_{2}^{2}}{2}} \nonumber \\
&& + 
\left( 
1-2e^{-\frac{h}{\hs}\tpsi}+e^{-2\frac{h}{\hs}\tpsi}
\right)
\int_{-\infty}^{\infty} \frac{dz_1}{\sqrt{2\pi}} e^{-\frac{z_{1}^{2}}{2}}
\int_{z_{1}}^{\infty} \frac{dz_2}{\sqrt{2\pi}} e^{-\frac{z_{2}^{2}}{2}}
 \left( 1-e^{-2X} \right)_{X=\frac{h}{\hs}(\tpsi-(z_{2}-z_{1}))}
\nonumber \\
&& +\int_{-\infty}^{\infty} \frac{dz_1}{\sqrt{2\pi}} e^{-\frac{z_{1}^{2}}{2}}
\int_{z_{1}}^{\infty} \frac{dz_2}{\sqrt{2\pi}} e^{-\frac{z_{2}^{2}}{2}}
\left[
-(1-e^{-2X})+\frac{4}{3}(1-e^{-3X})-\frac{1}{2}(1-e^{-4X})
\right]_{X=\frac{h}{\hs}(\tpsi-(z_{2}-z_{1}))}.
\eeqa

\section{Computations of sample-to-sample fluctuations of the 
thermal cumulants by a replica approach}
\label{appendix-sample-fluctuation-replica}

Here we report some details of the computations of
the moments of thermal cumulants $\kappa_{k}(Y)$ using
the prescription \eq{eq-kappa-n-p-replica}, which reads as,
\beqa
&& \sav{\kappa^{p}_{k}(Y)}= (\sqrt{N}\Dyeffs)^{kp}
\left.
\lim_{ n_{1},n_{2},\ldots,n_{p} \to 0} 
 \frac{1}{\prod_{r=1}^{p} n_{r}}
\frac{\partial^{k}}{\partial x_{1}^{k}}
\frac{\partial^{k}}{\partial x_{2}^{k}}
\cdots \frac{\partial^{k}}{\partial x_{p}^{k}}
\spsum
\exp \left[
\frac{1}{2}\sum_{r,s=1}^{p} x_{r}A_{rs}x_{s}
\right]
\right |_{x=0} \nonumber \\
&& = (\sqrt{N}\Dyeffs)^{kp}
\left.
\lim_{ n_{1},n_{2},\ldots,n_{p} \to 0} 
 \frac{1}{\prod_{r=1}^{p} n_{r}}
\frac{\partial^{k}}{\partial x_{1}^{k}}
\frac{\partial^{k}}{\partial x_{2}^{k}}
\cdots \frac{\partial^{k}}{\partial x_{p}^{k}}
\spav{
\exp \left[
\frac{1}{2}\sum_{r,s=1}^{p} x_{r}A_{rs}x_{s}
\right]
}\right |_{x=0}
 \label{eq-kappa-n-p-replica-appendix}
\eeqa
where
\beq
A_{rs}\equiv  \sum_{\alpha=(r,1)}^{(r,n_{r})} \sum_{\beta=(s,1)}^{(s,n_{s})}
\delta_{i_{\alpha},i_{\beta}}
\eeq
In  \eq{eq-kappa-n-p-replica-appendix}
the sum $\spsum$ stands for summation over all possible
partitions of $\nt$ replicas into $\nt/m$ clusters of 
size $m$.


In the 2nd equation $\spav{\ldots}$ is the average
over the all possible 1 step RSB solutions obtained by permutations of replicas
, i.e.,
\beq
\spav{\ldots} \equiv \frac{\spsum \ldots }{\spsum 1}
\label{eq-def-spav}
\eeq
where $\spsum 1=\frac{n!}{(m!)^{n/m}} \xrightarrow[ n \to 0]{} 1 $.

\begin{figure}[h]
\begin{center}
\includegraphics[width=0.8\textwidth]{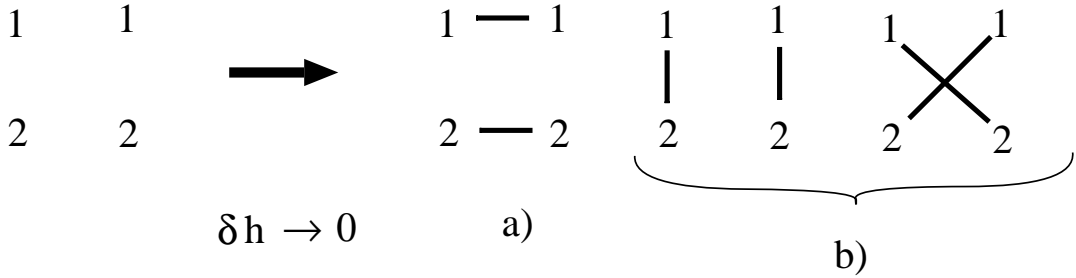}
\end{center}
\caption{A diagrammatic representation of terms which survive
after $\delta h \to 0$ among all terms
generated by the differentiations $\partial/\partial h$s 
in \eq{eq-kappa-n-p-replica-appendix}.
}
\label{fig-replica-graph-1}
\end{figure}

The computations of the moments
based on the formula \eq{eq-kappa-n-p-replica-appendix}
can be done diagrammatically. For instance let us consider
the case of $\sav{\kappa_{2}^{2}(Y)}$, i.e $k=2$ and $p=2$.
In this case we have $4$ derivatives in \eq{eq-kappa-n-p-replica-appendix}
among which $2$ of them are associated with the group-1 of replicas
and the other $2$ are associated with the group-2 of replicas. 
\begin{enumerate}
\item First we represent them by 'isolated indexes'
as in the left hand side of Fig.~\ref{fig-replica-graph-1}.
'1' means that it stands for a derivative associated with the group-1
of replicas and '2' means that it is associated with the group-2 of replicas.
\item  All terms generated by the differentiations 
$\partial/\partial \delta h$ 
at $\delta h=0$ in \eq{eq-kappa-n-p-replica-appendix} can be
enumerated by making pairs of indexes as in the right hand 
side of Fig.~\ref{fig-replica-graph-1}. As the result we find
\bmat
\sav{\kappa^{2}_{2}(Y)}= (N \Dyeff^{2})^{2}\cdot  \lim_{n_{1},n_{2} \to 0} 
\frac{1}{n_{1}n_{2}}(\spav{A_{11}A_{22}}+2\spav{A^{2}_{12}}).
\emat
Here $A_{11}A_{22}$ is due to the graph a) and $2A^{2}_{12}$
is due to the two graphs of type b).

\end{enumerate}

\begin{figure}[h]
\begin{center}
\includegraphics[width=0.9\textwidth]{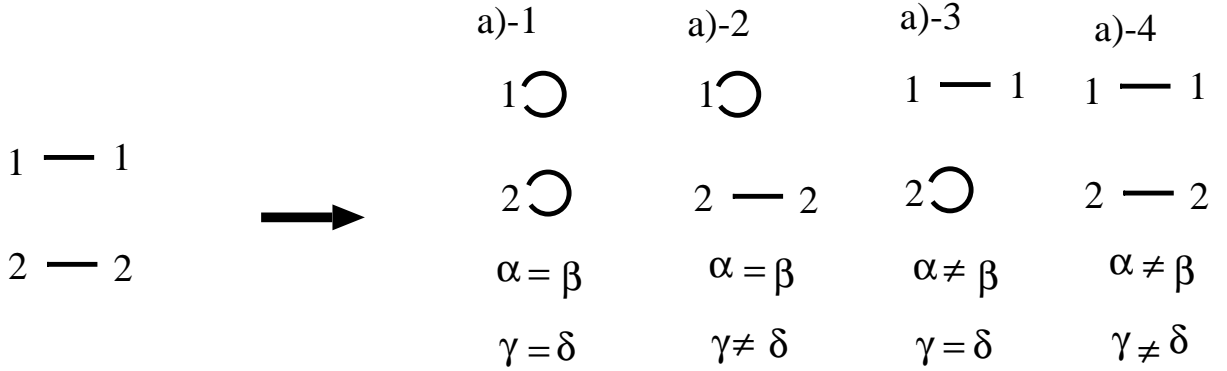}
\end{center}
\caption{Classification of 
the terms in $\spav{A_{11}A_{22}}$ represented by the diagram
on the left hand side. Indexes on the diagrams
a)-1 to a)-4 represent
non-identical replicas and their numbers represent the groups to which
they belongs to.} 
\label{fig-replica-graph-2}
\end{figure}

\begin{enumerate}
\setcounter{enumi}{2}
\item Now we have to evaluate
$\spav{A_{11}A_{22}}$ 
and $\spav{A_{12}^{2}}$. Let us take the case of $\spav{A_{11}A_{22}}$
which is represented by a) in Fig.~\ref{fig-replica-graph-1}.

Using \eq{eq-def-A}, $\spav{A_{11}A_{22}}$ can be written explicitly 
as,
\bmat
\spav{A_{11}A_{22}}=\sum_{\alpha=(1,1)}^{(1,n_{1})}\sum_{\beta=(1,1)}^{(1,n_{1})}\sum_{\gamma=(2,1)}^{(2,n_{2})}\sum_{\delta=(2,1)}^{(2,n_{2})}\spav{\delta_{i_{\alpha},i_{\beta}}\delta_{i_{\gamma},i_{\delta}}}.
\emat
Here replicas $\alpha$ and $\beta$ 
belong to the group-1 while $\gamma$ and $\delta$ belong to the group-2.

To proceed further, we classify
the terms in the multiple free-sums over replica indexes 
into some subsets as represented by 
the diagrams in Fig.~\ref{fig-replica-graph-2}.
Different diagrams represent different ways to contract the indexes.
From now on indexes represent replicas 
{\it which do not overlap with each other}.
Each bond represent a Kronecker-delta.
A diagram consists of one or several sets of 'connected bonds'.
A sum over replicas is associated with each index, under
the constraint that replicas associated with different 
indexes do not overlap with each other.

Thus the diagrams are identified as follows
\beqa
&& \mbox{a)-1} \hspace*{1cm}\to\hspace*{1cm} \ssum_{\alpha;\gamma} \spav{1} \nonumber \\
&& \mbox{a)-2} \hspace*{1cm}\to\hspace*{1cm} \ssum_{\alpha; \gamma, \delta} \spav{\delta_{i_{\gamma},i_{\delta}}}\nonumber \\
&& \mbox{a)-3} \hspace*{1cm}\to\hspace*{1cm}
\ssum_{\alpha , \beta ; \gamma} \spav{\delta_{i_{\alpha},i_{\beta}}} \nonumber \\
&& \mbox{a)-4} \hspace*{1cm}\to\hspace*{1cm}
\ssum_{\alpha , \beta ; \gamma , \delta} \spav{\delta_{i_{\alpha},i_{\beta}}\delta_{i_{\gamma},i_{\delta}}}
\eeqa
Here $\alpha$ and $\beta$ belong to group-1 while $\gamma$ and $\delta$
belong to group-2. We use the symbol $\ssum$ to mean that the running
indexes do not overlap with each other;
\bmat
\ssum_{\alpha,\beta;\gamma,\delta}\ldots  \equiv 2\sum_{\alpha=1}^{n_{1}}\sum_{\beta=\alpha+1}^{n_{1}} \cdot\; 2\sum_{\gamma=1}^{n_{2}}\sum_{\delta=\gamma+1}^{n_{2}} \ldots.
\emat
\end{enumerate}

\begin{figure}[t]
\begin{center}
\includegraphics[width=0.7\textwidth]{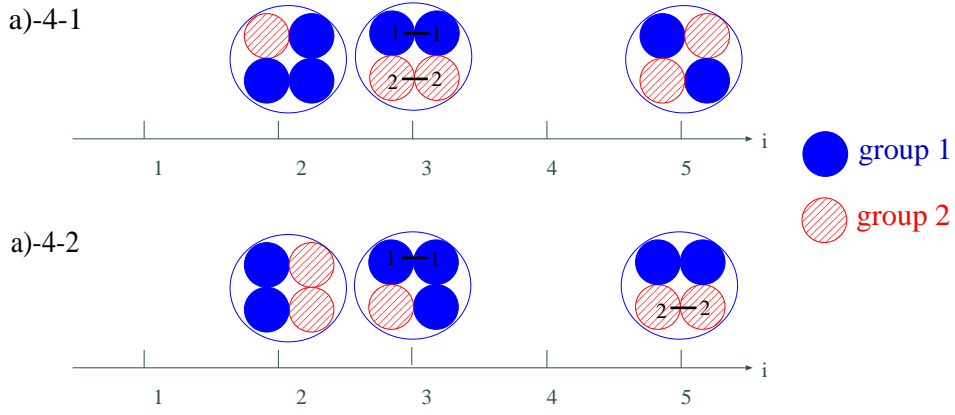}
\end{center}
\caption{A schematic representation of some 1RSB solutions. 
Closed circles represent clusters of the 1RSB 
solutions. $n_{1}$ replicas belong to group $1$ (blue) and $n_{2}$ 
replicas belong to group $2$ (red).
The replicas of the two groups $\nt=n_{1}+n_{2}$
are mixed together and divided into clusters of size $m$.
Here two kinds of solutions which 
make non-zero contributions for the diagram a)-4 are shown:
the two sets of 'connected bonds' $1-1$' and '$2-2$'are contained in a
single cluster a)-4-1 or in different clusters a)-4-2.
}
\label{fig-replica-graph-3}
\end{figure}

\begin{enumerate}
\setcounter{enumi}{3}
\item Now we are left to evaluate expectation values $\spav{\ldots}$
(See \eq{eq-def-spav}) of the diagrams. An 1RSB solution is
represented by a configuration of clusters (See Fig.~\ref{fig-m-atom}).

Note for instance 
that $\delta_{i_{\alpha},i_{\beta}}$ is $0$ in a solution where
$\alpha$ and $\beta$ belong to different clusters. Thus
we just need to count numbers of 1RSB solutions in which 
all replicas belonging to a set of 'connected bonds'
are contained in a common cluster.

For example let us consider the case of a)-4;
\bmat
\ssum_{\alpha,\beta;\gamma,\delta}
\spav{\delta_{i_{\alpha},i_{\beta}}\delta_{i_{\gamma},i_{\delta}}}
\emat
There are two
possible configurations of clusters which make non-zero contributions,
say a)-4-1 and a)-4-2 as shown in Fig.~\ref{fig-replica-graph-3}.

The case of a)-4-1 is evaluated as follows,
\bmat
n_{1}(n_{1}-1) \times n_{2}(n_{2}-1) \times
\frac{\nt}{m} \frac{m (m-1)}{\nt (\nt-1)} \times 
\frac{(m-2)(m-3)}{(\nt-2)(\nt-3)}
\emat
The first two factors $n_{1}(n_{1}-1)$ and  $n_{2}(n_{2}-1)$
represent number of ways to choose the $4$ different replicas
associated with the $4$ running indexes in a)-4:
$2$ replicas are chosen out of $n_{1}$
replicas in group-1 and $2$ replicas are chosen out of $n_{2}$ replicas
in group-2. The last two factors
\bmat
\frac{\nt}{m} \frac{m (m-1)}{\nt (\nt-1)}\times 
\frac{(m-2)(m-3)}{(\nt-2)(\nt-3)}
\emat
are the probability to have a solution like a)-4-1
in Fig.~\ref{fig-replica-graph-3} given $\nt$ replicas. 

The probability is obtained by
counting the number of such solutions and dividing the result
by the total number of solutions $\nt!$.
In the case a)-4-1 all $4$ replicas
associated with the 'connected bonds' are put in a single cluster.
Thus first we note that 
there are $\nt/m$ possible ways to choose the cluster among 
$\nt/m$ clusters. Then we note that 
there are $m(m-1)(m-2)(m-3)$ different ways to choose 
$4$ distinct replicas for the running $4$ indexes 
out of $m$ replicas in the cluster.
Last we note that there are $(\nt -4)!$ ways to choose the rest of replicas.
Thus the number of solutions we wanted is
$\nt/m \times m(m-1)(m-2)(m-3) \times (\nt -4)!$. Then dividing the
latter by the total number of solutions $\nt !$ we obtain the factor
displayed above.

The case of a)-4-2 is evaluated in the same way as follows,
\bmat
n_{1}(n_{1}-1) \times n_{2}(n_{2}-1) \times
\frac{\nt}{m} 
\frac{m (m-1)}{\nt (\nt-1)} \times 
\left (\frac{\nt}{m}-1 \right)
\frac{m(m-1)}{(\nt-2)(\nt-3)}.
\emat
Here the factor $\nt/m-1$ is the number of ways to choose a '2nd cluster'.

Summing up the results of a)-4-1 and a)-4-2 explained above, we obtain
\beqa
&& \lim_{n_{1},n_{2} \to 0}\frac{1}{n_{1}n_{2}}
\ssum_{\alpha,\beta;\gamma,\delta} \spav{\delta_{i_{\alpha},i_{\beta}}\delta_{i_{\gamma},i_{\delta}}} \nonumber \\
&& = \lim_{n_{1},n_{2} \to 0}\frac{1}{n_{1}n_{2}} 
n_{1}(n_{1}-1)n_{2}(n_{2}-1) 
\left \{  \frac{\nt}{m} \frac{m (m-1)}{\nt (\nt-1)}
\frac{(m-2)(m-3)}{(\nt-2)(\nt-3)}\right.\nonumber \\
&& 
+ \left. \frac{\nt}{m} 
\frac{m (m-1)}{\nt (\nt-1)} 
\left (\frac{\nt}{m}-1 \right)
\frac{m(m-1)}{(\nt-2)(\nt-3)}
\right \} \nonumber \\
&& = \frac{(1-m)(2-m)(3-m)}{3!}+ \frac{m(1-m)^{2}}{3!}
 \nonumber 
\eeqa
\end{enumerate}

\subsubsection{Linear susceptibility}
\label{subsec-linear-sus-replica-appendix}

\begin{figure}[b]
\begin{center}
\includegraphics[width=0.6\textwidth]{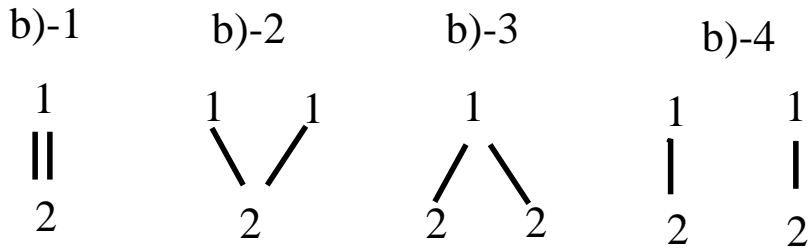}
\end{center}
\caption{
Diagrams contributing to  $\spav{A^{2}_{12}}$.
Indexes on the diagrams represent non-identical replicas and their numbers 
represent the groups to which they belongs to.
}
\label{fig-replica-graph-2b}
\end{figure}

The linear susceptibility is related to the
2nd thermal cumulant as $\chi_{1}=\beta \kappa_{2}(Y)$.
In section \ref{subsec-thermal-fluctuation-replica} we present the computation
of the sample-average of its 1st moment $\sav{\kappa_{2}(Y)}$.

In the following we describe computation of the 2nd moment
$\sav{\kappa^{2}_{2}(Y)}$.
In the previous subsection sec. \ref{subsec-generating-functional}
we have already found that
\beq
\sav{\kappa_{2}^{2}(Y)}= N \Dyeff^{2}\cdot  \lim_{n_{1},n_{2} \to 0} 
\frac{1}{n_{1}n_{2}}(\spav{A_{11}A_{22}}+2\spav{A^{2}_{12}}).
\label{eq-kappa-2-2-appendix-0}
\eeq
This is obtained by analysing possible pairings of the $4$ 'running
indexes' as in Fig.~\ref{fig-replica-graph-1}.  Next 
$\spav{A_{11}A_{22}}$ and $\spav{A^{2}_{12}}$ must be evaluated.

In the previous section we have evaluated $\spav{A_{11}A_{22}}$
to a certain extent. First we noted that it can be considered
as sum over contributions from diagrams a)-1 to a)-4 shown in
Fig.~\ref{fig-replica-graph-2}. An important point there is that
the multiple 'running indexes' do not overlap with each other.
We have evaluated the contribution from diagram a)-4.

In similar ways we obtain contributions from other diagrams a)-1 to a)-3.
From diagram a)-1 we simply find
\bmat
\lim_{n_{1},n_{2} \to 0}\frac{1}{n_{1}n_{2}}
\ssum_{\alpha;\gamma}1=1.
\emat
From diagram a)-2 we find,
\beqa
&& \lim_{n_{1},n_{2} \to 0}\frac{1}{n_{1}n_{2}}
\ssum_{\alpha;\gamma,\delta} \spav{\delta_{i_{\gamma},i_{\delta}}}
\nonumber \\
&& = \lim_{n_{2} \to 0}\frac{1}{n_{2}} 
n_{2}(n_{2}-1) 
 \frac{\nt}{m} 
\frac{m (m-1)}{\nt (\nt-1)} 
\nonumber \\
&& = -(1-m) \nonumber.
\eeqa
From diagram a)-3 we obtain the same result.
Summing up these results we obtain,
\beq
\lim_{n_{1},n_{2} \to 0}\frac{1}{n_{1}n_{2}}
\spav{A_{11}A_{22}}=1-2(1-m)+
\left \{ \frac{(1-m)(2-m)(3-m)}{3!}+ \frac{m(1-m)^{2}}{3!} \right \}.
\label{eq-a11a22}
\eeq

The evaluation of $\spav{A^{2}_{11}}$ which reads as
\bmat
\spav{A^{2}_{12}}=\sum_{\alpha=(1,1)}^{(1,n_{1})}\sum_{\beta=(2,1)}^{(2,n_{1})}\sum_{\gamma=(1,1)}^{(1,n_{2})}\sum_{\delta=(2,1)}^{(2,n_{2})}\spav{\delta_{i_{\alpha},i_{\beta}}\delta_{i_{\gamma},i_{\delta}}}.
\emat
can be done in the same way.
Diagrams corresponding to those in Fig.~\ref{fig-replica-graph-2}
are the diagrams b)-1 to b)-4 shown in Fig.~\ref{fig-replica-graph-2b}.
The diagrams are evaluated as the following,
\beqa
&& \mbox{b)-1} \hspace*{1cm}\to\hspace*{1cm} \lim_{n_{1},n_{2} \to 0}\frac{1}{n_{1}n_{2}} 
\ssum_{\alpha ;\gamma} \spav{\delta_{\alpha,\gamma}}=1-m \nonumber \\
&& \mbox{b)-2} \hspace*{1cm}\to\hspace*{1cm} \lim_{n_{1},n_{2} \to 0}\frac{1}{n_{1}n_{2}}
\ssum_{\alpha,\beta ;\gamma} \spav{\delta_{\alpha,\gamma}\delta_{\beta,\gamma}}
=-\frac{(1-m)(2-m)}{2} \nonumber \\
&& \mbox{b)-3} \hspace*{1cm}\to\hspace*{1cm} \lim_{n_{1},n_{2} \to 0}\frac{1}{n_{1}n_{2}}
\ssum_{\alpha ;\gamma,\delta} \spav{\delta_{\alpha,\gamma}\delta_{\alpha,\delta}}
=-\frac{(1-m)(2-m)}{2} \nonumber \\
&& \mbox{b)-4} \hspace*{1cm}\to\hspace*{1cm} \lim_{n_{1},n_{2} \to 0}\frac{1}{n_{1}n_{2}}
\ssum_{\alpha, \beta ;\gamma, \delta} \spav{\delta_{\alpha,\gamma}\delta_{\beta,\delta}}
= \frac{(1-m)(2-m)(3-m)}{3!}+ \frac{m(1-m)^{2}}{3!}\nonumber 
\eeqa
Note that b)-4 is equivalent to a)-4 so that one must consider two kinds of solutions with
different patterns of clustering as those shown in Fig.~\ref{fig-replica-graph-3}.
Summing up the above results we obtain
\beqa
\lim_{n_{1},n_{2} \to 0}\frac{1}{n_{1}n_{2}}
\spav{A^{2}_{12}} &=& 1-m-2\frac{(1-m)(2-m)}{2}+
\left\{\frac{(1-m)(2-m)(3-m)}{3!}+ \frac{m(1-m)^{2}}{3!} \right\} \nonumber \\
&=& \frac{1}{3}m(1-m).
\label{eq-a12a12}
\eeqa

Using \eq{eq-a11a22} and \eq{eq-a12a12} in \eq{eq-kappa-2-2-appendix-0} we finally obtain  \eq{eq-kappa-2-2-replica},
\beq
\sav{\kappa_{2}^{2}(Y)}= (N \Dyeff^{2})^{2} m .
\label{eq-kappa-2-2-appendix}
\eeq

\subsubsection{Non-linear susceptibility}
\label{subsec-non-linear-sus-replica-appendix}

The non-linear susceptibility is related to the
2nd thermal cumulant as $\chi_{2}=(\beta^{2}/2!) \kappa_{3}(Y)$.
Due to a simple symmetry reason sample-average 
of the 1st moment $\sav{\kappa_{3}(Y)}$ is zero.
We examine sample-average of the 2nd moment $\sav{\kappa^{2}_{3}(Y)}$ below.
Following the steps 1,2 in sec. \ref{subsec-generating-functional} we obtain,
\beq
\sav{\kappa^{2}_{3}(Y)}= 
(N \Dyeff^{2})^{3}\lim_{n_{1},n_{2} \to 0}\frac{1}{n_{1}n_{2}} 
(9\spav{A_{11}A_{12}A_{22}} + \spav{A^{3}_{12}}).
\eeq

\begin{figure}[t]
\begin{center}
\includegraphics[width=\textwidth]{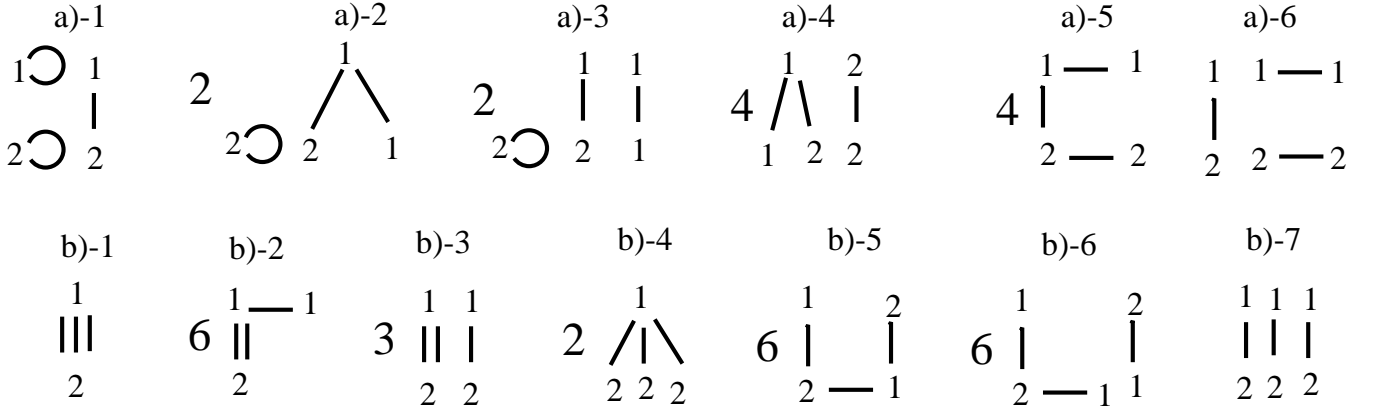}
\end{center}
\caption{
Diagrams contributing to $\sav{\kappa^{2}_{3}(Y)}$.
The upper $6$ diagrams contribute to $\spav{A_{12}A_{22}A_{12}}$
and the lower $7$ diagrams contribute to $\spav{A^{3}_{12}}$.
The numbers of the left hand side of the diagrams represent 
numbers of the graphs of the same geometry.
}
\label{fig-replica-graph-4}
\end{figure}

First we evaluate,
\beqa
&& \lim_{n_{1},n_{2} \to 0}\frac{1}{n_{1}n_{2}} \spav{A_{11}A_{22}A_{12}}\nonumber \\
&& = \lim_{n_{1},n_{2} \to 0}\frac{1}{n_{1}n_{2}}
 \sum_{\alpha_{1}=(1,1)}^{(1,n_{1})}
\sum_{\alpha_{2}=(1,1)}^{(1,n_{1})}
\sum_{\alpha_{3}=(1,1)}^{(1,n_{1})}
\sum_{\beta_{1}=(2,1)}^{(2,n_{2})}
\sum_{\beta_{2}=(2,1)}^{(2,n_{2})}
\sum_{\beta_{3}=(2,1)}^{(2,n_{2})}
\spav{
\delta_{i_{\alpha_{1}},i_{\alpha_{2}}}
\delta_{i_{\beta_{1}},i_{\beta_{2}}}
\delta_{i_{\alpha_{3}},i_{\beta_{3}}}
}.
\eeqa
Following the step 3 in sec. \ref{subsec-generating-functional}
the sum with $6$ running indexes can be divided into contributions
from the diagrams a)-1 to a)-6 shown in Fig.~\ref{fig-replica-graph-4}.
Then by the step 4 they are evaluated as the following, 
\beqa
&& \mbox{a)-1} \hspace*{1cm}\to\hspace*{1cm} \lim_{n_{1},n_{2} \to 0}\frac{1}{n_{1}n_{2}} 
\ssum_{\alpha_{1},\alpha_{2} ;\beta_{1},\beta_{2}} \spav{\delta_{\alpha_{2},\beta_{2}}}=
\lim_{n_{1},n_{2} \to 0} n_{1}n_{2}(1-m)=0 \nonumber \\
&& \mbox{a)-2} \hspace*{1cm}\to\hspace*{1cm} \lim_{n_{1},n_{2} \to 0}\frac{1}{n_{1}n_{2}} 
\ssum_{\alpha_{1},\alpha_{2};\beta_{1},\beta_{2}} 
\spav{\delta_{\alpha_{1},\alpha_{2}\delta_{\alpha_{2},\beta_{1}}}}=
\lim_{n_{1},n_{2} \to 0} O(n_{1}) =0 \nonumber \\
&& \mbox{a)-3} \hspace*{1cm}\to\hspace*{1cm} \lim_{n_{1},n_{2} \to 0}\frac{1}{n_{1}n_{2}} 
\ssum_{\alpha_{1},\alpha_{2},\alpha_{3};\beta_{1},\beta_{2}} 
\spav{\delta_{\alpha_{1},\alpha_{2}\delta_{\alpha_{3},\beta_{3}}}}=
\lim_{n_{1},n_{2} \to 0} O(n_{1}) =0 \nonumber \\
&& \mbox{a)-4} \hspace*{1cm}\to\hspace*{1cm} \lim_{n_{1},n_{2} \to 0}\frac{1}{n_{1}n_{2}} 
\ssum_{\alpha_{1},\alpha_{2};\beta_{1},\beta_{2},\beta_{3}} 
\spav{\delta_{\alpha_{1},\alpha_{2}}\delta_{\alpha_{2},\beta_{1}}\delta_{\beta_{2},\beta_{3}}} \nonumber \\
&&  \hspace*{4cm}=-2! \left\{\frac{m(1-m)^{2}(2-m)}{4!}+\frac{(1-m)(2-m)(3-m)(4-m)}{4!} \right\} \nonumber \\
&& \mbox{a)-5} \hspace*{1cm}\to\hspace*{1cm} \lim_{n_{1},n_{2} \to 0}\frac{1}{n_{1}n_{2}} 
\ssum_{\alpha_{1},\alpha_{2};\beta_{1},\beta_{2}} 
\spav{\delta_{\alpha_{1},\alpha_{2}}\delta_{\alpha_{2},\beta_{1}}\delta_{\beta_{1},\beta_{2}}} 
= \frac{(1-m)(2-m)(3-m)}{3!}\nonumber \\
&& \mbox{a)-6} \hspace*{1cm}\to\hspace*{1cm} \lim_{n_{1},n_{2} \to 0}\frac{1}{n_{1}n_{2}} 
\ssum_{\alpha_{1},\alpha_{2},\alpha_{3};\beta_{1},\beta_{2},\beta_{3}} 
\spav{\delta_{\alpha_{1},\alpha_{2}}\delta_{\beta_{1},\beta_{2}}\delta_{\alpha_{3},\beta_{3}}} \nonumber \\
&& \hspace*{4cm} =(2!)^{2} \left\{
\frac{2! m^{2}(1-m)^{3}}{5!}+3 \frac{m(1-m)^{2}(2-m)(3-m)}{5!} \right. \nonumber \\
&& \hspace*{5cm} \left. +\frac{(1-m)(2-m)(3-m)(4-m)(5-m)}{5!}
\right \}
\eeqa

Next we evaluate,
\beqa
&& \lim_{n_{1},n_{2} \to 0}\frac{1}{n_{1}n_{2}} \spav{A^{3}_{12}}\nonumber \\
&& = \lim_{n_{1},n_{2} \to 0}\frac{1}{n_{1}n_{2}}
 \sum_{\alpha_{1}=(1,1)}^{(1,n_{1})}
\sum_{\alpha_{2}=(1,1)}^{(1,n_{1})}
\sum_{\alpha_{3}=(1,1)}^{(1,n_{1})}
\sum_{\beta_{1}=(2,1)}^{(2,n_{2})}
\sum_{\beta_{2}=(2,1)}^{(2,n_{2})}
\sum_{\beta_{3}=(2,1)}^{(2,n_{2})}
\spav{
\delta_{i_{\alpha_{1}},i_{\beta_{1}}}
\delta_{i_{\alpha_{2}},i_{\beta_{2}}}
\delta_{i_{\alpha_{3}},i_{\beta_{3}}}
}.
\eeqa
The sum can be divided into contributions
from the diagrams b)-1 to b)-7 shown in Fig.~\ref{fig-replica-graph-4},
which are evaluated as,
\beqa
&& \mbox{b)-1} \hspace*{1cm}\to\hspace*{1cm} \lim_{n_{1},n_{2} \to 0}\frac{1}{n_{1}n_{2}} 
\ssum_{\alpha_{1} ;\beta_{1}} \spav{\delta_{\alpha_{1},\beta_{1}}}=1-m \nonumber \\
&& \mbox{b)-2} \hspace*{1cm}\to\hspace*{1cm} \lim_{n_{1},n_{2} \to 0}\frac{1}{n_{1}n_{2}} 
\ssum_{\alpha_{1},\alpha_{2};\beta_{1}} \spav{\delta_{\alpha_{1},\alpha_{2}}\delta_{\alpha_{1},\beta_{1}}}
=-\frac{(1-m)(2-m)}{2!} \nonumber \\
&& \mbox{b)-3} \hspace*{1cm}\to\hspace*{1cm} \lim_{n_{1},n_{2} \to 0}\frac{1}{n_{1}n_{2}} 
\ssum_{\alpha_{1},\alpha_{2} ;\beta_{1},\beta_{2}}
\spav{\delta_{\alpha_{1},\alpha_{2}}\delta_{\beta_{1},\beta_{2}}}
=\frac{m(1-m)^{2}}{3!}+\frac{(1-m)(2-m)(3-m)}{3!}\nonumber \\
&& \mbox{b)-4} \hspace*{1cm}\to\hspace*{1cm} \lim_{n_{1},n_{2} \to 0}\frac{1}{n_{1}n_{2}} 
\ssum_{\alpha_{1};\beta_{1},\beta_{2},\beta_{3}}
\spav{\delta_{\alpha_{1},\beta_{1}}\delta_{\alpha_{1},\beta_{2}}\delta_{\alpha_{1},\beta_{3}}}
=2! \frac{(1-m)(2-m)(3-m)}{3!}\nonumber \\
&& \mbox{b)-5} \hspace*{1cm}\to\hspace*{1cm} \lim_{n_{1},n_{2} \to 0}\frac{1}{n_{1}n_{2}} 
\ssum_{\alpha_{1},\alpha_{2};\beta_{1},\beta_{2}}
\spav{\delta_{\alpha_{1},\beta_{1}}\delta_{\alpha_{2},\beta_{1}}\delta_{\alpha_{2},\beta_{2}}}
=\frac{(1-m)(2-m)(3-m)}{3!}\nonumber \\
&& \mbox{b)-6} \hspace*{1cm}\to\hspace*{1cm} \lim_{n_{1},n_{2} \to 0}\frac{1}{n_{1}n_{2}} 
\ssum_{\alpha_{1},\alpha_{2},\alpha_{3};\beta_{1},\beta_{2}}
\spav{\delta_{\alpha_{1},\beta_{1}}\delta_{\alpha_{2},\beta_{1}}\delta_{\alpha_{3},\beta_{2}}} \nonumber \\
&& \hspace*{4cm}=-2! \left\{\frac{m(1-m)^{2}(2-m)}{4!}+\frac{(1-m)(2-m)(3-m)(4-m)}{4!}\right\}\nonumber \\
&& \mbox{b)-7} \hspace*{1cm}\to\hspace*{1cm} \lim_{n_{1},n_{2} \to 0}\frac{1}{n_{1}n_{2}} 
\ssum_{\alpha_{1},\alpha_{2},\alpha_{3};\beta_{1},\beta_{2},\beta_{3}}
\spav{\delta_{\alpha_{1},\beta_{1}}\delta_{\alpha_{2},\beta_{2}}\delta_{\alpha_{3},\beta_{3}}} \nonumber \\
&& \hspace*{4cm}=2! \frac{m^{2}(1-m)^{3}}{5!}+3\frac{m(1-m)^{2}(2-m)(3-m)}{5!} \nonumber \\
&& \hspace*{4cm} +\frac{(1-m)(2-m)(3-m)(4-m)(5-m)}{5!}.
\eeqa

Summing up the above results we finally obtain \eq{eq-kappa-3-2-replica},
\beq
\sav{\kappa^{2}_{3}(Y)}=(N \Dyeff^{2})^{3} 2m(1-m).
\label{eq-kappa-3-2-replica-appendix}
\eeq

\subsubsection{Other moments}
\label{subsec-other-moments-replica}

\begin{figure}[t]
\begin{center}
\includegraphics[width=0.35\textwidth]{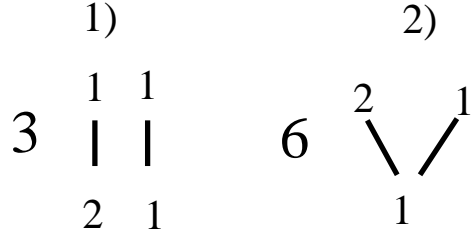}
\end{center}
\caption{
Diagrams contributing to $\sav{\kappa_{3}(Y)\kappa_{1}(Y)}$.
The numbers of the left hand side of the diagrams represent 
numbers of the graphs of the same geometry.
}
\label{fig-replica-graph-5}
\end{figure}

Here we present results of some other moments needed in the present paper.

The 2nd moment of the 1st thermal cumulant is evaluated as,
\beqa
&& \sav{\kappa^{2}_{1}(Y)}=(\sqrt{N} \Dyeffs)^{2}
\lim_{n_{1},n_{2} \to 0} \frac{1}{n_{1}n_{2}}
\frac{\partial}{\partial x_{1}}\frac{\partial}{\partial x_{2}}
\left.
\spav{\exp \left (\frac{1}{2}\sum_{r,s=1}^{2} x_{r}A_{rs}x_{s} \right)}
\right |_{x=0} \nonumber \\
&& =(\sqrt{N} \Dyeffs)^{2} \spav{A_{12}}
 =(\sqrt{N} \Dyeffs)^{2} \spav{\ssum_{\alpha;\beta} \delta_{i_{\alpha},i_{\beta}}} \nonumber \\
&& =(\sqrt{N} \Dyeffs)^{2}(1-m).
\eeqa

We also need $\sav{\kappa_{3}(Y)\kappa_{Y}}$ which is evaluated as,
\beqa
\sav{\kappa_{3}(Y)\kappa_{1}(Y)}= &&
(\sqrt{N} \Dyeffs )^{4} 
\lim_{n_{1} \to 0}\lim_{n_{2} \to 0}\frac{1}{n_{1}n_{2}}
\frac{\partial^{3}}{\partial x_{1}^{3}}
\frac{\partial}{\partial x_{2}}
\left. 
\spav{
\exp \left (\frac{1}{2}\sum_{r,s=1}^{2} x_{r}A_{rs}x_{s} \right)
}
\right |_{x=0} \nonumber \\
&& = (\sqrt{N} \Dyeffs )^{4}  
\lim_{n_{1} \to 0}\lim_{n_{2} \to 0}\frac{1}{n_{1}n_{2}}
3 \spav{A_{11}A_{12}} \nonumber \\
&& = -(\sqrt{N} \Dyeffs )^{4} \lim_{n_{1},n_{2} \to 0}
\frac{1}{n_{1}n_{2}}
\left ( 
3 \ssum_{\alpha,\beta,\gamma;\delta} \spav{\delta_{i_{\alpha},i_{\beta}}
\delta_{i_{\gamma},i_{\delta}}}
+6 \ssum_{\alpha,\beta;\delta} \spav{\delta_{i_{\alpha},i_{\beta}}
\delta_{i_{\beta},i_{\delta}}}
\right) \nonumber \\
&& = -(\sqrt{N} \Dyeffs )^{4}  m(1-m).
\eeqa
The 1st and 2nd terms in the brace $(\ldots)$ in the 2nd equation from below
correspond to the two diagrams presented in Fig. \ref{fig-replica-graph-5}
which are evaluated as the follwoing.
Diagrams corresponding to 1) in Fig. \ref{fig-replica-graph-5} yields,
\beq
\lim_{n_{1},n_{2} \to 0}
\frac{1}{n_{1}n_{2}}
\ssum_{\alpha,\beta,\gamma;\delta} \spav{\delta_{i_{\alpha},i_{\beta}}
\delta_{i_{\gamma},i_{\delta}}}=2\frac{(1-m)(2-m)(3-m)}{3!}
+2 \frac{m(1-m)^{2}}{3!}.
\eeq
The 1st/2nd term on r.h.s is due to the cases that '1-1' and '1-2'
belong to the same/different clusters. On the other hand diagrams
corresponding to 2) in Fig. \ref{fig-replica-graph-5} yields,
\beq
\lim_{n_{1},n_{2} \to 0}
\frac{1}{n_{1}n_{2}}
\ssum_{\alpha,\beta;\delta} \spav{\delta_{i_{\alpha},i_{\beta}}
\delta_{i_{\beta},i_{\delta}}}=-\frac{(1-m)(2-m)}{2!}.
\eeq

\section{Analysis of the cusp by a replica approach}
\label{appendix-replica-cusp}

The solution can be obtained
by closely following the approach of \cite{BMP} which we outline below.
First let us slightly modify how to take the summation over the solutions $\sum_{\rm SP}$.
In our original representation \eq{eq-overlap-replica-zero} there are $n/2$ replicas belonging 
to group '1' and the other $n/2$ replicas belonging to group '2'. 
Then 1 step RSB solution 
is given by mixing up all replicas and then deviding them into clusters of size $m$. 
We note that it is equivalent to say that
there are $n/m$ clusters of size $m$ and 1 step RSB solution is specified by deciding 
which group ('1' or '2') each replica belongs to. Using this mapping we can rewrite
the summation over the solutions in \eq{eq-overlap-replica-zero} as,
\beqa
&& \sum_{\rm SP}
\exp \left(\frac{1}{2m^{2}}\sum_{r,s=1}^{2}  \frac{\delta h_{r}}{\hs} \frac{\delta h_{s}}{\hs}
\sum_{\alpha=(1,1)}^{(1,n/2)}\sum_{\beta=(2,1)}^{(2,n/2)} \delta_{i_{\alpha},i_{\beta}}
\right) \nonumber \\
&& =\sideset{}{'}{\sum}_{\sigma}\exp\left(\frac{1}{2m^{2}} \sum_{a,b}I_{ab} 
\left[\frac{1-\sigma_{a}}{2}\frac{\delta h_{1}}{\hs}+\frac{1+\sigma_{a}}{2}\frac{\delta h_{1}}{\hs}\right]
\left[\frac{1-\sigma_{b}}{2}\frac{\delta h_{2}}{\hs}+\frac{1+\sigma_{b}}{2}\frac{\delta h_{2}}{\hs}\right]
\right).
\eeqa
Here $I_{ab}$ is a standard Parisi's matrix for 1 step RSB 
which is block-diagonal: block size is $m$ and 
the elements are $I=1$ in the block-diagonal part and $I=0$ for the 
rest (See Fig.~\ref{fig-1rsb-matrix}). We also introduced 'Ising variables' $\sigma_{a}= \pm 1$
for $a=1..n$. The summation $\sum_{\sigma}$ means to a trace over the Ising variables 
under the constraint that $\sum_{a=1}^{n}\sigma_{a}=0$.


To simplify notations we introduce $\tilde{h}_{1}=\delta h_{1}/\hs$ and
$\tilde{h}_{2}=\delta h_{2}/\hs$. Using a Hubbard-Stratonovich transformation
$e^{\frac{A}{2}\sum_{a,b} \sigma_{a}\sigma_{b}}=
\int_{-\infty}^{\infty}
\frac{d\lambda}{\sqrt{2\pi/A}}e^{-\frac{\lambda^{2}}{2A}}e^{\sum_{a}\sigma}
$ with $A>0$, we obtain,
\beqa
&& \sideset{}{'}{\sum}_{\sigma} \exp\left(\frac{1}{2m^{2}} \sum_{a,b}I_{ab} 
\left[
 \frac{1-\sigma_{a}}{2}\tilde{h}_{1}
+\frac{1+\sigma_{a}}{2}\tilde{h}_{2}\right]
\left[
 \frac{1-\sigma_{b}}{2}\tilde{h}_{1}
+\frac{1+\sigma_{b}}{2}\tilde{h}_{2}
\right]
\right) \nonumber \\
&& = e^{\frac{n}{m}\frac{(\tilde{h}_{1}+\tilde{h}_{2})^{2}}{8}}
\sideset{}{'}{\sum}_{\sigma}  \prod_{{\cal C}=1}^{n/m} \int_{-\infty}^{\infty}  
\frac{d \lambda_{\cal C}}{\sqrt{2\pi/A}}
\exp \left[ -\frac{\lambda_{\cal C}^{2}}{2A}+
\left (
-\frac{\tilde{h}_{1}^{2}-\tilde{h}_{2}^{2}}{4m}+\lambda_{\cal C} 
\right )
\sum_{a \in {\cal C}}\sigma_{a} \right] \nonumber \\
&& = e^{\frac{n}{m}\frac{(\tilde{h}_{1}+\tilde{h}_{2})^{2}}{8}}
\frac{2^{n+1}}{B(-n/2,-n/2)}
\int_{-\infty}^{\infty} dy
\left[ \int_{-\infty}^{\infty}  
\frac{d \lambda'}{\sqrt{2\pi}}
e^{-\frac{\lambda'^{2}}{2}}
\cosh (y+\sqrt{A}\lambda')\right]^{n/m}
\eeqa
with
\beq
A=\frac{(\tilde{h}_{1}-\tilde{h}_{2})^{2}}{4m^{2}}.
\eeq
Here ${\cal C}=1,2,\ldots,n/m$ are the blocks of the block-diagonal 
matirx $I_{ab}$ and $\sum_{a \in {\cal C}}$ denotes a summation over
replicas beloging to a block ${\cal C}$. To obtain the last equation we used 
the constraint $\sum_{a=1}^{n}\sigma_{a}=0$ and the identity \cite{BMP},
\beq
\sideset{}{'}{\sum}_{\sigma} e^{\sum_{a}h_{a}\sigma_{a}}
=\frac{2^{n+1}}{B(-n/2,-n/2)}\int_{-\infty}^{\infty} dy \cosh (y+ h_{a}).
\eeq
Chaging variablies with $\tilde{h}={\rm sgn}(\tilde{h}_{1}-\tilde{h}_{2})\lambda'$ and $\mu=e^{-2y}$ we can rewrite the integrals as,
\beq
\int_{-\infty}^{\infty} dy
\left[ \int_{-\infty}^{\infty}  
\frac{d \lambda'}{\sqrt{2\pi}}
e^{-\frac{\lambda'^{2}}{2}}
\cosh (y+\sqrt{A}\lambda')\right]^{n/m} \nonumber \\
= e^{\frac{n}{m}\frac{(\tilde{h}_{1}-\tilde{h}_{2})^{2}}{8}}
\frac{1}{2^{n+1}} \int_{0}^{\infty}\frac{d\mu}{\mu}\mu^{-n/2}
\left [
\int_{-\infty}^{\infty}\frac{d\tilde{h}}{\sqrt{2\pi}}
\left \{
e^{-\frac{(\tilde{h}-\tilde{h}_{1})^{2}}{2m}}
+\mu e^{-\frac{(\tilde{h}-\tilde{h}_{2})^{2}}{2m}} \right \}^{m}
\right ]^{n/m}
\eeq

Summing up the above results and using $\lim_{n \to 0} nB(n/2,n/2)=4$
we obtain,
\beqa
&& \lim_{T \to 0}\frac{\tav{Y(h_{0}+\delta h_{1})Y(h_{0}+\delta h_{2})}}
{(\sqrt{N}\Dyeffs(h_{0}))^{2}} \nonumber \\
&& = - \lim_{m \to 0}  \lim_{n \to 0}
\frac{1}{n}
\frac{\partial^{2}}{\partial \left( \frac{\delta h_{1}}{\hs}\right) \partial \left( \frac{\delta  h_{2}}{\hs}\right)}  e^{\frac{n}{4m} (\tilde{h}_{1}^{2}+\tilde{h}_{2}^{2})}
\int_{0}^{\infty}\frac{d\mu}{\mu}\mu^{-n/2}
\left [
\int_{-\infty}^{\infty}\frac{d\tilde{h}}{\sqrt{2\pi}}
\left \{
e^{-\frac{(\tilde{h}-\tilde{h}_{1})^{2}}{2m}}
+\mu e^{-\frac{(\tilde{h}-\tilde{h}_{2})^{2}}{2m}} \right \}^{m}
\right ]^{n/m} \nonumber \\
&& = g_{11}+g_{12}
\label{eq-overlap-replica-result}
\eeqa
with
\beq
g_{11}=\sqrt{\frac{2}{\pi}}e^{-\frac{l^{2}}{8}}
\int_{0}^{\infty}dx 
\frac{e^{-\frac{x^{2}}{2}} (x^{2}-\frac{l^{2}}{4})}{
e^{-\frac{|l|}{2}}M_{0}( x-\frac{|l|}{2})+
e^{ \frac{|l|}{2}}M_{0}(-x-\frac{|l|}{2})
}=1-\frac{3|l|}{2\sqrt{\pi}} + O(|l|^{2})
\eeq
and
\beq
g_{12}=-\frac{|l|}{\pi}e^{-\frac{l^{2}}{4}}\int_{0}^{\infty}dx 
\frac{e^{-x^{2}} e^{-x|l|}}{
\left\{
e^{-\frac{|l|}{2}}M_{0}( x-\frac{|l|}{2})+
e^{ \frac{|l|}{2}}M_{0}(-x-\frac{|l|}{2})
\right\}^{2}
}
=-\frac{|l|}{2\sqrt{\pi}}+ O(|l|^{2})
\eeq
with 
\beq
M_{k}(x)=\int_{x}^{\infty} \frac{dz}{\sqrt{2\pi}}z^{k}e^{-\frac{z^{2}}{2}}
\eeq
and
\beq
l=\frac{\delta h_{1}-\delta h_{2}}{\hs}.
\eeq
In the above equations we changed the integration 
variable using $\mu \equiv e^{-x \frac{|l|}{m}}$.
For the details of the evaluations of the $m \to 0$ limit of the integrals
see \cite{BMP}. 
In order to obtain the last equation of \eq{eq-cusp-replica},
we expanded the integrals by $|l|$ and repeated some partial integrations
and some Gaussian integrals.

\section{Fluctuations of clusters in one-step RSB solutions}
\label{appendix-m-fluctuation}

Let us recalled that a given 1RSB solution is
described as a configuration of $n$ replicas grouped into 
$n/m$ clusters of a unique size $m$. (See Fig.~\ref{fig-m-atom}.)
Being motivated by a work by Campellone et al \cite{CPV}, we now
consider a class of more generic solutions where the sizes of clusters 
can be inhomogeneous. As noted in sec \ref{sec-replica} this
is necessary to evaluate the parameter $\Dyeffs$ 
in the generating functional \eq{eq-generating-functional-p-spin} properly.

The need to take into account the fluctuation of $m$ has been pointed
out by several authros. Especially it has been pointed out  \cite{ThN,FGR} that
if one computes the heat capacity from the thermal fluctuations of energy $E$
via FDT: $(1/N)\beta^{2} \sav{ \tav{E^{2}}-\tav{E}^{2}}$
neglecting fluctuations from the usual 1RSB saddle points,
one does not recover the correct thermodynamic heat-capacity that one obtains
by taking two time derivatives of the thermodynamic free-energy
by temperature $T$. It has been suggested \cite{ThN,FGR,CPV} 
that one then must consider fluctuations from the saddle point, 
in particular, the size of clusters $m$ explicitly. 

Fluctuations of the size of clusters, 
seem rather different in nature from the usual small fluctuations 
\cite{TDK} of the amplitude of each element of the Parisi matrix.
In particular in the case of REM the overlap between the states 
can be either $0$ or $1$ by the construction of the model so that
fluctuation of the size of clusters amount to very large fluctuations
of the amplitudes of the elements of the Parisi matrix \cite{CPV}.

For simplicity we consider below an effective REM with the 
simple generalized complexity given by \eq{eq-gc-KM}.

\subsection{Fluctuations of clusters}

In order to obtain some insights let us consider first disorder-average 
of the replicated partition function given by \eq{eq-replicated-z-KM},
\beq
\sav{Z^{n}}=\sum_{i_{1},i_{2},\ldots,i_{n}=1}^{M=e^{cN}}
\exp \left[ cN  (\beta \Tc(h) )^{2} 
\sum_{\alpha,\beta =1}^{n} \delta_{i_{\alpha},i_{\beta}} 
\right].
\label{eq-pfunc-n}
\eeq

A given configuration of $n$ replicas of the REM can be viewed as
a collection {\it clusters}. A cluster is a  set of replicas sitting
on the same microscopic state among $M=e^{cN}$ possible states.
We label them as ${\cal C}_{1},\ldots,{\cal C}_{l}$ where $l$ is the
total number of clusters. We denote the size of $I$th cluster
, namely the number of replicas belonging to $C_{I}$, as $m_{I}$.

Now using an identity
\bmat
\sum_{\alpha,\beta=1}^{n}\delta_{i_{\alpha},i_{\beta}} = \sum_{I=1}^{l}m_{I}^{2}.
\emat
we can rewrite the replicated partition function as
\beq
\sav{Z^{n}}=\sum_{l} \int c N d\lambda 
\sum_{\{ m_{I}\}} \frac{n!}{\prod_{I=1}^{l}m_{I}! l!}
e^{-NG(\{m_{I}\},\lambda,l)}
\eeq
with
\beq
-G(\{m_{I}\},\lambda,l)=c\sum_{I=1}^{l} \left \{
1 + (\beta \Tc(h) )^{2} m_{I}^{2}  +\lambda m_{I} 
\right \}
-c\lambda n
\label{eq-pfunc-n-2}
\eeq
here we imposed the constraint
\beq
\sum_{I=1}^{l} m_{I}=n
\label{eq-constraint}
\eeq
by introducing a
Lagrange multiplier $\lambda$. The effective action $G$ can be cast
into the following form,
\beq
-\frac{G(\{m_{I}\},\lambda,l)}{c}=
l+(\beta \Tc(h))^{2}\frac{n}{l} n
-\frac{l}{4(\beta \Tc(h))^{2}} (\lambda-\lambda^{*}(l))^{2}
+(\beta \Tc(h))^{2} \sum_{I=1}^{l} (m_{I}-m^{*}(\lambda))^{2}
\eeq
with
\beq
m^{*}(\lambda)=-\frac{\lambda}{2(\beta \Tc(h))^{2}}
\qquad \lambda^{*}(l)=-\frac{n}{l} 2(\beta \Tc(h))^2
\eeq
Thus we find
\beq
\sav{Z^{n}}=\sum_{l} e^{c N [l +\frac{n}{l}n  (\beta \Tc(h))^2]}
\int c N d\lambda e^{-\frac{c N l}{4 (\beta \Tc(h))^2}(\lambda-\lambda^{*}(l))^2}
\sum_{\{ m_{I}\}} 
\frac{n!}{\prod_{I=1}^{l}m_{I}! l!}
\prod_{I=1}^{l} e^{cN (\beta \Tc(h))^2(m_{I}-m^{*}(\lambda)^{2}}.
\label{eq-pfunc-n-3}
\eeq
Here the order of the summations and integrals is crucial. 
In order to
enforce the constraint \eq{eq-constraint}, $G$ must be maximized
with respect to $\lambda$. On the other hand, $G$ is {\it minimized}
with respect to $\{m_{I}\}$ at $m_{I}=m^{*}(\lambda)$ and $l=l^*$ given 
by 
\beq
l^{*}=n \frac{T}{\Tc(h)}.
\eeq
If we disregard fluctuations completely we recover \eq{eq-m},
\beq
m_{i} \simeq m^{*}(\lambda^{*}(l^{*}))= \frac{T}{\Tc(h)}
\label{eq-m-2}
\eeq
which is the well known result. Note that usual extremization
of the uniform cluster size $m$ (See \eq{eq-extremization-m} 
is equivalent here to extremization of the number of clusters $l$.

In the following we denote fluctuation of the size 
of cluster ${\cal C}_{I}$ as 
\beq
\delta m_{I}=m_{I}-m^{*}(\lambda).
\eeq
From \eq{eq-pfunc-n-3} we find that 
measure for small fluctuations of the sizes of individual clusters 
$\delta m_{I}$ is given by,
\beq
\prod_{I=1}^{l}\int dm_{I}  e^{cN (\beta \Tc(h))^2 \delta m_{I}^{2}} \ldots ..
\label{eq-measure-fluc-m}
\eeq
The saddle point $\delta m_{I}=0$ is the {\it minimum} which is a strange
feature in the $n \to 0$ limit \cite{REM,GM}.

\subsection{Generating functional for correlation functions of energy}

Based on the above observations, let us now construct a generating 
functional for the correlation functions of energy $E$.
Let us consider $p$ real replicas $r=1,\ldots,p$, which are 
replicated further into $n_{r}$ replicas,  at inverse temperatures 
$\beta+\delta \beta_{r}$.   Here $\delta \beta_{r}$ is considered
as small 'probing fields'. The disorder-averaged partition function 
of the whole system is given by,
\beq
\sav{Z_{1}^{n_{1}}\cdots Z_{p}^{n_{p}}}=
\sum_{i_{(1,1)},\ldots,i_{(p,n_{p})}=1}^{M=e^{cN}}
\exp \left[ 
cN  \sum_{i=1}^{M}
\left(\sum_{r=1}^{p}
(\beta+\delta \beta_{r}) \Tc(h)
\sum_{\alpha=(r,1)}^{(r,n_{r})}
\delta_{i,i_{\alpha}}
\right)^{2}
\right]
\eeq

In a given configuration of the system, $n_{\rm tot}=n_{1}+\ldots n_{p}$ replicas
are divided into $l$ clusters of sizes $m_{I}$ $(I=1,\ldots,l)$.
Then we find the exponent in the r.h.s. of the last equation can be expanded as,
\beqa
&& cN\sum_{i=1}^{M}
\left(\sum_{r=1}^{p}
(\beta+\delta \beta_{r}) \Tc(h)
\sum_{\alpha=(r,1)}^{r,n_{r}}
\delta_{i,i_{\alpha}}
\right)^{2}= cN
 (\beta \Tc(h))^{2} \sum_{I=1}^{l}m_{I}^{2} \nonumber \\
&& + 2cN (\beta \Tc(h))\sum_{r=1}^{p} \delta \beta_{r} \Tc(h) \sum_{\alpha=(r,1)}^{(r,n_{r})}
m_{I(\alpha)}
+cN\sum_{r,s=1}^{p}(\delta \beta_{r} \Tc(h))(\delta \beta_{s} \Tc(h))
\sum_{\alpha=(r,1)}^{(r,n_{r})}\sum_{\beta=(s,1)}^{(s,n_{s})}
\delta_{i_{\alpha},i_{\beta}}
\label{eq-exponent-expansion}
\eeqa
Here $I(\alpha)$ used in the 2nd term on the r.h.s. 
denotes the label of the cluster to which the replica $\alpha$ belong to. 

Apparently the 3rd term on the r.h.s. of the above equation generates 
a 2nd order term $O((\delta \beta)^{2})$ 
in the effective action of the generating
functional quite similar to the one in \eq{eq-generating-functional-p-spin}.
It can yield a spurious finite heat capacity 
and a spurious temperature-chaos below $\Tc(h)$.  
However we know for sure that $O(N)$ heat capacity is $0$ (See \eq{eq-c-0})
and temperature-chaos is absent in REM since temperature-changes
cannot induce any level crossings of free-energies in the origianl
version of the REM (See \cite{kurkova} for a related discussion). 
Thus in the present REM the $O((\delta \beta)^{2})$ term  must be cancelled.

Now we consider effects of fluctuations of the size of clusters. 
Note that the 2nd term on the r.h.s. of \eq{eq-exponent-expansion}  can be
decomposed as,
\beq
2cN (\beta \Tc(h))\sum_{r=1}^{p} \delta \beta_{r} \Tc(h) \sum_{\alpha=(r,1)}^{(r,n_{r})}
m_{I(\alpha)}=
2cN (\beta \Tc(h))\sum_{r=1}^{p} \delta \beta_{r} \Tc(h) n_{r}
m^{*}(\lambda)
+2cN (\beta \Tc(h))\sum_{I=1}^{l} \delta m_{I}
\sum_{\alpha \in {\cal C}_{I}} \delta \beta_{\alpha} \Tc(h).
\label{eq-coupling-m-fluctuation}
\eeq
where $\delta \beta_{\alpha}$ is the {\it probing field} applied to 
replica $\alpha$.
Here we find the  probing field 
$\delta \beta_{r}$ in a cluster $C_{I}$ is linearly coupled 
to fluctuation of $m_{I}$.

Exponentiating the 2nd term in the r.h.s. of \eq{eq-coupling-m-fluctuation} 
and applying the measure for the fluctuation of $\delta m_{I}$ given in
\eq{eq-measure-fluc-m} we obtain,
\beqa
&& \prod_{I=1}^{l} \int d \delta m_{I}  e^{cN (\beta \Tc(h))^2 \delta m_{I}^{2}} 
\exp \left [ 
2 cN(\beta \Tc(h))\sum_{I=1}^{l} \delta m_{I}
\sum_{\alpha \in {\cal C}_{I}} \delta \beta_{\alpha} \Tc(h)  
\right] \nonumber \\
&& = \exp \left[ 
-cN\sum_{I=1}^{l}\sum_{\alpha \in C_{I}, \beta \in C_{I}} (\delta \beta_{\alpha}\Tc(h))(\delta \beta_{\beta}\Tc(h))\right] 
\prod_{I=1}^{l} \int d \delta m'_{I}  e^{cN (\beta \Tc(h))^2 
(\delta m'_{I})^{2}}  \nonumber \\
&& =\exp \left[ 
-cN \sum_{r,s=1}^{p} (\delta \beta_{r}\Tc(h))(\delta \beta_{s}\Tc(h))
\sum_{\alpha=(r,1)}^{(r,n_{r})}\sum_{\beta=(s,1)}^{(s,n_{s})}
\delta _{i_{\alpha},i_\beta}
\right]\prod_{I=1}^{l} \int d \delta m'_{I}  
e^{cN (\beta \Tc(h))^2 (\delta m'_{I})^{2}}  
\eeqa
in the last equations we changed the integration variable
to $\delta m'_{I}=\delta m_{I}+\sum_{\alpha \in C_{I}} 
\frac{\delta \beta_{\alpha}}{\beta}$.
The last result implies we have another 2nd order term in the action
\bmat
-cN \sum_{r,s=1}^{p} (\delta \beta_{r}\Tc(h))(\delta \beta_{s}\Tc(h))
\sum_{\alpha=(r,1)}^{(r,n_{r})}\sum_{\beta=(s,1)}^{(s,n_{s})}
\delta _{i_{\alpha},i_\beta}
\emat
which exactly cancels the other 
2nd order term in the action, i.e. 3rd term on the r.h.s. of 
\eq{eq-exponent-expansion}.  In another word, the effective variable
\eq{eq-gc-around-saddle} of the energies is
\beq
\Dyeffs=0.
\eeq

As the result we obtain the generating function for the fluctuations
of energies as,
\beqa
 \sav{Z_{1}^{n_{1}}Z_{2}^{n_{2}}\ldots Z_{p}^{n_{p}}}= &&
\exp \left [cN \frac{\nt}{m} \left\{ 1+ \left(\frac{\Tc(h)}{T/m}\right)^{2} \right\} \right]  
\spsum \exp \left[
2cN \sum_{r=1}^{p} \delta \beta_{r} \Tc(h) n_{r}
\right].
\label{eq-zp-1rsb-appendix-ene}
\eeqa
where the sum $\spsum$ stands for summation over all saddle points
where $\nt$ replicas are grouped into $\nt/m$ clusters of size $m$.
Here $m$ is the saddle point value given in \eq{eq-m}
(See also \eq{eq-m-2}). The $O(\delta \beta)$ term in the 2nd exponent is
due to the 1st term in the r.h.s. of \eq{eq-coupling-m-fluctuation}.

It can be seen easily that the $O(\delta \beta)$ term in the action 
generates the correct average energy 
\beq
E/N= \frac{\partial \beta F}{\partial \beta}=-2c \Tc(h)
\eeq
in agreement with the one obtained from the thermodynamic free-energy
(See \eq{eq-f-glass}). The absence of the 2nd order term $O(\delta \beta)^{2}$
in the action means that heat capacity is indeed zero at $O(N)$ and also 
absence of temperature-chaos, as it should be in REM.

\subsection{Generating functional for correlation functions of $Y$}

Now let us consider fluctuations of the variable $Y$ and construct
a generating functional for its correlation functions. 

We consider again $p$ real replicas $r=1,\ldots,p$, which are 
replicated further into $n_{r}$ replicas,  at the same temperature but
at slightly different fields $h+\delta h_{r}$.
\beq
\sav{Z_{1}^{n_{1}}\cdots Z_{p}^{n_{p}}}=
\sum_{i_{(1,1)},\ldots,i_{(p,n_{p})}=1}^{M=e^{cN}}
\exp \left[ 
cN  (\beta \Tc^{0})^{2} \sum_{\alpha,\beta} \delta_{i_{\alpha},i_{\beta}}
+\frac{N \Dy^{2}}{2} \sum_{i=1}^{M}
\left(\sum_{r=1}^{p}
\beta( h + \delta h_{r}) 
\sum_{\alpha=(r,1)}^{(r,n_{r})}
\delta_{i,i_{\alpha}}
\right)^{2}
\right]
\eeq
Note $\Tc^{0}$ is the critical temperature at $h=0$ given by \eq{eq-Tc}
and $\Dy$ is the variance of the original $Y$ variable.

The 2nd term in the exponent of the r.h.s of the last equation 
can be expanded as,
\beqa
&& \frac{N \Dy^{2}}{2}\sum_{i=1}^{M}
\left(\sum_{r=1}^{p}
\beta(h+\delta h_{r})
\sum_{\alpha=(r,1)}^{r,n_{r}}
\delta_{i,i_{\alpha}}
\right)^{2}= \frac{N \Dy^{2}}{2}
 (\beta h)^{2} \sum_{I=1}^{l}m_{I}^{2} \nonumber \\
&& + N \Dy^{2} (\beta h)\sum_{r=1}^{p} (\beta \delta h_{r}) \sum_{\alpha=(r,1)}^{(r,n_{r})}
m_{I(\alpha)}
+\frac{N \Dy^{2}}{2}
\sum_{r,s=1}^{p}(\beta \delta h_{r} )(\beta  \delta h_{s})
\sum_{\alpha=(r,1)}^{(r,n_{r})}\sum_{\beta=(s,1)}^{(s,n_{s})}
\delta_{i_{\alpha},i_{\beta}}
\label{eq-exponent-expansion-y}
\eeqa
Note that the 2nd term on the r.h.s. of \eq{eq-exponent-expansion-y} can be
decomposed as,
\beq
N \Dy^{2}(\beta h)
\sum_{r=1}^{p} \beta \delta h_{r} \sum_{\alpha=(r,1)}^{(r,n_{r})}
m_{I(\alpha)}=
N \Dy^{2}(\beta h)\sum_{r=1}^{p} \delta \beta_{r} n_{r}
m^{*}(\lambda)
+N \Dy^{2}(\beta h)\sum_{I=1}^{l} \delta m_{I}
\sum_{\alpha \in {\cal C}_{I}} \beta \delta h_{\alpha}.
\label{eq-coupling-m-fluctuation-y}
\eeq
where $\delta h_{\alpha}$ is the probing field applied to 
replica $\alpha$. Here again we find the probing field 
$\delta h_{r}$ in a cluster $C_{I}$ is linearly coupled 
to fluctuation of $m_{I}$. 

It can be seen that the coupling between the fluctuations of the
size of clusters and $\delta h_{r}$ is absent {\it only}
in the special case that the applied external field $h=0$.
Note also that such a situation never happens for the
case of energies (See \eq{eq-coupling-m-fluctuation-y}) in the 
whole glass phase $T < \Tc(h)$.

Exponentiating the 2nd term in the r.h.s. of \eq{eq-coupling-m-fluctuation-y} 
and applying the measure for the fluctuation of $\delta m_{I}$ given in
\eq{eq-measure-fluc-m} we obtain,
\beqa
&& \prod_{I=1}^{l} \int d \delta m_{I} 
e^{cN (\beta \Tc(h))^2 \delta m_{I}^{2}} 
\exp \left [ 
N \Dy^{2} (\beta h)\sum_{I=1}^{l} \delta m_{I}
\sum_{\alpha \in {\cal C}_{I}} \beta \delta h_{\alpha}
\right] \nonumber \\
&& = \exp \left[ 
-\frac{N \Dy^{4}}{4c} \left ( \frac{h}{\Tc(h)} \right)^{2}
\sum_{I=1}^{l}\sum_{\alpha \in C_{I}, \beta \in C_{I}} (\beta \delta h_{\alpha})(\beta \delta h_{\beta})\right] 
\prod_{I=1}^{l} \int d \delta m'_{I}  e^{cN (\beta \Tc(h))^2 
(\delta m'_{I})^{2}}  \nonumber \\
&& =\exp \left[ 
-\frac{N \Dy^{4}}{4c} \left ( \frac{h}{\Tc(h)} \right)^{2}
\sum_{r,s=1}^{p} (\beta \delta h_{r})(\beta \delta h_{s})
\sum_{\alpha=(r,1)}^{(r,n_{r})}\sum_{\beta=(s,1)}^{(s,n_{s})}
\delta _{i_{\alpha},i_\beta}
\right]\prod_{I=1}^{l} \int d \delta m'_{I}  
e^{cN (\beta \Tc(h))^2 (\delta m'_{I})^{2}}  
\eeqa
in the last equations we changed the integration variable
to $\delta m'_{I}=\delta m_{I}+\sum_{\alpha \in C_{I}} 
\frac{\Dy^{2}}{2c(\beta \Tc(h))^{2}} (\beta h) (\beta \delta h_{\alpha})$.
The last result implies we have another 2nd order term in the action
\bmat
-\frac{N \Dy^{4}}{4c} \left ( \frac{h}{\Tc(h)} \right)^{2}
\sum_{r,s=1}^{p} (\beta \delta h_{r})(\beta \delta h_{s})
\sum_{\alpha=(r,1)}^{(r,n_{r})}\sum_{\beta=(s,1)}^{(s,n_{s})}
\delta _{i_{\alpha},i_\beta}
\emat
Combining with the 3rd term on the r.h.s. of \eq{eq-exponent-expansion-y}
and using \eq{eq-Tc}
we obtain the total $O(\delta h)^{2}$ term to be,
\bmat
\frac{N \Dyeff^{2}}{2} \sum_{r,s=1}^{p} (\beta \delta h_{r})(\beta \delta h_{s})
\sum_{\alpha=(r,1)}^{(r,n_{r})}\sum_{\beta=(s,1)}^{(s,n_{s})}
\delta _{i_{\alpha},i_\beta}
\emat
with ``renormalized variance''
\beq
\Dyeffs \equiv 
\frac{\Dy}{\sqrt{1+  (h \Dy})^{2}}
\eeq
Note that this coincides with \eq{eq-dy-eff}.

Now collecting also the 1st term in the r.h.s of 
\eq{eq-coupling-m-fluctuation-y} we finally obtain,
\beqa
 \sav{Z_{1}^{n_{1}}Z_{2}^{n_{2}}\ldots Z_{p}^{n_{p}}}= &&
\exp \left [cN \frac{\nt}{m} \left\{ 1+ \left(\frac{\Tc(h)}{T/m}\right)^{2} \right\} \right]  \nonumber \\
&& \sum_{\rm SP} \exp \left[
N \Dy^{2}m(\beta h)\sum_{r=1}^{p} \delta \beta_{r} n_{r} 
+
\frac{N}{2} \Dyeff^{2} \sum_{r,s=1}^{p} \beta \delta h_{r} \beta \delta h_{s}
\sum_{\alpha=(r,1)}^{(r,n_{r})}\sum_{\beta=(s,1)}^{(s,n_{s})}
\delta_{i_{\alpha},i_{\beta}}
\right].
\label{eq-zp-1rsb-appendix-y}
\eeqa

It can be seen easily that the $O(\delta h)$ term in the action 
generates the average 
\beq
\sav{\tav{Y}}= -\frac{\partial F}{\partial h}=
N \Dy^{2}\frac{T}{\Tc(h)}
\eeq
with $\Dy$ being the variance of the original $Y$ variable.
It agrees precisely with the thermodynamic one \eq{eq-y-av-glass}).
Moreover the $O((\delta h)^{2})$ generates the average linear susceptibility
which also agrees with the thermodynamic one \eq{eq-chi-thermodynamic}.


\end{document}